\documentclass[a4paper,11pt]{article}
\pdfoutput=1
\usepackage{jheppub}

\newcommand{\ii}{\mathrm{i}}
\newcommand{\rme}{\mathrm{e}}

\newcommand{\vev}[1]{\langle #1 \rangle}
\newcommand{\Tr}{\mathrm{Tr}\,}

\newcommand{\cN}{{\mathcal{N}}}

\newcommand{\one}{{\rm 1\kern -.9mm l}}

\newcommand{\ft}[2]{{\textstyle\frac{#1}{#2}}}

\newcommand{\be}{\begin{equation}}
\newcommand{\ee}{\end{equation}}
\newcommand{\p}{\partial}

\newcommand\undermat[2]{%
\makebox[0pt][l]{$\smash{\underbrace{\phantom{%
    \begin{matrix}#2\end{matrix}}}_{\text{$#1$}}}$}#2}
\newdimen\tableauside\tableauside=1.0ex
\newdimen\tableaurule\tableaurule=0.4pt
\newdimen\tableaustep
\def\phantomhrule#1{\hbox{\vbox to0pt{\hrule height\tableaurule
width#1\vss}}}
\def\phantomvrule#1{\vbox{\hbox to0pt{\vrule width\tableaurule
height#1\hss}}}
\def\sqr{\vbox{%
  \phantomhrule\tableaustep
\hbox{\phantomvrule\tableaustep\kern\tableaustep\phantomvrule\tableaustep}%
  \hbox{\vbox{\phantomhrule\tableauside}\kern-\tableaurule}}}
\def\squares#1{\hbox{\count0=#1\noindent\loop\sqr
  \advance\count0 by-1 \ifnum\count0>0\repeat}}
\def\tableau#1{\vcenter{\offinterlineskip
  \tableaustep=\tableauside\advance\tableaustep by-\tableaurule
  \kern\normallineskip\hbox
    {\kern\normallineskip\vbox
      {\gettableau#1 0 }%
     \kern\normallineskip\kern\tableaurule}%
  \kern\normallineskip\kern\tableaurule}}
\def\gettableau#1 {\ifnum#1=0\let\next=\null\else
  \squares{#1}\let\next=\gettableau\fi\next}
\newcommand{\Yfund}{\tableau{1}}
\newcommand{\Ysymm}{\tableau{2}}
\newcommand{\Yasymm}{\tableau{1 1}}

\def\XXint#1#2#3{{\setbox0=\hbox{$#1{#2#3}{\int}$}
     \vcenter{\hbox{$#2#3$}}\kern-.5\wd0}}

\title{\boldmath Modular and duality properties of surface operators in $\mathcal{N}=2^{\star}$ gauge theories}

\author[a,b]{S.~K.~Ashok,}
\affiliation[a]{Institute of Mathematical Sciences \\
   C.~I.~T.~Campus, Taramani\\
   Chennai, India 600113}
   
   \affiliation[b]{Homi Bhabha National Institute.\\
Training School Complex, Anushakti Nagar,\\ 
Mumbai, India 400085}
\emailAdd{sashok@imsc.res.in}

\author[c]{M.~Bill\`o,}
\affiliation[c]{Universit\`a di Torino, Dipartimento di Fisica
\\ and I.~N.~F.~N.~- sezione di Torino, 
Via P. Giuria 1, I-10125 Torino, Italy}
\emailAdd{billo@to.infn.it}

\author[a,b]{E.~Dell'Aquila,}
\emailAdd{edellaquila@imsc.res.in}

\author[c]{M.~Frau,}
\emailAdd{frau@to.infn.it}

\author[a,b]{R.~R.~John,}
\emailAdd{renjan@imsc.res.in}

\author[d]{and A.~Lerda}
\affiliation[d]{Universit\`a del Piemonte Orientale, Dipartimento di Scienze e Innovazione Tecnologica, \\
and I.~N.~F.~N.~- sezione di Torino, Via P. Giuria 1, I-10125 Torino, Italy}
\emailAdd{lerda@to.infn.it}

\abstract{
We calculate the instanton partition function of the four-dimensional $\cN=2^{\star}$ SU$(N)$ gauge theory
in the presence of a generic surface operator, using equivariant localization.
By analyzing the constraints that arise from S-duality, we show that the effective twisted 
superpotential, which governs the infrared dynamics of the 
two-dimensional theory on the surface operator, satisfies a modular anomaly equation.
Exploiting the localization results, we solve this equation in terms of elliptic and quasi-modular forms 
which resum all non-perturbative corrections. 
We also show that our results, derived for monodromy defects in the four-dimensional theory, match the effective twisted superpotential describing the infrared properties of certain two-dimensional 
sigma models coupled either to pure ${\mathcal N} = 2$ or to ${\mathcal N} = 2^{\star}$ gauge theories. 
}


\preprint{ }

\begin{document}
\maketitle
\flushbottom
\section{Introduction}
\label{intro}
The study of how a quantum field theory responds to the presence of defects is a very important subject, which
has received much attention in recent years especially in the context of supersymmetric gauge theories.
In this paper we study a class of two-dimensional defects, also known as surface operators,
on the Coulomb branch of the ${\mathcal N}=2^{\star}$ SU$(N)$ gauge theory in four 
dimensions\,\footnote{For a review of surface operators see \cite{Gukov:2014gja}.}.
Such surface operators can be introduced and analyzed in different ways. 
They can be defined by the transverse singularities they induce in the four-dimensional 
fields \cite{Gukov:2006jk, Gukov:2008sn}, or can be
characterized by the two-dimensional theory they support on their world-volume
\cite{Gaiotto:2009fs,Gaiotto:2013sma}. 

A convenient way to describe four-dimensional gauge theories with $\mathcal{N}=2$ supersymmetry is to
consider M5 branes wrapped on a  punctured Riemann surface \cite{Witten:1997sc, Gaiotto:2009we}.
{From} the point of view of the six-dimensional $(2,0)$ theory on the M5 branes, surface 
operators can be realized by means of either M$5^\prime$ or M2 branes giving rise, respectively, 
to codimension-2 and codimension-4 defects. 
While a codimension-2 operator extends over the Riemann surface wrapped by the M5 brane realizing the gauge 
theory, a codimension-4 operator intersects the Riemann surface at a point.
Codimension-2 surface operators were systematically studied in \cite{Alday:2010vg} where, 
in the context of the of the $4d$/$2d$ correspondence \cite{Alday:2009aq}, 
the instanton partition functions of $\mathcal{N}=2$ SU(2) super-conformal quiver theories with
surface operators were mapped to the conformal blocks of a two-dimensional conformal field theory with 
an affine sl(2) symmetry. These studies were later extended to SU($N$) quiver theories 
whose instanton partition functions in the presence of surface operators were related to conformal field 
theories with an affine sl($N$) symmetry \cite{Kozcaz:2010yp}. The study of codimension-4 surface 
operators was pioneered in \cite{Alday:2009fs} where the instanton partition function of the 
conformal SU(2) theory with a surface operator was mapped to the Virasoro blocks of the Liouville 
theory, augmented by the insertion of a degenerate primary field. Many generalizations and 
extensions of this have been considered in the last 
few years \cite{Taki:2009zd,Marshakov:2010fx,Kozcaz:2010af,Dimofte:2010tz,Maruyoshi:2010iu,Taki:2010bj,Bullimore:2014awa,Assel:2016wcr}. 

Here we study ${\mathcal N}=2^{\star}$ theories in the presence of surface operators. The 
low-energy effective dynamics of the bulk four-dimensional theory is completely encoded in 
the holomorphic prepotential which at the non-perturbative level can be very efficiently determined 
using localization \cite{Nekrasov:2002qd} along with the constraints 
that arise from S-duality. The latter turn out to imply \cite{Billo:2013fi,Billo:2013jba} a 
modular anomaly equation \cite{Minahan:1997if} for the prepotential, which is intimately related 
to the holomorphic anomaly equation occurring in topological string theories on local 
Calabi-Yau manifolds 
\cite{Bershadsky:1993ta,Witten:1993ed,Aganagic:2006wq,Gunaydin:2006bz}\,\footnote{Modular anomaly equations have been studied in various contexts, such as 
the $\Omega$-background \cite{Billo:2013fi,
Billo:2013jba,Huang:2006si,Grimm:2007tm,Huang:2009md,Huang:2010kf,Galakhov:2012gw,
Nemkov:2013qma,Billo:2014bja}, the $4d$/$2d$ 
correspondence \cite{KashaniPoor:2012wb,Kashani-Poor:2013oza,Kashani-Poor:2014mua}, SQCD theories 
with fundamental matter \cite{Billo:2013fi,Billo:2013jba,Ashok:2015cba,Ashok:2016oyh,Beccaria:2016vxq} 
and in ${\mathcal N}=2^{\star}$ theories \cite{Billo:2013fi, Billo:2013jba, Billo':2015ria,
Billo':2015jta,Beccaria:2016nnb,Ashok:2016ewb}.}. 
Working perturbatively in the mass of the adjoint hypermultiplet, the modular anomaly equation allows 
one to resum all instanton corrections to the prepotential into (quasi)-modular forms, and to 
write the dependence on the Coulomb branch parameters in terms of particular sums over the roots 
of the gauge group, thus making it possible to treat any semi-simple algebra \cite{Billo':2015ria, Billo':2015jta}. 

In this paper we apply the same approach to study the effective twisted superpotential which governs 
the infrared dynamics on the world-volume of the two-dimensional surface operator in the 
$\mathcal{N}=2^\star$ theory. For simplicity, we limit ourselves to SU($N$) gauge groups 
and consider half-BPS surface defects that, from the six-dimensional point of view, are codimension-2 operators. 
These defects introduce singularities characterized by the pattern of gauge symmetry breaking, {\it{i.e.}} 
by a Levi decomposition of SU($N$), and also by a set of continuous (complex) parameters. 
In \cite{Kanno:2011fw} it has been shown that the effect of these surface operators on the instanton 
moduli action is equivalent to a suitable orbifold projection which produces structures known 
as ramified instantons \cite{mehta1980,biswas1997,Kanno:2011fw}.
Actually, the moduli spaces of these ramified instantons were already studied in \cite{Feigin} 
from a mathematical point of view in terms of representations of a quiver that can be 
obtained by performing an orbifold projection of the usual ADHM moduli space of the standard instantons.
In Section~\ref{SOmicro} we explicitly implement such an orbifold procedure on the 
non-perturbative sectors of the theory realized by means of systems of D3 and D($-1$)
branes \cite{Douglas:1995bn,Billo:2002hm}.
In Section~\ref{app:micro} we carry out the integration on the ramified instanton moduli via equivariant 
localization. The logarithm of the resulting partition function exhibits both a $4d$ and a 
$2d$ singularity in the limit of vanishing $\Omega$ deformations\,\footnote{We actually calculate the effective superpotential in the Nekrasov-Shatashvili 
limit \cite{Nekrasov:2009rc} in which only one of the $\Omega$-deformation parameters is turned on.}. 
The corresponding residues are regular in this limit and encode,
respectively, the prepotential $\mathcal{F}$ and the twisted superpotential $\mathcal{W}$. 
The latter depends, in addition to the Coulomb vacuum expectation values and the adjoint mass, 
on the continuous parameters of the defect. 

In Section~\ref{secn:modaneq} we show that, as it happens for the prepotential, the constraints 
arising from S-duality lead to a modular anomaly equation for $\mathcal{W}$. 
In Section~\ref{SU2}, we solve this equation explicitly for the SU$(2)$ theory and prove that the 
resulting $\mathcal{W}$ agrees with the twisted superpotential obtained in \cite{KashaniPoor:2012wb}
in the framework of the  $4d$/$2d$ correspondence with the insertion of 
a degenerate field in the Liouville theory. 
Since this procedure is appropriate for codimension-4 defects \cite{Alday:2009fs}, the agreement we find
supports the proposal of a duality between the two classes of defects recently 
put forward in \cite{Frenkel:2015rda}.
In Section~\ref{SUN}, we turn our attention to generic surface operators in the SU$(N)$ 
theory and again, order by order in the adjoint mass, solve the modular anomaly equations in terms of 
quasi-modular elliptic functions and sums over the root lattice. 

We also consider the relation between our findings and what is known for surface defects defined 
through the two-dimensional theory they support on their world-volume.
In \cite{Gaiotto:2013sma} the coupling of the sigma-models defined on such defects to a large class 
of four-dimensional gauge theories was investigated and the twisted superpotential governing 
their dynamics was obtained. 
Simple examples for pure $\mathcal{N}=2$ SU$(N)$ gauge theory include the linear 
sigma-model on $\mathbb{C P}^{N-1}$, that corresponds to the so-called simple defects
with Levi decomposition of type $\{1,N-1\}$, and sigma-models on Grassmannian manifolds corresponding 
to defects of type $\{p,N-p\}$.
The main result of \cite{Gaiotto:2013sma} is that
the Seiberg-Witten geometry of the four-dimensional theory can be recovered by analyzing how the vacuum 
structure of these sigma-models is fibered over the Coulomb moduli space.
Independent analyses based on the $4d$/$2d$ correspondence 
also show that the twisted  superpotential for the simple surface operator is related to the line integral of 
the Seiberg-Witten differential over the punctured Riemann surface \cite{Alday:2009fs}. 
In Section~\ref{secn:IRduality}, we test this claim in detail by considering first the 
pure ${\mathcal N}=2$ gauge theory. Since this theory can be recovered 
upon decoupling the massive adjoint hypermultiplet, we take the decoupling limit on our 
$\mathcal{N}=2^\star$ results for $\mathcal{W}$ and precisely reproduce those findings. 
Furthermore, we show that for simple surface defects 
the relation between the twisted superpotential and the line integral of the Seiberg-Witten differential 
holds prior to the decoupling limit, {\it{i.e.}} in the $\mathcal{N}=2^{\star}$ theory itself.
The agreement we find provides evidence for the proposed duality between the two types of 
descriptions of the surface operators.

Finally, in Section~\ref{secn:concl} we present our conclusions and discuss possible future perspectives.
Some useful technical details are provided in four appendices.

\vspace{0.5cm}
\section{Instantons and surface operators in $\mathcal{N}=2^\star$ SU($N$) gauge theories}
\label{SOmicro}

The $\mathcal{N}=2^\star$ theory is a four-dimensional gauge theory with $\mathcal{N}=2$ supersymmetry 
that describes the dynamics of a vector multiplet and a massive hypermultiplet in the adjoint representation. 
It interpolates between the $\mathcal{N}=4$ super Yang-Mills theory, to which it 
reduces in the massless limit, and the pure $\mathcal{N}=2$ theory, which is recovered 
by decoupling the matter hypermultiplet.
In this paper, we will consider for simplicity only special unitary gauge groups SU$(N)$.
As is customary, we combine the Yang-Mills coupling constant $g$ 
and the vacuum angle $\theta$ into the complex coupling
\begin{equation}
\tau = \frac{\theta}{2\pi}+\ii\,\frac{4\pi}{g^2}~,
\label{tau}
\end{equation}
on which the modular group SL($2,\mathbb{Z}$) acts in the standard fashion:
\begin{equation}
\tau~~\to~~\frac{a\tau+b}{c\tau+d}
\label{modular}
\end{equation}
with $a,b,c,d \in \mathbb{Z}$ and $ad-bc=1$. In particular under S-duality we have
\begin{equation}
S(\tau) = -\frac{1}{\tau}~.
\label{Sontau}
\end{equation}

The Coulomb branch of the theory is parametrized by the vacuum expectation value of the adjoint 
scalar field $\phi$ in the vector multiplet, which we take to be of the form
\begin{equation}
\langle \phi \rangle = \mathrm{diag}(a_1,a_2,\cdots,a_N)\quad\quad\text{with}~~~\sum_{u=1}^Na_u=0~.
\label{vev}
\end{equation}
The low-energy effective dynamics on the Coulomb branch is entirely described by a single 
holomorphic function $\mathcal{F}$, called the prepotential, which contains a classical 
term, a perturbative 1-loop contribution and a tail of instanton corrections. The latter 
can be obtained from the instanton partition function
\begin{equation}
Z_{\text{inst}} =\sum_{k=0}^{\infty} q^k\, Z_{k}
\label{Zinst}
\end{equation}
where 
\begin{equation}
q= \rme^{2\pi\ii\tau}
\label{q}
\end{equation}
and $Z_k$ is the partition function in the $k$-instanton sector that can be explicitly 
computed using localization methods\,\footnote{Our conventions are such that $Z_0=1$.}.
For later purposes, it is useful to recall that the weight $q^k$ in (\ref{Zinst}) originates
from the classical instanton action
\begin{equation}
S_{\text{inst}}= -2\pi\ii\tau\left(\frac{1}{8\pi^2}\int_{\mathbb{R}^4}\Tr F\wedge F\right) = -2\pi\ii\tau\,k
\label{Sinst}
\end{equation}
where in the last step we used the fact that the second Chern class of the gauge field strength $F$ equals
the instanton charge $k$. Hence, the weight $q^k$ is simply $\rme^{-S_{\text{inst}}} $.

Let us now introduce a surface operator which we view as a non-local defect $D$ supported 
on a two-dimensional plane inside the four-dimensional (Euclidean) space-time 
(see Appendix \ref{flux_cartan} for more details).
In particular, we parametrize $\mathbb{R}^4\simeq \mathbb{C}^2$ by two complex variables 
$(z_1,z_2)$, and place $D$ at $z_2=0$, filling the $z_1$-plane. The presence of the surface 
operator induces a singular behavior in the gauge connection $A$, which has the following 
generic form \cite{Alday:2010vg,Kanno:2011fw}:
\begin{equation}
\label{Asing}
A=A_{\mu}\, dx^{\mu}\,\simeq\,-\,\text{diag}
\left(
\begin{array}{cccccccc}
\undermat{n_1}{\gamma_1,\cdots,\gamma_1},\undermat{n_2}{\gamma_2,\cdots,\gamma_2},\cdots, 
\undermat{n_M}{\gamma_M,\cdots,\gamma_M}  
\end{array}
\right)\,d\theta
\vspace{.5cm}
\end{equation}
as $r\to0$. Here $(r,\theta)$ denotes the set of polar coordinates in the $z_2$-plane, and the $\gamma_I$'s are 
constant parameters, where $I=1,\cdots,M$. The $M$ integers $n_I$ satisfy 
\begin{equation}
\sum_{I=1}^M n_I = N
\end{equation}
and define a vector $\vec n$ that identifies the type of the surface operator. This vector
is related to the breaking pattern of the gauge group (or Levi decomposition) felt on the two-dimensional
defect $D$, namely
\begin{equation}
\label{split}
\mathrm{SU}(N) \to \mathrm{S}\big[\mathrm{U}(n_1)\times \mathrm{U}(n_2)\times \cdots \times \mathrm{U}(n_M)\big]~.
\end{equation} 
The type $\vec n=\{1,1,\cdots,1\}$ corresponds to what are called full surface operators, originally considered in 
\cite{Alday:2010vg}. The type $\vec n=\{1,N-1\}$ corresponds to simple 
surface operators,
while the type $\vec n=\{N\}$ corresponds to no surface operators and hence will not be considered. 

In the presence of a surface operator, one can turn on magnetic fluxes for each factor of the
gauge group (\ref{split}) and thus the instanton action can receive
contributions also from the corresponding first Chern classes.
This means that (\ref{Sinst}) is replaced by \cite{Gukov:2006jk,Alday:2009fs, Alday:2010vg,Kanno:2011fw}
\begin{equation}
S_{\text{inst}}[\vec{n}]= -2\pi\ii\tau\left(\frac{1}{8\pi^2}\int_{\mathbb{R}^4}\Tr F\wedge F\right)
-2\pi\ii\,\sum_{I=1}^M\eta_I\left(\frac{1}{2\pi}\int_D \Tr F_{\mathrm{U}(n_I)}\right)
\end{equation}
where $\eta_I$ are constant parameters. As shown in detail in Appendix \ref{flux_cartan}, given the
behavior (\ref{Asing}) of the gauge connection near the surface operator, one has
\begin{equation}
\begin{aligned}
&\frac{1}{8\pi^2}\int_{\mathbb{R}^4}\Tr F\wedge F = k+\sum_{I=1}^M \gamma_I \,m_I~,\\
&\frac{1}{2\pi}\int_D \Tr F_{\mathrm{U}(n_I)} = m_I
\end{aligned}
\label{classes}
\end{equation}
with $m_I\in\mathbb{Z}$. As is clear from the second line in the above equation, each $m_I$ represents 
the flux of the U$(1)$ factor in each subgroup U$(n_I)$ in the Levi decomposition (\ref{split});
furthermore, these fluxes satisfy the constraint 
\begin{equation}
\sum_{I=1}^M m_I=0~.
\label{suN}
\end{equation}
Using (\ref{classes}), we easily find
\begin{equation}
S_{\text{inst}}[\vec{n}]=-2\pi\ii\tau\,k-2\pi\ii\sum_{I=1}^M\big(\eta_I+\tau\,\gamma_I\big)m_I=
-2\pi\ii\tau\,k-2\pi\ii\,\vec{t}\cdot \vec m
\label{Sinst0}
\end{equation}
where in the last step we have combined the electric and magnetic parameters $(\eta_I, \gamma_I)$
to form the $M$-dimensional vector
\begin{equation}
\vec{t}= \{t_I\}=\{\eta_I+\tau\,\gamma_I\}~.
\label{tIdef}
\end{equation}
This combination has simple duality transformation properties under SL($2,\mathbb{Z}$).
Indeed, as shown in \cite{Gukov:2006jk}, given an element $\mathcal{M}$ of the modular group the
electro-magnetic parameters transform as
\begin{equation}
\big(\gamma_I,\eta_I\big)~~\to~~\big(\gamma_I,\eta_I\big)
\mathcal{M}^{-1}=\big(d\,\gamma_I-c\,\eta_I,a\,\eta_I-b\,\gamma_I\big)~.
\end{equation}
Combining this with the modular transformation (\ref{modular})
of the coupling constant, it is easy to show that
\begin{equation}
t_I~~\to~~\frac{t_I}{c\tau+d}~.
\label{modulart}
\end{equation}
In particular under S-duality we have
\begin{equation}
S(t_I)=-\frac{t_I}{\tau}~.
\label{Sont}
\end{equation}

Using (\ref{Sinst0}), we deduce that the weight of an instanton configuration in 
the presence of a surface operator of type $\vec{n}$ is
\begin{equation}
\rme^{-S_{\text{inst}}[\vec{n}]}= q^k\,\rme^{2\pi\ii\,\vec{t}\cdot\vec{m}}~,
\end{equation}
so that the instanton partition function can be written as
\begin{equation}
\label{Zinstmiti}
Z_{\text{inst}}[\vec{n}]= \sum_{k,\vec{m}}
\,q^k\, \rme^{2\pi\ii \,\vec{t}\cdot \vec{m}} \, Z_{k, \vec{m}}[\vec{n}] ~.
\end{equation}
In the next section, we will describe the computation of $Z_{k, \vec{m}}[\vec{n}]$ using equivariant localization.

\vspace{0.5cm}
\section{Partition functions for ramified instantons}
\label{app:micro}
As discussed in \cite{Kanno:2011fw}, the $\mathcal{N}=2^*$ theory with 
a surface defect of type $\vec n=\{n_1,\cdots,n_M\}$, which has a six-dimensional
representation as a codimension-2 surface operator, can be realized
with a system of D3-branes in the orbifold background
\begin{equation}
\label{background}
\mathbb{C}\times \mathbb{C}^2/\mathbb{Z}_M \times \mathbb{C} \times \mathbb{C}
\end{equation} 
with coordinates $(z_1,z_2,z_3,z_4,v)$ on which the $\mathbb{Z}_M$-orbifold acts as
\begin{equation}
\label{orbact}
(z_2,z_3) \to (\omega \,z_2,\omega^{-1}\,z_3)~,~~~\text{where } \omega = \mathrm{e}^{\frac{2\pi\ii}{M}}~.
\end{equation}
Like in the previous section, the complex coordinates $z_1$ and $z_2$ span the 
four-dimensional space-time where the gauge theory is defined (namely the world-volume 
of the D3-branes), while the $z_1$-plane is the world-sheet of the surface operator $D$ that 
sits at the orbifold fixed point $z_2=0$. The (massive) deformation which leads from 
the $\mathcal{N}=4$ to the $\mathcal{N}=2^*$ theory takes place in the $(z_3,z_4)$-directions. 
Finally, the $v$-plane corresponds to the Coulomb moduli space of the gauge theory.

Without the $\mathbb{Z}_M$-orbifold projection, the isometry group of the ten-dimensional background is
$\mathrm{SO}(4)\times\mathrm{SO}(4)\times\mathrm{U}(1)$,
since the D3-branes are extended in the first four directions and are moved in the last two when 
the vacuum expectation values (\ref{vev}) are turned on. In the presence of the surface operator and hence of 
the $\mathbb{Z}_M$-orbifold in the $(z_2,z_3)$-directions, this group is broken to
\begin{equation}
\mathrm{U}(1)\times\mathrm{U}(1)\times\mathrm{U}(1)\times\mathrm{U}(1)\times\mathrm{U}(1)~.
\label{U1s}
\end{equation}
In the following we will focus only on the first four U(1) factors, since it is in the first four complex directions
that we will introduce equivariant deformations to apply localization methods. 
We parameterize a transformation of this $\mathrm{U}(1)^4$ group by the vector
\begin{equation}
\vec\epsilon=\{\epsilon_1,\frac{\epsilon_2}{M},\frac{\epsilon_3}{M},\epsilon_4\}
=\{\epsilon_1,\hat\epsilon_2,\hat\epsilon_3,\epsilon_4\}
\end{equation}
where the $1/M$ rescalings in the second and third entry, suggested by the orbifold projection, 
are made for later convenience. If we denote by 
\begin{equation}
\vec l=\{l_1,l_2,l_3,l_4\}
\end{equation}
the weight vector of a given state of the theory, then under $\mathrm{U}(1)^4$ such a state transforms 
with a phase given by $\rme^{2\pi\ii\,\vec l\cdot\vec\epsilon}$, 
while the $\mathbb{Z}_M$-action produces a phase $\omega^{l_2 - l_3}$.

On top of this, we also have to consider the action of the orbifold group on the Chan-Paton factors
carried by the open string states stretching between the D-branes.
There are different types of D-branes depending on the
irreducible representation of $\mathbb{Z}_M$ in which this action takes place.
Since there are $M$ such representations, we have $M$ types of D-branes, which we label
with the index $I$ already used before. 
On a D-brane of type $I$, the generator of $\mathbb{Z}_M$ acts as $\omega^I$, and thus
the Chan-Paton factor of a string stretching between a D-brane of type $I$ and a D-brane of type $J$ 
transforms with a phase $\omega^{I-J}$
under the action of the orbifold generator. 

In order to realize the split of the gauge group in (\ref{split}), we consider $M$ stacks 
of $n_I$ D3-branes of type $I$, and in order to introduce non-perturbative effects we
add on top of the D3's $M$ stacks of $d_I$ D-instantons of type $I$. The latter 
support an auxiliary ADHM group which is
\begin{equation}
\mathrm{U}(d_1) \times \mathrm{U}(d_2) \times \cdots \times \mathrm{U}(d_M)~.
\label{ADHMgroup}
\end{equation} 
In the resulting D3/D$(-1)$-brane systems there are many different sectors of open strings depending
on the different types of branes to which they are attached. 
Here we focus only on the states of open strings with at least one end-point on the D-instantons, because
they represent the instanton moduli \cite{Douglas:1995bn,Billo:2002hm} on which one eventually has 
to integrate in order to obtain the instanton partition function.

Let us first consider the neutral states, corresponding to strings stretched between two D-instantons.
In the bosonic Neveu-Schwarz sector one finds states with $\mathrm{U}(1)^4$ weight vectors
\begin{equation}
\{\pm1,0,0,0\}_{0}~,~~\{0,\pm1,0,0\}_{0}~,~~\{0,0\pm1,0\}_{0}~,~~\{0,0,0\pm1\}_{0}~,~~
\{0,0,0,0\}_{\pm1}~,
\label{wevec}
\end{equation}
where the subscripts denote the charge under the last U(1) factor of (\ref{U1s}). They
correspond to space-time vectors along the directions $z_1$, $z_2$, $z_3$, $z_4$ and $v$, respectively.
In the fermionic Ramond sector one finds states with weight vectors
\begin{equation}
\big\{\!\!\pm\!\ft12,\pm\ft12,\pm\ft12,\pm\ft12\big\}_{\pm\frac12}
\label{wespin}
\end{equation}
with a total odd number of minus signs due to the GSO projection. They correspond to anti-chiral 
space-time spinors\,\footnote{Of course one could have chosen a 
GSO projection leading to chiral spinors, and the 
final results would have been the same.}.

It is clear from (\ref{wevec}) and (\ref{wespin}) that the orbifold phase $\omega^{l_2-l_3}$ takes the values
$\omega^0$, $\omega^{+1}$ or $\omega^{-1}$ and can be compensated only if one considers 
strings of type $I$-$I$, $I$-$(I+1)$ or $(I+1)$-$I$, respectively. Therefore, the $\mathbb{Z}_M$-invariant 
neutral moduli carry Chan-Paton factors that transform 
in the $(\mathbf{d}_I, \mathbf{\bar d}_I)$, $(\mathbf{d}_I, \mathbf{\bar d}_{I+1})$
or $(\mathbf{d}_{I+1}, \mathbf{\bar d}_I)$ representations of the ADHM group (\ref{ADHMgroup}).

Let us now consider the colored states, corresponding to strings stretched between a D-instanton and
a D3-brane or vice versa. Due to the twisted boundary conditions in the first two complex space-time directions, 
the weight vectors of the bosonic states in the Neveu-Schwarz sector are
\begin{equation}
\big\{\!\!\pm\!\ft12,\pm\ft12,0,0\big\}_0
\label{vectw}
\end{equation}
while those of the fermionic states in the Ramond sector are
\begin{equation}
\big\{0,0,\pm\ft12,\pm\ft12\big\}_{\pm\frac{1}{2}}~.
\label{vectmu}
\end{equation}
Assigning a negative intrinsic parity to the twisted vacuum, both in (\ref{vectw}) and
in (\ref{vectmu}) the GSO-projection selects only those vectors with an even number of minus signs. Moreover,
since the orbifold acts on two of the twisted directions, the vacuum carries also
an intrinsic $\mathbb{Z}_M$-weight. We take this to be $\omega^{-\frac{1}{2}}$
when the strings are stretched between a D3-brane and a D-instanton, and $\omega^{+\frac{1}{2}}$
for strings with opposite orientation.
Then, with this choice we find $\mathbb{Z}_M$-invariant bosonic and fermionic states either 
from the $3/(-1)$ strings of type $I$-$I$, whose Chan-Paton factors transform in the 
$(\mathbf{n}_I, \mathbf{\bar d}_I)$ representation of the gauge and ADHM groups, 
or from the $(-1)/3$ strings of type $I$-$(I+1)$, whose Chan-Paton factors transform 
in the $(\mathbf{d}_I, \mathbf{\bar n}_{I+1})$ representation, plus of course the corresponding 
states arising from the strings with opposite orientation.

In Appendix \ref{secn:moduli} we provide a detailed account of all moduli, both neutral and colored, and 
of their properties in the various sectors. It turns out that the moduli action, which can be 
derived from the interactions of the moduli on disks with at least a part of their boundary attached to 
the D-instantons \cite{Billo:2002hm}, is exact with respect to the supersymmetry charge $Q$ of weight
\begin{equation}
\big\{\!\!+\!\ft12,+\ft12,+\ft12,+\ft12\big\}_{-\frac{1}{2}}~.
\label{weightQ}
\end{equation}
Therefore $Q$ can be used as the equivariant BRST-charge to localize the integral over the moduli 
space provided one considers $\mathrm{U}(1)^4$ transformations under which it is invariant. This corresponds
to requiring that
\begin{equation}
\label{sumepsi}
\epsilon_1+\hat\epsilon_2+\hat\epsilon_3+\epsilon_4=0~.  
\end{equation}
Thus we are left with three equivariant parameters, say $\epsilon_1$, $\hat\epsilon_2$ and 
$\epsilon_4$; as we will see, the latter is related to the (equivariant) mass $m$ of the adjoint hypermultiplet 
of $\mathcal{N}=2^*$ theory. 

As shown in Appendix \ref{secn:moduli}, all instanton moduli can be paired in $Q$-doublets
of the type $(\varphi_\alpha,\psi_\alpha)$ such that
\begin{equation}
Q\,\varphi_\alpha=\psi_\alpha~,~~~Q\,\psi_\alpha=Q^2\varphi_\alpha=\lambda_\alpha\,\varphi_\alpha
\end{equation}
where $\lambda_\alpha$ are the eigenvalues of $Q^2$, determined by the action of the Cartan subgroup 
of the full symmetry group of the theory, namely the gauge group (\ref{split}), the ADHM group (\ref{ADHMgroup}),
and the residual isometry group $\mathrm{U}(1)^4$ with parameters satisfying (\ref{sumepsi}) in such 
a way that the invariant points in the moduli space are finite and isolated. The only exception to this structure
of $Q$-doublets is represented by the neutral bosonic moduli with weight
\begin{equation}
\{0,0,0,0\}_{-1}
\end{equation}
transforming in the adjoint representation 
$(\mathbf{d}_I, \mathbf{\bar d}_I)$ of the ADHM group U$(d_I)$, which remain unpaired. 
We denote them as $\chi_I$, and in order to obtain the instanton partition function
we must integrate over them. In doing so, we can exploit the $\mathrm{U}(d_I)$ symmetry to rotate $\chi_I$
into the maximal torus and write it in terms of the eigenvalues $\chi_{I,\sigma}$, with $\sigma=1,\cdots,d_I$,
which represent the positions of the $D$-instantons of type $I$ in the $v$-plane. In this way we are 
left with the integration over all the $\chi_{I,\sigma}$'s and a 
Cauchy-Vandermonde determinant 
\begin{equation}
\label{vander}
\mathcal{V}=
\prod_{I=1}^{M}\prod_{\sigma,\tau=1}^{d_I} (\chi_{I,\sigma} - \chi_{I,\tau}+ \delta_{\sigma\tau}) ~.
\end{equation}
More precisely, the instanton partition function in the presence of a surface operator of type $\vec{n}$ 
is defined by
\begin{equation}
\label{Zso}
Z_{\text{inst}}[\vec{n}]
= \sum_{\{d_I\}} \prod_{I=1}^M q_I^{d_I}\,Z_{\{d_I\}}[\vec{n}]\quad
\mbox{with}~~~Z_{\{d_I\}}[\vec{n}]=\frac{1}{d_I!}
\,\int \prod_{\sigma=1}^{d_I} \frac{d\chi_{I,\sigma}}{2\pi\ii}~
z_{\{d_I\}}
\end{equation}
where $z_{\{d_I\}}$ is the result of the integration over all $Q$-doublets which localizes on the fixed
points of $Q^2$, and $q_I$ is the counting parameter associated to the D-instantons of type $I$.
With the convention that $z_{\{d_I=0\}}=1$, we find
\begin{equation}
\label{zdI}
z_{\{d_I\}}= \mathcal{V}\, \prod_\alpha \big[\lambda_\alpha\big]^{(-)^{F_\alpha+1}}~,
\end{equation}
where the index $\alpha$ labels the $Q$-doublets and $\lambda_\alpha$ denotes the corresponding 
eigenvalue of $Q^2$. This contribution goes to the denominator or to the numerator depending upon the bosonic or
fermionic statistics ($F_\alpha=0$ or $1$, respectively) 
of the first component of the doublet. Explicitly, using the data in Tab.\,\ref{tab:moduli} of 
Appendix \ref{secn:moduli} and the determinant (\ref{vander}), we find
\begin{equation}
\begin{aligned}
z_{\{d_I\}} & = \,\prod_{I=1}^M \prod_{\sigma,\tau=1}^{d_I}\,
\frac{\left(\chi_{I,\sigma} - \chi_{I,\tau} + \delta_{\sigma,\tau}\right)
\left(\chi_{I,\sigma} - \chi_{I,\tau} + \epsilon_1+\epsilon_4\right)
}{\left(\chi_{I,\sigma} - \chi_{I,\tau} + \epsilon_4\right)\left(\chi_{I,\sigma} - \chi_{I,\tau} 
+ \epsilon_1\right)}\\
&~~\times\prod_{I=1}^M \prod_{\sigma=1}^{d_I}\prod_{\rho=1}^{d_{I+1}}\,
\frac{\left(\chi_{I,\sigma} - \chi_{I+1,\rho} + \epsilon_1 + \hat\epsilon_2\right)
\left(\chi_{I,\sigma} - \chi_{I+1,\rho} +\hat\epsilon_2+ \epsilon_4\right)}
{\left(\chi_{I,\sigma} - \chi_{I+1,\rho} -\hat\epsilon_3\right)\left(\chi_{I,\sigma} - \chi_{I+1,\rho} 
+ \hat\epsilon_2\right)}
\\
& ~~\times
\prod_{I=1}^M \prod_{\sigma=1}^{d_I} \prod_{s=1}^{n_I} \,
\frac{\left(a_{I,s} - \chi_{I,\sigma} + \frac 12 (\epsilon_1 + \hat\epsilon_2)
+\epsilon_4\right)}{\left(a_{I,s}-\chi_{I,\sigma} + \frac 12 (\epsilon_1 + \hat\epsilon_2)\right)}
\\
&~~\times
\prod_{I=1}^M \prod_{\sigma=1}^{d_I}\prod_{t=1}^{n_{I+1}}\,
\frac{\left(\chi_{I,\sigma} - a_{I+1,t} + \frac 12 (\epsilon_1 + \hat\epsilon_2)
+\epsilon_4\right)}{\left(\chi_{I,\sigma} - a_{I+1,t} + \frac 12 (\epsilon_1 + \hat\epsilon_2)\right)}
\end{aligned}
\label{zexplicit}
\end{equation}
where $d_{M+1}=d_1$, $n_{M+1}=n_1$ and $a_{M+1,t}=a_{1,t}$. 
The integrations in (\ref{Zso}) must be suitably defined and regularized. The standard 
prescription \cite{Moore:1998et,Billo':2015ria,Billo':2015jta} is
to consider $a_{I,s}$ to be real and close the contours in the upper-half $\chi_{I,\sigma}\,$-planes 
with the choice
\begin{equation}
\mathrm{Im}\, \epsilon_4 \gg \mathrm{Im}\, \hat\epsilon_3 \gg \mathrm{Im}\, \hat\epsilon_2\gg \mathrm{Im}\, 
\epsilon_1 > 0~,
\label{prescription}
\end{equation}
and enforce (\ref{sumepsi}) at the very end of the calculations. 

In this way one finds that these integrals receive contributions from the poles of $z_{\{d_I\}}$, 
which are in fact the critical points of $Q^2$. Such poles can be put in one-to-one 
correspondence with a set of $N$ 
Young tableaux $Y = \{Y_{I,s}\}$, with $I=1,\cdots,M$ and $s=1,\cdots n_I$, in the sense that 
the box in the $i$-th row and $j$-th column of the tableau $Y_{I,s}$ represents 
one component of the critical value:
\begin{equation}
\label{chicritY}
\chi_{I+(j-1)\mathrm{mod} M,\sigma} = a_{I,s} + \left((i-1) + \ft12\right)\epsilon_1 + \left((j-1)+\ft12\right)
 \hat\epsilon_2~.
\end{equation}
Note that in this correspondence, a single tableau accounts for $d_I!$ equivalent ways of
relabeling $\chi_{I,\sigma}$.

\subsection{Summing over fixed points and characters}
Summing over the Young tableaux collections $Y$ we get all the non-trivial critical points corresponding 
to all possible values of $\{d_I\}$. Eq. (\ref{chicritY}) tells us that we get a 
distinct $\chi_{I,\sigma}$ for each box in the $j$-th column of the tableau 
$Y_{I+1-j\,\mathrm{mod}\,M,s}$. Relabeling the index $j$ as
\begin{equation}
\label{jtoJ}
j \to J + j \, M~,
\end{equation}
with $J=1,\ldots M$, we have
\begin{equation}
\label{dIY}
d_I(Y) = \sum_{J=1}^M  \sum_{s=1}^{n_{I+1-J}}\sum_j Y_{I+1-J,s} ^{(J+ j M)}~,
\end{equation}
where $Y_{I,s} ^{(j)}$ denotes the height of the $j$-th column of the tableau $Y_{I,s}$, and the subscript
index $I+1-J$ is understood modulo $M$.

The instanton partition function (\ref{Zso}) can thus be rewritten as a sum over Young tableaux as follows
\begin{equation}
\label{ZsoY}
Z_{\text{inst}}[\vec{n}]= \sum_Y \prod_{I=1}^M q_I^{d_I(Y)}\, Z(Y)
\end{equation}
where $Z(Y)$ is the residue of $z_{\{d_I\}}$ at the critical point $Y$. This is obtained by deleting in 
(\ref{zexplicit}) the denominator factors that yield the identifications (\ref{chicritY}), and performing 
these identifications in the other factors. In other terms,
\begin{equation}
\label{zY}
Z(Y) = \mathcal{V}(Y) \prod_{\alpha\,:\,\lambda_\alpha(Y)\not=0} [\lambda_\alpha(Y)]^{(-)^{F_\alpha+1}}~,
\end{equation}
where $\mathcal{V}(Y)$ and $\lambda_\alpha(Y)$ are the Vandermonde determinant and the eigenvalues
of $Q^2$ evaluated on (\ref{chicritY}).

A more efficient way to encode the eigenvalues $\lambda_\alpha(Y)$ is to employ the character of 
the action of $Q^2$, which is defined as follows
\begin{equation} 
\label{def_char}
X_{\{d_I\}} = \sum_\alpha (-)^{F_\alpha} \mathrm{e}^{\ii \lambda_\alpha}~.
\end{equation} 
If we introduce
\begin{equation}
\label{VIWIdef}
V_I = \sum_{\sigma=1}^{d_I} \mathrm{e}^{\ii\chi_{I,\sigma} - \frac{\ii}{2} (\epsilon_1 + \hat\epsilon_2)}~,~~~
W_I = \sum_{s=1}^{n_I} \mathrm{e}^{\ii a_{I,s}}
\end{equation} 
and
\begin{equation}
\label{Tidef}
T_1 = \mathrm{e}^{\ii \epsilon_1}~,~~~
T_2 = \mathrm{e}^{\ii \hat\epsilon_2}~,~~~
T_4 = \mathrm{e}^{\ii \epsilon_4}~,
\end{equation}
we can write the contributions to the character from the various $Q$-doublets as in the last 
column of Tab.\,\ref{tab:moduli} in Appendix \ref{secn:moduli}. 
Then, by summing over all doublets and adding also 
the contribution of the Vandermonde determinant, we obtain 
\begin{equation}
\label{Xchar}
X_{\{d_I\}} = (1 - T_4) \sum_{I=1}^M \left[-(1- T_1) V_I^* V_I + (1 - T_1) V^*_{I+1} V_I T_2 + 
V^*_I W_I + W^*_{I+1} V_I T_1 T_2\right]~.
\end{equation}
As we have seen before, through (\ref{dIY}) and (\ref{chicritY}) each set $Y$ determines 
both the dimensions $d_I(Y)$ and the eigenvalues $\lambda_\alpha(Y)$. Thus, the character 
$X(Y)$ associated to a set of Young tableaux is obtained from $X_{\{d_I\}}$ by 
substituting (\ref{chicritY}) into the definitions of $V_I$, namely
\begin{equation}
\label{VIWIsub}
V_I  = \sum_{J=1}^M \sum_{s=1}^{n_{I+1-J}} \mathrm{e}^{\ii a_{I+1-J,s}} T_2^J 
\sum_{(i,J+jM)\in Y_{I+1-J,s}} T_1^{i-1} T_2^{jM - 1}~.
\end{equation}
By analyzing $X(Y)$ obtained in this way we can extract the explicit expression for the eigenvalues 
$\lambda_s(Y)$ and finally write the instanton partition function. This procedure is easily implemented 
in a computer program, and yields the results we will use in the next sections. In Appendix (\ref{secn:su2app}),
as an example, we illustrate these computations for the SU(2) gauge theory.

In our analysis we worked with the moduli action that describes D-branes probing the orbifold geometry. 
An alternative approach works with the resolution of the orbifold geometry \cite{Bonelli:2013mma, Nawata:2014nca}. 
This involves analyzing a gauged linear sigma-model that describes a system of D1 and D5-branes in the 
background $\mathbb{C}\times \mathbb{C}/\mathbb{Z}_M \times T^{\star}S^2 \times \mathbb{R}^2$. One then uses 
the localization formulas for supersymmetric field theories defined on the 2-sphere
\cite{Doroud:2012xw,Benini:2012ui} to obtain exact results. This is a very powerful approach since it also 
includes inherently stringy corrections to the partition function arising from world-sheet instantons
\cite{Bonelli:2013mma}. The results for the instanton partition function of the ${\mathcal N}=2^{\star}$ 
theory in the presence of surface operators obtained in \cite{Nawata:2014nca} are equivalent to 
our results in (\ref{zexplicit}).

\subsection{Map between parameters}
One of the key points that needs to be clarified is the map between the microscopic 
counting parameters $q_I$ which appear in (\ref{ZsoY}), and the parameters $(q, t_I)$
which were introduced in Section \ref{SOmicro} in discussing SU$(N)$ gauge theories with surface operators.
To describe this map, we start by rewriting the partition function (\ref{Zso}) in terms of the total 
instanton number $k$ and the magnetic fluxes $m_I$ of the gauge groups on the surface operator which are 
related to the parameters $d_I$ as follows \cite{Alday:2010vg, Kanno:2011fw}:
\begin{equation}
d_1 = k~,\quad d_{I+1} = d_{I}+m_{I+1}~.
\end{equation}
Therefore, instead of summing over $\{d_I\}$ we can sum over $k$ and $\vec{m}$ and find
\begin{equation}
Z_{\mathrm{inst}}[\vec{n}] = \sum_{k, \vec{m}}\, (q_1\cdots q_M)^k \,
(q_2\cdots q_M)^{m_2}\,(q_3\cdots q_M)^{m_3} \cdots (q_{M})^{m_{M}}\, Z_{k, \vec{m}}[\vec{n}] 
\end{equation}
Furthermore, if we set
\begin{equation}
\label{paramap}
\begin{aligned}
q_I &= \rme^{2\pi\ii(t_{I}-t_{I+1})} \quad\text{for}\quad I \in \{2, \ldots M-1 \}~,\\
q_M &=\rme^{2\pi\ii(t_{M}-t_{1})}\quad\text{and}\quad q = \prod_{I=1}^M q_I ~,
\end{aligned}
\end{equation}
we easily get
\begin{equation}
\label{Ztdiff}
\begin{aligned}
Z_{\mathrm{inst}}[\vec{n}] &= \sum_{k, \vec{m}}\,q^k \rme^{2\pi\ii \sum_{I=2}^{M} m_I ( t_I- t_1)}
\,Z_{k, \vec{m}} 
= \sum_{k,\vec{m}}
\,q^k\, \rme^{2\pi\ii \,\vec{t}\cdot \vec{m}} \, Z_{k, \vec{m}}[\vec{n}] 
\end{aligned}
\end{equation}
where in the last step we introduced $m_1$ such that that $\sum_I m_I=0$ (see (\ref{suN})) in order to write
the result in a symmetric form.
This is precisely the expected expression of the partition function in the presence of a surface operator 
as shown in (\ref{Zinstmiti}) and justifies the map (\ref{paramap}) between the parameters of the two descriptions. 
{From} (\ref{Ztdiff}) we see that only differences of the parameters $t_I$ appear in the partition function 
so that it may be convenient to use as independent parameters $q$ and the $(M-1)$ variables
\begin{equation}
\label{zdeft}
z_J = t_J - t_1  \quad\text{for}\quad J \in \{2, \ldots M \} ~.
\end{equation}
This is indeed what we are going to see in the next sections where we will show how to extract 
relevant information from the the instanton partition functions described above.

\subsection{Extracting the prepotential and the twisted superpotential}
The effective dynamics on the Coulomb branch of the four-dimensional $\mathcal{N}=2^\star$ gauge theory 
is described by the prepotential $\mathcal{F}$, while the infrared physics of the two-dimensional theory defined
on the world-sheet of the surface operator is governed by the twisted superpotential $\mathcal{W}$.
The non-perturbative terms of both $\mathcal{F}$ and $\mathcal{W}$ can be derived from the
instanton partition function previously discussed, by considering its
behavior for small deformation parameters $\epsilon_1$ and
$\epsilon_2$ and, in particular, in the so-called Nekrasov-Shatashvili (NS) limit \cite{Nekrasov:2009rc}.

To make precise contact with the gauge theory quantities, we set
\begin{equation}
\label{e4}
\epsilon_4=-m-\frac{\epsilon_1}{2}
\end{equation}
where $m$ is the mass of the adjoint hypermultiplet, and then take the limit for small $\epsilon_{1}$ and
$\epsilon_2$. In this way we find \cite{Alday:2010vg}:
\begin{equation}
\log Z_{\mathrm{inst}}[\vec{n}] ~\simeq~
-\frac{\mathcal{F}_{\mathrm{inst}}(\epsilon_1)}{\epsilon_1\epsilon_2}
+\frac{\mathcal{W}_{\mathrm{inst}}(\epsilon_1)}{\epsilon_1}
+\mathcal{O}(\epsilon_2)~.
\label{FW}
\end{equation}
The two leading singular 
contributions arise, respectively, from the (regularized) equivariant volume parts coming from the 
four-dimensional gauge theory and from the 
two-dimensional degrees of freedom supported on the surface defect $D$. 
This can be understood from the fact that, in the $\Omega$-deformed theory, the respective super-volumes 
are finite and given by \cite{Gukov:2014gja,Billo:2006jm}:
\begin{equation}
\int_{\mathbb{R}^4_{\epsilon_1, \epsilon_2}} \!\!\!\!d^4x\,d^4\theta \,\longrightarrow \,
\frac{1}{\epsilon_1\epsilon_2} \qquad \mbox{and} \qquad\int_{\mathbb{R}^2_{\epsilon_1}} \!\!\!\!d^2x\,d^2\theta 
\,\longrightarrow\, \frac{1}{\epsilon_1} ~. 
\end{equation}
The non-trivial result is that the functions 
$\mathcal{F}_{\mathrm{inst}}$ and $\mathcal{W}_{\mathrm{inst}}$ defined in this way are analytic in the 
neighborhood of $\epsilon_1=0$. 
As an illustrative example, we now describe in some detail the SU(2) theory.

\subsubsection*{SU(2)}
When the gauge group is SU(2), the only surface operators are of type $\vec{n}=\{1,1\}$, 
the Coulomb branch is parameterized by
\begin{equation}
\langle \phi \rangle = \mathrm{diag}(a,-a)~,
\label{vevSU2}
\end{equation}
and the map (\ref{paramap}) can be written as
\begin{equation}
q_1=\frac{q}{x}~,~~~q_2=x=\rme^{2\pi\ii\,z}
\label{mapSU2}
\end{equation}
where, for later convenience, we have defined $z=(t_2-t_1)$. 
Using the results presented in Appendix \ref{secn:su2app} and their extension to higher orders, 
it is possible to check that the instanton prepotential arising from (\ref{FW}), namely
\begin{equation}
\mathcal{F}_{\mathrm{inst}}=-\lim_{\epsilon_2\to0}\Big(\epsilon_1\epsilon_2
\log Z_{\mathrm{inst}}[1,1]\Big)
\label{FSU2}
\end{equation}
is, as expected, a function only of the instanton counting parameter $q$ and not of $x$. Expanding in inverse
powers of $a$, we have
\begin{equation}
\mathcal{F}_{\mathrm{inst}}=\sum_{\ell=1}^\infty f_{\ell}^{\mathrm{inst}}
\end{equation}
where $f_\ell \sim a^{2-\ell}$. The first few coefficients of this expansion are
\begin{equation}
\begin{aligned}
f_{2\ell+1}^{\mathrm{inst}}&=0\quad\mbox{for}~\ell=0,1,\cdots\phantom{\Big|}~,\\
f_2^{\mathrm{inst}}&=-\Big(m^2-\frac{\epsilon_1^2}{4}\Big)\Big(2q+3q^2+\frac{8}{3}q^3+\cdots
\Big)~,\\
f_4^{\mathrm{inst}}&=\frac{1}{2a^2}\,\Big(m^2-\frac{\epsilon_1^2}{4}\Big)^2\Big(q+3q^2+
4q^3+\cdots\Big)~,\\
f_6^{\mathrm{inst}}&=\frac{1}{16a^4}\,\Big(m^2-\frac{\epsilon_1^2}{4}\Big)^2\Big(2\epsilon_1^2\,q
-3\big(4m^2-7\epsilon_1^2\big)q^2-8 \big(8m^2-9 \epsilon_1^2\big)q^3+\cdots\Big)~.\\
\end{aligned}
\label{fellinst}
\end{equation}
One can check that this precisely agrees with the NS limit of the prepotential derived for example in
\cite{Billo:2013fi,Billo:2013jba}. This complete match is a strong and non-trivial check on the correctness
and consistency of the whole construction.

Let us now consider the non-perturbative superpotential, which according to (\ref{FW}) is
\begin{equation}
\mathcal{W}_{\mathrm{inst}}=\lim_{\epsilon_2\to0}\Big(\epsilon_1
\log Z_{\mathrm{inst}}[1,1]+\frac{\mathcal{F}_{\mathrm{inst}}}{\epsilon_2}\Big)~.
\end{equation}
Differently from the prepotential, $\mathcal{W}_{\mathrm{inst}}$ is, as expected, 
a function both of $q$ and $x$. If we expand it as
\begin{equation}
\mathcal{W}_{\mathrm{inst}}= \sum_{\ell=1}^\infty w_\ell^{\mathrm{inst}}
\end{equation}
with $w_\ell^{\mathrm{inst}}\sim a^{1-\ell}$, using the results of Appendix \ref{secn:su2app} we find
\begin{subequations}
\begin{align}
w_1^{\mathrm{inst}}&=-\Big(m-\frac{\epsilon_1}{2}\Big)\left[\Big(x + \frac{x^2}{2} +\frac{x^3}{3} 
+\frac{x^4}{4}+\cdots\Big)+\Big(\frac{1}{x}+2+x+\cdots\Big)q
\right.\notag\\
&\qquad\qquad\qquad\qquad\quad\left.+
\Big(\frac{1}{2x^2}+\frac{1}{x}+3 +\cdots\Big)q^2+\cdots\right]~,
\label{w1}\\
w_2^{\mathrm{inst}}&=-\frac{1}{a}\Big(m^2-\frac{\epsilon_1^2}{4}\Big)\left[
\Big(\frac{x}{2}+\frac{x^2}{2}+\frac{x^3}{2}+\frac{x^4}{2}+\cdots\Big)+
\Big(\frac{x}{2}-\frac{1}{2x}+\cdots\Big)q \right.\notag\\
&\qquad\qquad\qquad\qquad\quad\left.-\Big( \frac{1}{2 x^2} + \frac{1}{2 x}+\cdots\Big)q^2+\cdots \right]~,
\label{w2}\\
w_3^{\mathrm{inst}}&=-\frac{\epsilon_1}{a^2}\Big(m^2-\frac{\epsilon_1^2}{4}\Big)\left[
\Big(\frac{x}{4} +\frac{x^2}{2} +\frac{3 x^3}{4}+ x^4+\cdots\Big)+
\Big(\frac{1}{4x}+\frac{x}{4}+\cdots\Big)q\right.\notag\\
&\qquad\qquad\qquad\qquad\quad\left.+\Big(\frac{1}{2x^2}+\frac{1}{4x}+\cdots\Big)q^2+\cdots\right]~,
\label{w3}
\end{align}
\label{well}
\end{subequations}
\hspace{-0.12cm}and so on.
For later convenience we explicitly write down the logarithmic derivatives with respect to $x$, namely
\begin{subequations}
\begin{align}
w^\prime_1&=-\Big(m-\frac{\epsilon_1}{2}\Big)\left[\Big(x+x^2+x^3 
+x^4+\cdots\Big)-\Big(\frac{1}{x}-x+\cdots\Big)q
\right.\notag\\
&\qquad\qquad\qquad\qquad\quad\left.
-\Big(\frac{1}{x^2}+\frac{1}{x}+\cdots\Big)q^2+\cdots\right]~,\label{wprime1}\\
w^\prime_2&=-\frac{1}{a}\Big(m^2-\frac{\epsilon_1^2}{4}\Big)\left[
\Big(\frac{x}{2}+x^2+\frac{3x^3}{2}+2x^4+\cdots\Big)+
\Big(\frac{x}{2}+\frac{1}{2x}+\cdots\Big)q \right.\notag\\
&\qquad\qquad\qquad\qquad\quad\left.+\Big( \frac{1}{x^2} + \frac{1}{2 x}+\cdots\Big)q^2+\cdots \right]~,
\label{wprime2}\\
w^\prime_3&=-\frac{\epsilon_1}{a^2}\Big(m^2-\frac{\epsilon_1^2}{4}\Big)\left[
\Big(\frac{x}{4} +x^2 +\frac{9 x^3}{4}+ 4x^4+\cdots\Big)-
\Big(\frac{1}{4x}-\frac{x}{4}+\cdots\Big)q\right.\notag\\
&\qquad\qquad\qquad\qquad\quad\left.-\Big(\frac{1}{x^2}+\frac{1}{4x}+\cdots\Big)q^2+\cdots\right]
\label{wprime3}
\end{align}
\label{wprimeell}
\end{subequations}
\hspace{-0.12cm}where $w^\prime_\ell:= x\,\frac{\partial}{\partial x}\big(w_\ell^{\mathrm{inst}}\big)$.
In the coming sections we will show that these expressions are the weak-coupling expansions of combinations of
elliptic and quasi-modular forms of the modular group SL($2,\mathbb{Z}$).

\vspace{0.5cm}
\section{Modular anomaly equation for the twisted superpotential}
\label{secn:modaneq}
In \cite{Billo:2013fi,Billo:2013jba} it has been shown for the ${\mathcal N}=2^{\star}$ SU(2) 
theory that the instanton expansions of the prepotential coefficients (\ref{fellinst}) can be 
resummed in terms of (quasi-) modular forms of the duality group SL$(2, {\mathbb Z})$ and that 
the behavior under S-duality severely constrains the prepotential $\mathcal{F}$ 
which must satisfy a modular anomaly equation. This analysis has been
later extended to ${\mathcal N}=2^{\star}$ theories with arbitrary classical or exceptional gauge groups
\cite{Billo:2014bja,Billo':2015ria,Billo':2015jta}, and also to $\mathcal{N}=2$ SQCD theories 
with fundamental matter
\cite{Ashok:2015cba,Ashok:2016oyh}. In this section we use a similar approach to study how S-duality
constrains the form of the twisted superpotential ${\mathcal W}$. 

For simplicity and without loss of generality, in the following we consider a full surface operator of type 
$\vec{n}=\{1,1,\cdots,1\}$ with electro-magnetic parameters $\vec{t}=\{t_1,t_2,\cdots,t_N\}$. 
Indeed, surface operators of other type correspond to the case in which these parameters are not 
all different from each other and form $M$ distinct sets,
namely
\begin{equation}
\vec{t} =
\left\{
\begin{array}{ccccccccc}
\undermat{n_1}{t_1, & \ldots , & t_1,} & \undermat{n_2}{ t_2, & \ldots, &t_2,} &\cdots,& 
\undermat{n_M}{ t_M, &\ldots, t_M} 
\end{array}
\right\} ~.
\vspace{.5cm}
\end{equation}
Thus they can be simply recovered from the full ones with suitable identifications.

Before analyzing the S-duality constraints it is necessary to take into account the 
classical and the perturbative 1-loop contributions to the prepotential and the superpotential.

\subsection*{The classical contribution}
Introducing the notation $\vec{a}=\{a_1,a_2,\cdots,a_N\}$ for the vacuum expectation values, 
the classical contributions to the prepotential and the superpotential are given respectively by
\begin{equation}
\mathcal{F}_{\mathrm{class}} = \pi \ii \tau \,\vec{a}\cdot\vec{a}
\label{Fclass}
\end{equation}
and
\begin{equation}
{\mathcal W}_{\mathrm{class}} = 2\pi\ii\, \vec{t}\cdot \vec{a} ~.
\label{Wclassical}
\end{equation}
Note that if we use the 
tracelessness condition (\ref{vev}), ${\mathcal W}_{\mathrm{class}}$ can be rewritten as
\begin{equation}
{\mathcal W}_{\mathrm{class}} = 2\pi\ii\sum_{I=2}^N z_I \,a_I
\end{equation}
where $z_I$ is as defined in (\ref{zdeft}).

These classical contributions have very simple behavior under S-duality. Indeed
\begin{subequations}
\begin{align}
S\big({\mathcal F}_{\mathrm{class}} \big)&=-\,{\mathcal F}_{\mathrm{class}}~,\label{SonFclass}\\
S\big({\mathcal W}_{\mathrm{class}}\big)&=-\,{\mathcal W}_{\mathrm{class}}~.\label{SonWclass}
\end{align}
\label{SonFWclass}
\end{subequations} 
To show these relations one has to use the S-duality rules (\ref{Sontau}) and (\ref{Sont}), 
and recall that
\begin{equation}
S\big(\vec{a}\big)= \vec{a}_{\text{D}} := \frac{1}{2\pi\ii}\,
\frac{\partial{\mathcal F}}{\partial \vec{a}}\qquad\mbox{and}\qquad
S\big(\vec{a}_\text{D}\big)=-\vec{a}~,
\end{equation}
which for the classical prepotential simply yield $S(\vec{a})=\tau\,\vec{a}$.

\subsection*{The 1-loop contribution}

The 1-loop contribution to the partition function of the $\Omega$-deformed gauge 
theory in the presence of a full surface operator of type $\{1,1,\cdots,1\}$ can be 
written in terms of the function
\begin{equation}
\gamma(x):=\log\Gamma_2(x|\epsilon_1, \epsilon_2)= 
\left.\frac{d}{ds}\left(\frac{\Lambda^s}{\Gamma(s)} \int_0^{\infty} dt 
\frac{t^{s-1}e^{-tx}}{(e^{-\epsilon_1 t}-1)(e^{-\epsilon_2 t}-1)} \right)\right|_{s=0}~,
\end{equation}
where $\Gamma_2$ is the Barnes double $\Gamma$-function and $\Lambda$ an arbitrary scale. 
Indeed, as shown for example in \cite{Nawata:2014nca}, the perturbative contribution is 
\begin{equation}
\log Z_{\text{pert}}[1,1,\cdots,1] = \mathop{\sum_{u,v=1}^N}_{u\not=v}\Big[
\gamma\bigl
(a_{uv} + \big \lceil{\ft{v-u}{N}}\big \rceil\epsilon_2
\bigr)-\gamma\bigl(a_{uv} +m+\ft{\epsilon_1}{2}+  \big \lceil{\ft{v-u}{N}}\big 
\rceil \epsilon_2\bigr)\Big]
\label{Zpert}
\end{equation}
where $a_{uv}=a_u-a_v$, and the ceiling function $\lceil{y}\rceil$ denotes the smallest integer 
greater than or equal to $y$. The first term in (\ref{Zpert}) represents the contribution of the vector multiplet,
while the second term is the contribution of the massive hypermultiplet. 
Expanding (\ref{Zpert}) for small $\epsilon_{1,2}$ and using the same definitions 
(\ref{FW}) used for the instanton part, we obtain the perturbative contributions
to the prepotential and the superpotential in the NS limit:
\begin{equation}
\begin{aligned}
\mathcal{F}_{\mathrm{pert}}&=-\lim_{\epsilon_2\to0}\Big(\epsilon_1\epsilon_2
\log Z_{\mathrm{pert}}[1,1,\cdots,1]\Big)~,\\
\mathcal{W}_{\mathrm{pert}}&=\lim_{\epsilon_2\to0}\Big(\epsilon_1
\log Z_{\mathrm{pert}}[1,1,\cdots,1]+\frac{\mathcal{F}_{\mathrm{pert}}}{\epsilon_2}\Big)~.
\end{aligned}
\end{equation}
Exploiting the series expansion of the $\gamma$-function, one can explicitly compute these expressions
and show that $\mathcal{F}_{\mathrm{pert}}$ precisely matches the perturbative prepotential in the NS limit
obtained in \cite{Billo:2014bja,Billo':2015ria}, while the contribution to the superpotential is novel.
For example, in the case of the SU(2) theory we obtain
\begin{subequations}
\begin{align}
\label{Fpertexp}
\mathcal{F}_{\mathrm{pert}} &= \frac{1}{2}\Big(m^2-\frac{\epsilon_1^2}{4}\Big)
\log\frac{4a^2}{\Lambda^2} \!-\frac{1}{48a^2}\Big(m^2-\frac{\epsilon_1^2}{4}\Big)^2 
\!- \frac{1}{960a^4}\Big(m^2-\frac{\epsilon_1^2}{4}\Big)^2
\Big(m^2-\frac{3\epsilon_1^2}{4}\Big) \!+\cdots~,\\
\label{wpertexp}
{\mathcal W}_{\mathrm{pert}}&=- \frac{1}{4a}\Big(m^2-\frac{\epsilon_1^2}{4}\Big) \!
-\frac{1}{96a^3}\Big(m^2- \frac{\epsilon_1^2}{4}\Big)^2
\!-\frac{1}{960a^5}\Big(m^2- \frac{\epsilon_1^2}{4}\Big)^2\Big(m^2- \frac{3\epsilon_1^2}{4}\Big) \!+ \cdots~.
\end{align}
\end{subequations}
Note that, unlike the prepotential, the twisted superpotential has no logarithmic term\,\footnote{
This fact is due to the superconformal invariance, and is no longer true in 
the pure ${\mathcal N}=2$ SU(2) gauge theory, for which we find
\begin{equation*}
{\mathcal W}_{\text{pert}}  = -\Big(2-2\log\frac{2a}{\Lambda}\Big)a 
+ \frac{\epsilon_1^2}{24a}-\frac{\epsilon_1^4}{2880a^3}+\frac{\epsilon_1^6}{40320a^5} + \cdots~.
\end{equation*}
}.
Furthermore, it is interesting to observe that
\begin{equation}
{\mathcal W}_{\text{pert}} = -\frac{1}{4}\frac{\partial F_{\text{pert}}}{\partial a}~.
\end{equation}

\subsection{S-duality constraints}
We are now in a position to discuss the constraints on the twisted superpotential arising from S-duality. 
Adding the classical, the perturbative and the instanton terms described in the previous
sections, we write the complete prepotential and 
superpotential in the NS limit as
\begin{equation}
\begin{aligned}
\mathcal{F} &= \mathcal{F}_{\mathrm{class}}+\mathcal{F}_{\mathrm{pert}}
+\mathcal{F}_{\mathrm{inst}}
= \pi \ii \tau \,\vec{a}\cdot\vec{a} +\sum_{\ell=1}^\infty f_\ell(\tau,\vec{a})~,\\
\mathcal{W} &= \mathcal{W}_{\mathrm{class}}+\mathcal{W}_{\mathrm{pert}}
+\mathcal{W}_{\mathrm{inst}}
= 2\pi \ii \sum_{I=2}^Nz_I\,a_I +\sum_{\ell=1}^\infty w_\ell(\tau,z_I,\vec{a})\\
\end{aligned}
\label{FandW}
\end{equation}
where for later convenience,  we have kept the classical terms separate. The quantum coefficients
$f_\ell$ and $w_\ell$ scale as $a^{2-\ell}$ and $a^{1-\ell}$, respectively, and account for the perturbative
and instanton contributions. While $f_\ell$ depend on the coupling constant $\tau$, 
the superpotential coefficients $w_\ell$ are also functions of the surface operator variables $z_I$, as we have
explicitly seen in the SU(2) theory considered in the previous section.

The coefficients $f_{\ell}$ have been explicitly calculated in terms 
of quasi-modular forms in \cite{Billo:2014bja,Billo':2015ria} and we list 
the first few of them in Appendix \ref{Freview}. 
Their relevant properties can be summarized as follows: 
\begin{itemize}
\item All $f_\ell$ with $\ell$ odd vanish, while those with $\ell$ even are 
homogeneous functions of $\vec{a}$ and satisfy the scaling relation\,\footnote{To be precise, one 
should also scale $\Lambda\to \lambda\Lambda$ in the logarithmic term of $f_2$.}
\begin{equation}
f_{2\ell}(\tau,\lambda\,\vec{a}) = \lambda^{2-2\ell}\,f_{2\ell}(\tau,\vec{a})~.
\end{equation}
Since the prepotential has mass-dimension two, the $f_{2\ell}$ are homogeneous 
polynomials of degree $2\ell$, in $m$ and $\epsilon_1$.  
\item The coefficients 
$f_{2\ell}$ depend on the coupling constant $\tau$ only through the Eisenstein series $E_2(\tau)$,
$E_4(\tau)$ and $E_6(\tau)$, and are quasi-modular forms of SL(2,$\mathbb{Z}$) of weight $2\ell-2$, such that
\begin{equation}
f_{2\ell}\Bigl(\!-\ft{1}{\tau},\vec{a}\Bigr) = \tau^{2\ell-2}\,
f_{2\ell}(\tau,\vec{a})\Big|_{E_2\to E_2 + \delta}
\end{equation} 
where $\delta=\frac{6}{\pi \ii\tau}$. The shift $\delta$ in $E_2$ is due to 
the fact that the second Eisenstein series is a quasi-modular form with an 
anomalous modular transformation (see (\ref{SonE})).
\item The coefficients $f_{2\ell}$ satisfy a modular anomaly equation
\begin{equation}
\frac{\partial f_{2\ell}}{\partial E_2}+\frac{1}{24}\sum_{n=1}^{\ell-1} 
\frac{\partial f_{2n}}{\partial \vec{a}}\cdot\frac{\partial f_{2\ell - 2n}}{\partial \vec{a}}=0
\end{equation}
which can be solved iteratively. 
\end{itemize}
\noindent
Using the above properties, it is possible to show that S-duality acts on the 
prepotential $\mathcal{F}$ in the NS limit as a Legendre transform \cite{Billo':2015ria,Billo':2015jta}.

Let us now turn to the twisted superpotential ${\mathcal W}$. As we have seen in (\ref{SonFWclass}), 
S-duality acts very simply at the classical level but some subtleties arise in the quantum theory.
We now make a few important points, anticipating some results of the next sections. 
It turns out that $\mathcal{W}$ receives contributions so that the coefficients ${w}_{\ell}$ do not 
have a well-defined modular weight.
However, these anomalous terms depend only on the coupling constant $\tau$ and the vacuum expectation
values $\vec{a}$. In particular, they are independent of the continuous parameters $z_I$
that characterize the surface operator. For this reason it
is convenient to consider the $z_I$ derivatives of the superpotential: 
\begin{equation}
{\mathcal W}^{(I)}:=\,\frac{1}{2\pi\ii}\frac{\partial{\mathcal W}}{\partial z_I} = a_I + \sum_{\ell=1}^\infty
w_\ell^{(I)}(\tau,z_I,\vec{a})
\label{WI}
\end{equation}
where, of course, $w_\ell^{(I)}:= \frac{1}{2\pi\ii}\,\frac{\partial w_\ell}{\partial z_I}$.

Combining intuition from the classical S-duality transformation (\ref{SonWclass}) with the
fact that the $z_I$-derivative increases the modular weight by one, and introduces 
an extra factor of $(-\tau)$ under S-duality, we are naturally led to propose that
\begin{equation}
S\big({\mathcal W}^{(I)}\big) = \tau\,{\mathcal W}^{(I)}~.
\label{SonW}
\end{equation}
This constraint can be solved if we assume that the coefficients $w_\ell^{(I)}$ 
satisfy the following properties (which are simple generalizations of those satisfied by $f_{\ell}$):
\begin{itemize}
\item They are homogeneous functions of $\vec{a}$ and satisfy the scaling relation
\begin{equation}
w_\ell^{(I)}(\tau,z_I,\lambda\,\vec{a}) = \lambda^{1-\ell}\,w_\ell^{(I)}(\tau,z_I,\vec{a})~.
\label{gellscaling}
\end{equation}
Given that the twisted superpotential has mass-dimension one, it follows that $w_\ell^{(I)}$ must be
homogeneous polynomials of degree $\ell$ in $m$ and $\epsilon_1$.
\item
The dependence of $w_{\ell}^{(I)}$ on $\tau$ and $z_I$ is only through linear 
combinations of quasi-modular forms made up with the Eisenstein series and elliptic 
functions with total weight $\ell$, such that
\begin{equation}
w_{\ell}^{(I)}\Bigl(\!-\ft{1}{\tau},-\ft{z_I}{\tau},\vec{a}\Bigr) = \tau^{\ell}\,
w_{\ell}^{(I)}(\tau,z_I,\vec{a})\Big|_{E_2\to E_2 + \delta}~.
\label{Sonwell}
\end{equation}
\end{itemize}
We are now ready to discuss how S-duality acts on the superpotential coefficients $w_{\ell}^{(I)}$.
Recalling that
\begin{equation}
S(\vec{a})=\vec{a}_{\text{D}}:= \,\frac{1}{2\pi\ii}\,\frac{\partial\mathcal{F}}{\partial \vec{a}}=
\tau\,\vec{a}+\frac{1}{2\pi\ii}\,\frac{\partial f}{\partial \vec{a}}=
\tau\Big(\vec{a}+\frac{\delta}{12}\,\frac{\partial f }{\partial \vec{a}}\Big)
\label{adual}
\end{equation}
where $f=\mathcal{F}_{\mathrm{pert}}+\mathcal{F}_{\mathrm{inst}}$, we have
\begin{equation}
\begin{aligned}
S\big( w_\ell^{(I)}\big)&= w_{\ell}^{(I)}\Bigl(\!-\ft{1}{\tau},-\ft{z_I}{\tau},\vec{a}_{\text{D}}\Bigr)
=\tau^\ell \,w_{\ell}^{(I)}(\tau,z_I,\vec{a}_{\text{D}})\Big|_{E_2\to E_2 + \delta}\phantom{\Bigg|}\\
&=\tau\,w_{\ell}^{(I)}\Bigl(\tau,z_I, \vec{a}
+\ft{\delta}{12}\,\ft{\partial f }{\partial \vec{a}}\Bigr)\Big|_{E_2\to E_2 + \delta}
\end{aligned}
\end{equation}
where in the last step we exploited the scaling behavior (\ref{gellscaling}) together with (\ref{adual}). 
Using this result in (\ref{WI}) and formally expanding in $\delta$, we obtain
\begin{equation}
\begin{aligned}
\frac{1}{\tau}\,S\big({\mathcal W}^{(I)}\big)&= {\mathcal W}^{(I)}\Bigl(\tau,z_I, \vec{a}
+\ft{\delta}{12}\,\ft{\partial f }{\partial \vec{a}}\Bigr)\Big|_{E_2\to E_2 + \delta}\phantom{\Bigg|}\\
&={\mathcal W}^{(I)}+\delta\,
\bigg(\frac{\partial{\mathcal W}^{(I)}}{\partial E_2}
+\frac{1}{12}\frac{\partial{\mathcal W}^{(I)}}{\partial \vec{a}}\cdot
\frac{\partial f}{\partial \vec{a}}\bigg)+\mathcal{O}(\delta^2)~.
\end{aligned}
\end{equation}
The constraint (\ref{SonW}) is satisfied if 
\begin{equation}
\frac{\partial{\mathcal W}^{(I)}}{\partial E_2}+
\frac{1}{12}\frac{\partial{\mathcal W}^{(I)}}{\partial \vec{a}}\cdot\frac{\partial f}{\partial \vec{a}}=0~,
\label{mae}
\end{equation}
which also implies the vanishing of all terms of higher order in $\delta$. This modular anomaly equation
can be equivalently written as
\begin{equation}
\label{mod ano eqn}
\frac{\partial w^{(I)}_{\ell}}{\partial E_2}
+\frac{1}{12}\sum_{n=0}^{\ell-1}\frac{\partial f_{\ell-n}}{\partial \vec{a}}\cdot
\frac{\partial w^{(I)}_{n}}{\partial\vec{a}}=0
\end{equation}
where we have defined $w_0^{(I)}=a_I$.

In the next sections we will solve this modular anomaly equation and determine the superpotential coefficients
$w^{(I)}_{\ell}$ in terms of Eisenstein series and elliptic functions; we will also show that by 
considering the expansion of these quasi-modular functions we recover precisely all
instanton contributions computed using localization, 
thus providing a very strong and highly non-trivial consistency
check on our proposal (\ref{SonW}) and on our entire construction. Since the explicit results are quite involved 
in the general case, we will start by discussing the SU$(2)$ theory.

\vspace{0.5cm}
\section{Surface operators in $\mathcal{N}=2^\star$ SU(2) theory}
\label{SU2}

We now consider the simplest ${\mathcal N}=2^{\star}$ theory with gauge group SU(2) and solve in this case
the modular anomaly equation (\ref{mod ano eqn}). A slight modification from the earlier discussion is 
needed since for SU(2) the Coulomb vacuum expectation value of the adjoint scalar field takes the form
$\langle \phi \rangle=\mathrm{diag}(a,-a)$ and the index $I$ used in the previous section only takes one value,
namely $I=2$. Thus we have a single $z$-parameter, corresponding to the unique surface operator we can 
have in the theory, and (\ref{WI}) becomes
\begin{equation}
{\mathcal W}^{\,\prime}:=\,\frac{1}{2\pi\ii}\frac{\partial{\mathcal W}}{\partial z} = -a + \sum_{\ell=1}^\infty
w_\ell^{\prime}
\label{Wprime}
\end{equation}
with $w_\ell^{\prime}:= \frac{1}{2\pi\ii}\,\frac{\partial w_\ell}{\partial z}$, while the recurrence relation
(\ref{mod ano eqn}) becomes
\begin{equation}
\label{modanoeqnSU2}
\frac{\partial w_\ell^{\prime}}{\partial E_2}+\frac{1}{24}\sum_{n=0}^{\ell-1}
\frac{\partial f_{\ell-n}}{\partial a}\,\frac{\partial w_n^{\prime}}{\partial a}=0 
\end{equation}
with the initial condition $w_0^{\prime}=-a$. The coefficient $w_1$ and its $z$-derivative $w_1^{\prime}$ do
not depend on $a$ and are therefore irrelevant for the IR dynamics on the surface operator. Moreover, $w_1^{\prime}$
drops out of the anomaly equation and plays no role in determining $w_\ell^{\prime}$
for higher values of $\ell$.
Nevertheless, for completeness, we observe that if we use the elliptic 
function
\begin{equation}
\label{h1defn0}
h_1(z|\tau) = \frac{1}{2\pi\ii} \frac{\p}{\p z} \log\theta_1(z|\tau)
\end{equation}
where $\theta_1(z|\tau)$ is the first Jacobi $\theta$-function, and exploit the expansion
reported in (\ref{h1exp}), comparing with the instanton expansion (\ref{wprime1}) 
obtained from localization, we are immediately led to,
\begin{equation}
w_1^{\prime} = \Big(m-\frac{\epsilon_1}{2}\Big)\Big(h_1+ \frac{1}{2}\Big)~.
\label{w1prime}
\end{equation}
By expanding $h_1$ to higher orders one can ``predict'' all higher 
instanton contributions to $w_1^{\prime}$. We have checked that these predictions perfectly match 
the explicit results obtained from localization methods involving Young tableaux with up to six boxes.

The first case in which the modular anomaly equation (\ref{modanoeqnSU2}) shows its power is the case $\ell=2$.
Recalling that the prepotential coefficients $f_n$ with $n$ odd vanish, we have
\begin{equation}
\label{l=2recursion}
\frac{\partial w_2^{\prime}}{\partial E_2}+\frac{1}{24}\frac{\partial f_2}{\partial a}\,
\frac{\partial w_0^{\prime}}{\partial a}=0~.
\end{equation}
Using the initial condition $w_0^{\prime}=-a$, substituting the exact expression for $f_2$  
given in (\ref{f2app}) and then integrating, we get
\begin{equation}
w_2^{\prime}=\frac{1}{24a}\Big(m^2-\frac{\epsilon_1^2}{4}\Big)\,\big(E_2+\text{modular term}\big)~.
\label{w2prime0}
\end{equation}
At this juncture, it is important to observe that the elliptic and modular forms of SL$(2,\mathbb{Z})$, which 
are allowed to appear in the superpotential coefficients, are polynomials in the ring generated 
by the Weierstra\ss~function $\wp(z|\tau)$ and its $z$-derivative $\wp^\prime(z|\tau)$, and by
the Eisenstein series $E_4$ and $E_6$. These basis elements have weights $2,3,4$ and $6$ respectively. 
We refer to Appendix \ref{useful} for a collection of useful formulas for these elliptic and modular forms
and for their perturbative expansions. Since $w_2^{\prime}$ must have weight 2, the
modular term in (\ref{w2prime0}) is restricted to be proportional to the Weierstra\ss~function, namely
\begin{equation}
w_2^{\prime}=\frac{1}{24a}\Big(m^2-\frac{\epsilon_1^2}{4}\Big)\,\Big(E_2+\alpha\,\frac{\wp}{4\pi^2}\Big)
\label{w2prime1}
\end{equation}
where $\alpha$ is a constant. Therefore our proposal works only if by fixing a \emph{single} parameter
$\alpha$ we can match \emph{all} the microscopic contributions to $w_2^\prime$ 
computed in the previous sections. Given the many constraints that this requirement puts, it is not at all 
obvious that it works. But actually it does! Indeed, using the expansions
of $E_2$ and $\widetilde\wp=\frac{\wp}{4\pi^2}$ given in (\ref{Ekexp}) and (\ref{wpexp}) respectively,
and comparing with (\ref{wprime2}), one finds a perfect match if $\alpha=12$. Thus, the \emph{exact}
expression of $w_2^{\prime}$ is
\begin{equation}
w_2^{\prime}=\frac{1}{24a}\Big(m^2-\frac{\epsilon_1^2}{4}\Big)\,\Big(E_2+12\,\widetilde\wp\Big)~.
\label{w2exact}
\end{equation}
We have checked up to order six that the all instanton corrections predicted by this formula
completely agree with the microscopic results obtained from localization.

Let us now consider the modular anomaly equation (\ref{modanoeqnSU2}) for $\ell=3$. In this case since
$w_1^{\prime}$ is $a$-independent and the coefficients $f_n$ with $n$ odd vanish, we simply have
\begin{equation}
\frac{\partial w_3^{\prime}}{\partial E_2}=0
\label{modanom3}
\end{equation}
According to our proposal, $w_3^{\prime}$ must be an elliptic function with modular weight 3, and in 
view of (\ref{modanom3}), the only candidate is the derivative of the Weierstra\ss~function $\wp^\prime$.
By comparing the expansion (\ref{wprimeexp}) with the semi-classical results (\ref{wprime3}) we find a perfect
match and obtain
\begin{equation}
w_3^{\prime}=\frac{\epsilon_1}{4a^2}\Big(m^2-\frac{\epsilon_1^2}{4}\Big)\widetilde\wp^{\,\prime}~.
\label{w3exact}
\end{equation}
Again we have checked that the higher order instanton corrections predicted by this formula agree
with the localization results up to order six.

A similar analysis can done for higher values of $\ell$ without difficulty. Obtaining the anomalous behavior 
by integrating the modular anomaly equation, and fixing the coefficients of the modular terms by 
comparing with the localization results, after a bit of elementary algebra, we get
\begin{align}
w_4^{\prime}&=\frac{1}{1152 a^3}\Big(m^2-\frac{\epsilon_1^2}{4}\Big)
\bigg[\!\Big(m^2-\frac{\epsilon_1^2}{4}\Big)\big(2E_2^2-E_4+24E_2\,\widetilde\wp+144\widetilde\wp^2\big)
+6\,\epsilon_1^2\big(E_4-144\widetilde\wp^2\big)\!\bigg]\notag~,\\
\notag \\
w_5^{\prime}&=\frac{\epsilon_1}{48 a^4}\Big(m^2-\frac{\epsilon_1^2}{4}\Big)
\bigg[\Big(m^2-\frac{\epsilon_1^2}{4}\Big)\big(E_2+12\widetilde\wp\big)\widetilde\wp^{\,\prime}-36\,
\epsilon_1^2\,\widetilde\wp\,\widetilde\wp^{\,\prime}\bigg]\label{w5primeexact}~,\\
\notag\\
w_6^{\prime}&=\frac{1}{138240a^5}\Big(m^2-\frac{\epsilon_1^2}{4}\Big)\bigg[
\Big(m^2-\frac{\epsilon_1^2}{4}\Big)^2\big(20E_2^3-11E_2E_4-4E_6+240E_2^2\,\widetilde\wp
-60E_4\,\widetilde\wp\notag\\
&\qquad~~+2160E_2\,\widetilde\wp^2+8640\widetilde\wp^3\big)
+2\Big(m^2-\frac{\epsilon_1^2}{4}\Big)\epsilon_1^2\,
\big(39E_2E_4+56E_6+1440E_4\,\widetilde\wp\phantom{\bigg|}\notag\\
&\qquad~~-6480E_2\,\widetilde\wp^2-120960\widetilde\wp^3\big)-240\,\epsilon_1^4
\,\big(E_6+27E_4\,\widetilde\wp
-2160\widetilde\wp^3\big)\bigg]~,\notag
\end{align}
and so on. The complete agreement with the microscopic localization results of the above expressions
provides very strong and highly non-trivial evidence for the validity of the modular anomaly equation
and the S-duality properties of the superpotential, and hence of our entire construction.

Exploiting the properties of the function $h_1$ defined in (\ref{h1defn0}) and its relation with the
Weierstra\ss~function (see Appendix~\ref{useful}), 
it is possible to rewrite the above expressions as total $z$-derivatives. 
Indeed, we find
\begin{align}
w_2^{\prime}&=\frac{1}{2a}\Big(m^2-\frac{\epsilon_1^2}{4}\Big)h_1^\prime~,
\qquad\qquad
w_3^{\prime}=\frac{\epsilon_1}{4a^2}\Big(m^2-\frac{\epsilon_1^2}{4}\Big)h_1^{\prime\prime}~,\notag\\
w_4^{\prime}&=\frac{1}{48a^3}\Big(m^2-\frac{\epsilon_1^2}{4}\Big)
\bigg[\Big(m^2-\frac{\epsilon_1^2}{4}\Big)
\big(E_2\,h_1-h_1^{\prime\prime}\big)+
6\,\epsilon_1^2\,h_1^{\prime\prime}\bigg]^\prime~,\label{wnprime}\\
w_5^{\prime}&=\frac{\epsilon_1}{8a^4}\Big(m^2-\frac{\epsilon_1^2}{4}\Big)
\bigg[\Big(m^2-\frac{\epsilon_1^2}{4}\Big)(h_1^\prime)^2
+\frac{\epsilon_1^2}{2}\big(E_2-6 h_1^\prime\big)h_1^{\prime}
\bigg]^\prime~.\notag
\end{align}
We have checked that the same is also true for $w_6^{\prime}$ (and for a few higher coefficients as well),
which however we do not write explicitly for brevity.
Of course this is to be expected since they are the coefficients of the expansion of the 
derivative of the superpotential. The latter can then be simply obtained by integrating with respect to $z$
and fixing the integration constants  by comparing with the explicit localization results. In this way
we obtain\,\footnote{We neglect the $a$-independent terms 
originating from (\ref{w1prime}) since they are irrelevant 
for the infrared dynamics on the defect.}
\begin{align}
{\mathcal W}= - 2\pi\ii z\,a+ \sum_n w_n
\end{align}
with
\begin{align}
w_2&=\frac{1}{2a}\Big(m^2-\frac{\epsilon_1^2}{4}\Big)h_1~,\qquad\quad
w_3=\frac{\epsilon_1}{4a^2}\Big(m^2-\frac{\epsilon_1^2}{4}\Big)h_1^{\prime}
~,\label{Wnlocalization}\\
w_4&=\frac{1}{48a^3}\Big(m^2-\frac{\epsilon_1^2}{4}\Big)
\bigg[\Big(m^2-\frac{\epsilon_1^2}{4}\Big)
\big(E_2\,h_1-h_1^{\prime\prime}\big)+
6\,\epsilon_1^2\,h_1^{\prime\prime}
+\frac{1}{2}\Big(m^2-\frac{\epsilon_1^2}{4}\Big)\big(E_2-1)\bigg]~,\notag\\
w_5&=\frac{\epsilon_1}{8a^4}\Big(m^2-\frac{\epsilon_1^2}{4}\Big)
\bigg[\Big(m^2-\frac{\epsilon_1^2}{4}\Big)(h_1^\prime)^2
+\frac{\epsilon_1^2}{2}\big(E_2-6 h_1^\prime\big)h_1^{\prime}
+\frac{1}{96}\Big(m^2-\frac{9\epsilon_1^2}{4}\Big)\big(E_2^2-E_4\big)
\bigg]~,\notag
\end{align}
and so on. Note that, as anticipated in the previous section, the coefficients $w_{n}$ do not have a 
homogeneous modular weight.

\subsection{Relation to CFT results}
\label{secn:CFT}
So far we have studied the twisted superpotential and its $z$-derivative as semi-classical expansions 
for large $a$. 
However, we can also arrange these expansions in terms of the deformation parameter $\epsilon_1$.
For example, using the results in (\ref{w2exact}), (\ref{w3exact}) and (\ref{w5primeexact}), we 
obtain
\begin{equation}
{\mathcal W}^{\,\prime}=-a + \sum_{n=0}^{\infty} \epsilon_1^n\, {\mathcal W}^{\,\prime}_n
\label{wprimeexp1}
\end{equation}
where 
\begin{align}
\phantom{\bigg|}{\mathcal W}^{\,\prime}_0 &= \frac{m^2}{24 a} \big(E_2+12 \widetilde{\wp}\big)
+\frac{m^4}{1152 a^3} \big(2 E_2^2-E_4+24E_2\, \widetilde{\wp}+144 \widetilde{\wp}^2\big)
+\frac{m^6}{138240 a^5} \big(20 E_2^3\notag\\
&~~~-11 E_2 E_4-4 E_6+240 E_2^2 \,\widetilde{\wp}-60 E_4\,
\widetilde{\wp}+2160 E_2\, \widetilde{\wp}^2+8640 \widetilde{\wp}^3\big)
+\mathcal{O}\big(a^{-7}\big)~,\phantom{\bigg|}\notag\\
\phantom{\bigg|}{\mathcal W}^{\,\prime}_1 &= \frac{m^2}{4 a^2}\, \widetilde{\wp}^{\,\prime} 
+ \frac{m^4}{48 a^4}\big(E_2+12 \widetilde{\wp}\big)\widetilde{\wp}^{\,\prime} 
+\mathcal{O}\big(a^{-6}\big)~,\notag\\
\phantom{\bigg|}{\mathcal W}^{\,\prime}_2&= -\frac{1}{96 a}\big(E_2+12 \wp\big)
-\frac{m^2}{2304 a^3} \big(2E_2^2-13E_4+24 E_2\, \widetilde{\wp}+1872 \widetilde{\wp}^2\big)\\
&~~~\,-\frac{m^4}{110592 a^5} \big(12 E_2^3-69 E_2 E_4-92 E_6+144 E_2^2 \,\widetilde{\wp}
-2340 E_4\, \widetilde{\wp}\phantom{\bigg|}\notag\\
&\qquad\qquad\qquad\quad
+11664 {E_2}\, \widetilde{\wp}^2+198720 \widetilde{\wp}^3\big)+\mathcal{O}\big(a^{-7}\big)~,\phantom{\bigg|}\notag\\
\phantom{\bigg|}{\mathcal W}^{\,\prime}_3&= -\frac{1}{16 a^2}\,\widetilde{\wp}^{\,\prime}
-\frac{m^2}{96 a^4} \big(E_2+84 \widetilde{\wp}\big)\widetilde{\wp}^{\,\prime} 
+\mathcal{O}\big(a^{-6}\big)~,\notag
\end{align}
and so on. 
Quite remarkably, up to a sign flip $a\rightarrow -a$, these expressions precisely coincide 
with the results obtained in \cite{KashaniPoor:2012wb} from the null-vector decoupling 
equation for the toroidal 1-point conformal block in the Liouville theory.

We would like to elaborate a bit on this match. Let us first recall that in the so-called AGT 
correspondence \cite{Alday:2009aq} the toroidal 1-point conformal block of a Virasoro primary field $V$ 
in the Liouville theory is related to the Nekrasov partition function of the $\mathcal{N}=2^{\star}$ 
SU(2) gauge theory. In \cite{Alday:2009fs} it was shown that the insertion of 
the degenerate null-vector $V_{2,1}$ in the Liouville conformal block corresponds to the partition function 
of the SU(2) theory in the presence of a surface operator. In the semi-classical limit of the Liouville theory
(which corresponds to the NS limit $\epsilon_2\rightarrow 0$), one has \cite{Alday:2009fs,KashaniPoor:2012wb}
\begin{equation}
\vev{V(0) V_{2,1}(z)}_{\text{torus}} \simeq {\mathcal N}\, 
\exp\Big(\!\!-\frac{\mathcal{F}}{\epsilon_1\epsilon_2} 
+ \frac{{\mathcal W}(z)}{\epsilon_1}+\cdots \Big)~,
\end{equation}
where ${\mathcal N}$ is a suitable normalization factor. In \cite{KashaniPoor:2012wb} 
the null-vector decoupling equation satisfied by the degenerate conformal block was used to explicitly
calculate the prepotential $\mathcal{F}$ and the $z$-derivative of the twisted effective 
superpotential $\mathcal{W}^\prime$ for the ${\mathcal N}=2^{\star}$ SU(2) theory,
which fully agrees with the one we have obtained using the modular anomaly equation and localization
methods.  It is important to keep in mind that the insertion of the degenerate field $V_{2,1}$ in the
Liouville theory corresponds to the insertion of a surface operator of codimension-4 in 
the six-dimensional $(2,0)$ theory. In the brane picture, this defect corresponds to an M2 brane ending on the 
M5 branes that wrap a Riemann surface and support the gauge theory in four dimensions. On the other hand,
as explained in the introduction, the results we have obtained using the orbifold construction and
localization pertain to a surface operator of codimension-2 in the six dimensional theory, corresponding to
an M$5^\prime$ intersecting the original M5 branes. The equality between our results and those of
\cite{KashaniPoor:2012wb} supports 
the proposal of a duality between the two types of surface operators in \cite{Frenkel:2015rda}. 
This also supports the conjecture of \cite{Wyllard:2010vi}, based on 
\cite{Awata:2010bz,Kozcaz:2010yp,Wyllard:2010rp}, that in the presence of simple surface operators 
the instanton partition function is insensitive to whether they are realized as codimension-2 or 
codimension-4 operators. 
In Section~\ref{purecase} we will comment on such relations in the case of higher rank gauge groups 
and will also make contact with the results for the twisted chiral rings when the surface defect 
is realized by coupling two-dimensional sigma-models to pure ${\mathcal N}=2$ SU(N) gauge theory.

\vspace{0.5cm}
\section{Surface operators in $\mathcal{N}=2^\star$ SU($N$) theories}
\label{SUN}

We now generalize the previous analysis to SU($N$) gauge groups. 
As discussed in Section~\ref{SOmicro}, in the higher rank cases there are many types 
of surface operators corresponding to the different partitions of $N$.
We start our discussion from simple surface operators of type $\{1,(N-1)\}$.

\subsection{Simple surface operators}
\label{subsecn:simpleN}
In the case of the simple partition $\{1,(N-1)\}$, the vector $\vec{t}$ of the electro-magnetic parameters
characterizing the surface operator takes the form
\begin{equation}
\vec{t} =
\left\{
\begin{array}{cccc}
t_1,&  \undermat{N-1}{ t_2, & \ldots, &t_2}
\end{array}
\right\} ~.
\vspace{.5cm}
\label{tvec12}
\end{equation}
Correspondingly, the classical contribution to the twisted effective superpotential becomes
\begin{equation}
{\mathcal W}_{\text{class}} = 2\pi\ii\, \vec{t}\cdot\vec{a}=2\pi\ii \,
\Big(a_1 \,t_1 +t_2 \sum_{i=2}^N a_i \Big) = -2\pi\ii \, z\,a_1
\end{equation}
where we have used the tracelessness condition on the vacuum expectation values and, according to
(\ref{zdeft}), have defined $z = t_2-t_1$.

When quantum corrections are included, one finds that the coefficients $w_\ell^\prime$ of the 
$z$-derivative of the superpotential satisfy the modular anomaly equation (\ref{mod ano eqn}). 
The solution
of this equation proceeds along the same lines as in the SU(2) case, although new structures, 
involving the differences $a_{ij}=a_i-a_j$, appear. We omit details of the calculations and 
merely present the results. As 
for the SU(2) theory, the coefficients can be compactly written in terms of modular
and elliptic functions, particularly the second Eisenstein series and the function $h_1$ 
defined in (\ref{h1defn0}). For clarity, and also for later convenience, we indicate the dependence on $z$ but
understand the dependence on $\tau$ in $h_1$. The first few coefficients $w_\ell^\prime$ are
\begin{subequations}
\begin{align}
{\phantom{\Bigg|}}w_2^{\prime}&=\Big(m^2-\frac{\epsilon_1^2}{4}\Big)\sum_{j=2}^N
\frac{h_1^\prime(z)}{a_{1j}}~,\label{w2prime}\\
{\phantom{\Bigg|}}w_3^{\prime}&=\epsilon_1\Big(m^2-\frac{\epsilon_1^2}{4}\Big)
\sum_{j=2}^N\frac{h_1^{\prime\prime}(z)}{a_{1j}^2}+\frac{1}{2}
\Big(m^2-\frac{\epsilon_1^2}{4}\Big)\Big(m+\frac{\epsilon_1}{2}\Big)
\sum_{j\ne k =2}^N\frac{h_1^{\prime\prime}(z)}{a_{1j}\,a_{1k}}~,\label{w3prime}\\
{\phantom{\Bigg|}}w_4^{\prime}&=\frac{1}{6}\Big(m^2-\frac{\epsilon_1^2}{4}\Big)
\bigg[
\Big(m^2-\frac{\epsilon_1^2}{4}\Big)\big(E_2\,h_1^{\prime}(z)-h_1^{\prime\prime\prime}(z)\big)
+6\,\epsilon_1^2\,h_1^{\prime\prime\prime}(z)\bigg]\sum_{j=2}^N\frac{1}{a_{1j}^3}\notag\\
&\qquad+\epsilon_1\Big(m^2-\frac{\epsilon_1^2}{4}\Big)\Big(m+\frac{\epsilon_1}{2}\Big)\!\!
\sum_{j\ne k=2}^N\frac{h_1^{\prime\prime\prime}(z)}{a_{1j}^2\,a_{1k}}\label{w4prime}\\
&\qquad+\frac{1}{6}\Big(m^2-\frac{\epsilon_1^2}{4}\Big)
\Big(m+\frac{\epsilon_1}{2}\Big)^2\!\!\!
\sum_{j\ne k \ne \ell=2}^N\frac{h_1^{\prime\prime\prime}(z)}{a_{1j}\,
a_{1k}\,a_{1\ell}}~,\notag
\end{align}
\label{wellprime}
\end{subequations}
\hspace{-0.13cm}and so on. 
We have explicitly checked the above formulas against localization results up to SU(7) finding 
complete agreement.
It is easy to realize that for $N=2$ only the highest order poles contribute and the corresponding expressions
precisely coincide with the results in the previous section.
In the higher rank cases, there are also contributions from structures with lesser order poles 
that are made possible because of the larger number of Coulomb parameters. 
Furthermore, we observe that there is no pole when $a_j$ approaches $a_k$ with $j,k=2,\cdots\!,N$. 

It is interesting to observe that the above expressions can be rewritten in a suggestive form 
using the root system $\Phi$ of SU($N$). The key observation is that using the vector $\vec{t}$ 
defined in (\ref{tvec12}) 
we can select a subset of roots $\Psi\subset\Phi$ such that their scalar products with the vector 
$\vec{a}$ of the vacuum expectation values produce exactly all the factors of $a_{1j}$
in the denominators of (\ref{wellprime}). Defining
\begin{equation}
\Psi=\big\{\vec{\alpha}\in\Phi~\big|~\vec{\alpha}\cdot\vec{t}+z=0\big\}~,
\label{Psi}
\end{equation}
one can verify that for any $\vec{\alpha}\in\Psi$, the scalar product $\vec{\alpha}\cdot\vec{a}$
is of the form $a_{1j}$. Therefore, $w_2^\prime$ in (\ref{w2prime}) can be written
as
\begin{equation}
\begin{aligned}
w_2^{\prime}&=\Big(m^2-\frac{\epsilon_1^2}{4}\Big)\sum_{\vec{\alpha}\in\Psi}
\frac{h_1^\prime(-\vec{\alpha}\cdot\vec{t})}{{\phantom{\big|}}\vec{\alpha}\cdot\vec{a}}
=\Big(m^2-\frac{\epsilon_1^2}{4}\Big)\sum_{\vec{\alpha}\in\Psi}
\frac{h_1^\prime(\vec{\alpha}\cdot\vec{t})}{{\phantom{\big|}}\vec{\alpha}\cdot\vec{a}}
\end{aligned}
\label{w2prime2}
\end{equation}
where in the last step we used the fact that $h_1^\prime$ is an even function. Similarly the other 
coefficients in (\ref{wellprime}) can also be rewritten using the roots of SU($N$). Indeed, introducing the
subsets of $\Psi$ defined as\,\footnote{These definitions are analogous to the 
ones used in \cite{Billo':2015ria,Billo':2015jta} to define the
root lattice sums appearing in the prepotential; see also (\ref{Psialphaapp}).}
\begin{equation}
\begin{aligned}
&\Psi(\vec{\alpha})=\big\{\vec{\beta}\in\Psi~\big|~\vec{\alpha}\cdot\vec{\beta}=1\big\}~,\\
&\Psi(\vec{\alpha},\vec{\beta})=\big\{\vec{\gamma}\in\Psi~\big|~\vec{\alpha}\cdot\vec{\gamma}
=\vec{\beta}\cdot\vec{\gamma}=1\big\}~,
\end{aligned}
\label{Psialfa}
\end{equation}
we find that $w_3^\prime$ in (\ref{w3prime}) becomes
\begin{equation}
\begin{aligned}
w_3^{\prime}&=-\epsilon_1\Big(m^2-\frac{\epsilon_1^2}{4}\Big)
\sum_{\vec{\alpha}\in\Psi}
\frac{h_1^{\prime\prime}(\vec{\alpha}\cdot\vec{t}\,)}{{\phantom{\Big|}}(\vec{\alpha}\cdot\vec{a}\,)^2}
\\&\qquad
-\frac{1}{2}
\Big(m^2-\frac{\epsilon_1^2}{4}\Big)\Big(m+\frac{\epsilon_1}{2}\Big)
\sum_{\vec{\alpha}\in\Psi}
\sum_{\vec{\beta}\in\Psi(\vec{\alpha})}
\frac{h_1^{\prime\prime}(\vec{\alpha}\cdot\vec{t}\,)}{{\phantom{\Big|}}(\vec{\alpha}\cdot\vec{a}\,)
\,(\vec{\beta}\cdot\vec{a}\,)}~,
\end{aligned}
\label{w3prime2}
\end{equation}
while $w_4^\prime$ in (\ref{w4prime}) is
\begin{equation}
\begin{aligned}
w_4^{\prime}&=\frac{1}{6}\Big(m^2-\frac{\epsilon_1^2}{4}\Big)\bigg[
\Big(m^2-\frac{\epsilon_1^2}{4}\Big)
\sum_{\vec{\alpha}\in\Psi}\frac{E_2\,h_1^{\prime}(\vec{\alpha}\cdot\vec{t}\,)
-h_1^{\prime\prime\prime}(\vec{\alpha}\cdot\vec{t}\,)}{{\phantom{\Big|}}(\vec{\alpha}\cdot\vec{a}\,)^3}
+6\,\epsilon_1^2\sum_{\vec{\alpha}\in\Psi}
\frac{h_1^{\prime\prime\prime}(\vec{\alpha}\cdot\vec{t}\,)}{{\phantom{\Big|}}(\vec{\alpha}\cdot\vec{a}\,)^3}
~\bigg]\\
&\quad+\epsilon_1\Big(m^2-\frac{\epsilon_1^2}{4}\Big)\Big(m+\frac{\epsilon_1}{2}\Big)
\sum_{\vec{\alpha}\in\Psi}\sum_{\vec{\beta}\in\Psi(\vec{\alpha})}
\frac{h_1^{\prime\prime\prime}(\vec{\alpha}\cdot\vec{t}\,)}{{\phantom{\Big|}}(\vec{\alpha}\cdot\vec{a}\,)^2
\,(\vec{\beta}\cdot\vec{a}\,)}{\phantom{\bigg|}}\\
&\quad+\frac{1}{4}\Big(m^2-\frac{\epsilon_1^2}{4}\Big)
\Big(m+\frac{\epsilon_1}{2}\Big)^2\bigg[\sum_{\vec{\alpha}\in\Psi}
\sum_{\vec{\beta}\ne\vec{\gamma}\in\Psi(\vec{\alpha})}
\frac{h_1^{\prime\prime\prime}(\vec{\alpha}\cdot\vec{t}\,)}{{\phantom{\Big|}}(\vec{\alpha}\cdot\vec{a}\,)
\,(\vec{\beta}\cdot\vec{a}\,)\,(\vec{\gamma}\cdot\vec{a}\,)}\\
&\qquad\qquad\qquad\qquad\qquad\qquad\qquad-\frac{1}{3}
\sum_{\vec{\alpha}\in\Psi}\sum_{\vec{\beta}\in\Psi(\vec{\alpha})}
\sum_{\vec{\gamma}\in\Psi(\vec{\alpha},\vec{\beta})}
\frac{h_1^{\prime\prime\prime}(\vec{\alpha}\cdot\vec{t}\,)}{{\phantom{\Big|}}(\vec{\alpha}\cdot\vec{a}\,)
\,(\vec{\beta}\cdot\vec{a}\,)\,(\vec{\gamma}\cdot\vec{a}\,)}\bigg]~.
\end{aligned}
\label{w4prime2}
\end{equation}
We observe that the two sums in the last two lines of (\ref{w4prime2}) are actually equal to each other and 
exactly reproduce the last line of (\ref{w4prime}). 
However, for different sets of roots the two sums are different and lead to different structures. 
Thus, for reasons that will soon become clear, we have kept them separate even in this case.

\vspace{1cm}
\subsection{Surface operators of type $\{p,N-p\}$}

We now discuss a generalization of the simple surface operator in which we still have a single complex variable
$z$ as before, but the type is given by the following vector 
\begin{equation}
\vec{t} =
\left\{
\begin{array}{cccccc}
\undermat{p}{ t_1, & \ldots, &t_1},&  \undermat{N-p}{ t_2, & \ldots, &t_2}
\end{array}
\right\} ~.
\vspace{0.5cm}
\label{tvecpN}
\end{equation}
In this case, using the tracelessness condition on the vacuum expectation values, 
the classical contribution to the superpotential is 
\begin{equation}
{\mathcal W}_{\text{class}} = 2\pi\ii\bigg(t_1 \sum_{i=1}^p a_i  +t_2 \sum_{j=p+1}^N a_j \bigg)
= - 2\pi\ii \,z\,\sum_{i=1}^p a_i
\end{equation}
where again we have defined $z=t_2-t_1$.

It turns out that the quantum corrections to the $z$-derivatives of the
superpotential are given exactly by the \emph{same} formulas
(\ref{w2prime2}), (\ref{w3prime2}) and (\ref{w4prime2}) in which the only difference is in the subsets 
of the root system $\Phi$ that have to be considered in the lattice sums. 
These subsets are still defined as in (\ref{Psi}) and (\ref{Psialfa}) but with the 
vector $\vec{t}$ given by (\ref{tvecpN}). We observe that 
in this case the two last sums in (\ref{w4prime2}) are different.
We have verified these formulas against the localization results up to SU(7) finding perfect agreement.
The fact that the superpotential coefficients can be formally written in the same way for all unitary groups
and for all types with two entries, suggests that probably universal formulas should exist 
for surface operators with more than two distinct entries in the $\vec{t}$-vector.
This is indeed what happens as we will show in the next subsection.

\subsection{Surface operators of general type}
A surface operator of general type corresponds to splitting the SU($N$) gauge group as in (\ref{split}) which
leads to the following partition of the Coulomb parameters
\begin{equation}
\vec{a} =\left\{
\begin{array}{ccccccccc}
\undermat{n_1}{a_1, & \cdots & a_{n_1},} & \undermat{n_2}{ a_{n_1+1}, & \cdots&a_{n_1+n_2},} &\cdots, 
&\undermat{n_M}{ a_{N-n_M+1}, &\ldots a_{N} }
\end{array}
\right\} ~,
\label{agen}
\end{equation}

\noindent
and to the following $\vec{t}$-vector
\begin{equation}
\vec{t} =
\left\{
\begin{array}{ccccccccc}
\undermat{n_1}{t_1, & \cdots, & t_1,} & \undermat{n_2}{t_2, & \cdots,&t_2,}&\cdots ,
& \undermat{n_M}{ t_M, &\cdots, t_M }
\end{array}
\right\}
\end{equation}
with 
\begin{equation}
\sum_{I=1}^M n_I=N~.
\end{equation}
In this case we therefore have several variables $z_I$ defined as in (\ref{zdeft}), and several 
combinations of elliptic functions evaluated at different points. However, if we use the
root system $\Phi$ of SU($N$) the structure of the superpotential coefficients is very 
similar to what we have seen before in the simplest cases. 
To see this, let us first define the following subsets\,\footnote{When $J=1$ one must take
$z_1=0$.}
of $\Phi$:
\begin{equation}
\begin{aligned}
&\Psi_{IJ}=\big\{\vec{\alpha}\in\Phi~\big|~\vec{\alpha}\cdot\vec{t}+z_I-z_J=0\big\}~,\\
&\Psi_{IJ}(\vec{\alpha})=\big\{\vec{\beta}\in\Psi_{IJ}~\big|~\vec{\alpha}\cdot\vec{\beta}=1\big\}~,\\
&\Psi_{IJ}(\vec{\alpha},\vec{\beta})=\big\{\vec{\gamma}\in\Psi_{IJ}~\big|~\vec{\alpha}
\cdot\vec{\gamma}=\vec{\beta}\cdot\vec{\gamma}=1\big\}
\end{aligned}
\end{equation}
which are obvious generalizations of the definitions (\ref{Psi}) and (\ref{Psialfa}). Then, writing
\begin{equation}
{\mathcal{W}^{(I)}}=\frac{1}{2\pi\ii}\,\frac{\partial\mathcal{W}}{\partial z_I}= 
a_{I_1}+\cdots a_{I_{n_I}}+\sum_{\ell}w_\ell^{(I)}~,
\end{equation}
for $I=2,\cdots,M$, we find that the first few coefficients $w_\ell^{(I)}$ are given by
\begin{align}
{\phantom{\Bigg|}}w_2^{(I)}&=\Big(m^2-\frac{\epsilon_1^2}{4}\Big)\sum_{J\ne I}\,\sum_{\vec{\alpha}\in\Psi_{IJ}}
\frac{h_1^\prime(\vec{\alpha}\cdot\vec{t}\,)}{{\phantom{\big|}}\vec{\alpha}\cdot\vec{a}}~,\\
{\phantom{\Bigg|}}w_3^{(I)}&=-\epsilon_1\Big(m^2-\frac{\epsilon_1^2}{4}\Big)
\sum_{J\ne I}\,\sum_{\vec{\alpha}\in\Psi_{IJ}}
\frac{h_1^{\prime\prime}(\vec{\alpha}\cdot\vec{t}\,)}{{\phantom{\Big|}}(\vec{\alpha}\cdot\vec{a}\,)^2}
\notag\\
&\qquad-\frac{1}{2}
\Big(m^2-\frac{\epsilon_1^2}{4}\Big)\Big(m+\frac{\epsilon_1}{2}\Big)
\sum_{J\ne I}\,\sum_{\vec{\alpha}\in\Psi_{IJ}}
\sum_{\vec{\beta}\in\Psi_{IJ}(\vec{\alpha})}
\frac{h_1^{\prime\prime}(\vec{\alpha}\cdot\vec{t}\,)}{{\phantom{\Big|}}(\vec{\alpha}\cdot\vec{a}\,)
\,(\vec{\beta}\cdot\vec{a}\,)}~,\\
{\phantom{\Bigg|}}w_4^{(I)}&=\frac{1}{6}\Big(m^2-\frac{\epsilon_1^2}{4}\Big)\bigg[\!
\Big(m^2-\frac{\epsilon_1^2}{4}\Big)\!
\sum_{J\ne I}\sum_{\vec{\alpha}\in\Psi_{IJ}}\!\!\frac{E_2\,h_1^{\prime}(\vec{\alpha}\cdot\vec{t}\,)
-h_1^{\prime\prime\prime}(\vec{\alpha}\cdot\vec{t}\,)}{{\phantom{\Big|}}(\vec{\alpha}\cdot\vec{a}\,)^3}
\notag\\
&\qquad\qquad\qquad\qquad+6\,\epsilon_1^2\sum_{J\ne I}\sum_{\vec{\alpha}\in\Psi_{IJ}}\!\!
\frac{h_1^{\prime\prime\prime}(\vec{\alpha}\cdot\vec{t}\,)}{{\phantom{\Big|}}
(\vec{\alpha}\cdot\vec{a}\,)^3}\bigg]\notag\\
&\quad+\epsilon_1\Big(m^2-\frac{\epsilon_1^2}{4}\Big)\Big(m+\frac{\epsilon_1}{2}\Big)
\sum_{J\ne I}\sum_{\vec{\alpha}\in\Psi_{IJ}}\sum_{\vec{\beta}\in\Psi_{IJ}(\vec{\alpha})}\!
\frac{h_1^{\prime\prime\prime}(\vec{\alpha}\cdot\vec{t}\,)}{{\phantom{\Big|}}(\vec{\alpha}\cdot\vec{a}\,)^2
\,(\vec{\beta}\cdot\vec{a}\,)}{\phantom{\bigg|}}\notag\\
&\quad+\frac{1}{4}\Big(m^2-\frac{\epsilon_1^2}{4}\Big)
\Big(m+\frac{\epsilon_1}{2}\Big)^2\bigg[\sum_{J\ne I}\sum_{\vec{\alpha}\in\Psi_{IJ}}
\sum_{\vec{\beta}\ne\vec{\gamma}\in\Psi_{IJ}(\vec{\alpha})}
\frac{h_1^{\prime\prime\prime}(\vec{\alpha}\cdot\vec{t}\,)}{{\phantom{\Big|}}(\vec{\alpha}\cdot\vec{a}\,)
\,(\vec{\beta}\cdot\vec{a}\,)\,(\vec{\gamma}\cdot\vec{a}\,)}\label{w4primegen}\\
&\qquad\qquad\qquad\qquad-\frac{1}{3}\sum_{J\ne I}
\sum_{\vec{\alpha}\in\Psi_{IJ}}\sum_{\vec{\beta}\in\Psi_{IJ}(\vec{\alpha})}
\sum_{\vec{\gamma}\in\Psi_{IJ}(\vec{\alpha},\vec{\beta})}
\frac{h_1^{\prime\prime\prime}(\vec{\alpha}\cdot\vec{t}\,)}{{\phantom{\Big|}}(\vec{\alpha}\cdot\vec{a}\,)
\,(\vec{\beta}\cdot\vec{a}\,)\,(\vec{\gamma}\cdot\vec{a}\,)}\bigg]\notag\\
&\quad+\Big(m^2-\frac{\epsilon_1^2}{4}\Big)^2\sum_{J\ne K\ne I}\,
\sum_{\vec{\alpha}\in\Psi_{IJ}}\,\sum_{\vec{\beta}\in\Psi_{IK}(\vec{\alpha})}\frac{h_1^{\prime}
(\vec{\alpha}\cdot\vec{t}\,)\,h_1^{\prime}(\vec{\alpha}\cdot\vec{t}
-\vec{\beta}\cdot\vec{t}\,)}{{\phantom{\Big|}}(\vec{\alpha}\cdot\vec{a}\,)
\,(\vec{\beta}\cdot\vec{a}\,)\,(\vec{\alpha}\cdot\vec{a}-\vec{\beta}\cdot\vec{a}\,)}\notag
\end{align}
where the summation indices $J,K,\cdots,$ take integer values from 1 to $M$. One can explicitly 
check that these formulas reduce to those of the previous subsections if $M=2$
and that no singularity arises when two $a$'s belonging to the same subgroup in (\ref{agen}) approach each other. 
We have verified these expressions in many cases up to SU(7), always finding agreement with the explicit
localization results. Of course it is possible to write down similar expressions for the higher coefficients
$w_\ell^{(I)}$, which however become more and more cumbersome as $\ell$ increases.
Given the group theoretic structure of these formulas, it is tempting to speculate that 
they may be valid for the other simply laced groups of the ADE series as well, similarly to what happens
for the analogous expressions of the prepotential coefficients \cite{Billo':2015ria}. It would be interesting
to verify whether this happens or not.

\vspace{0.5cm}
\section{Duality between surface operators}
\label{secn:IRduality}

In this section we establish a relation between our localization results and those obtained when the 
surface defect is realized by coupling two-dimensional sigma-models to the four dimensional gauge theory. 
When the surface operators are realized in this way, the twisted chiral ring has been independently 
obtained by studying the two-dimensional $(2,2)$ theories \cite{Witten:1993yc,Hanany:1997vm} 
and related to the Seiberg-Witten geometry of the four dimensional gauge theory
\cite{Gaiotto:2009fs, Gaiotto:2013sma}. Building on these general results, 
we extract the semi-classical limit and compare it with the localization answer, finding agreement.

In order to be explicit, we will consider only gauge theories without $\Omega$-deformation, 
and begin our analysis by first discussing the pure $\mathcal{N}=2$ theory 
with gauge group SU($N$); in the end we will return to the ${\mathcal N}=2^{\star}$ theory.

\subsection{The pure ${\mathcal N}=2$ SU($N$) theory}
\label{purecase}

The pure ${\mathcal N}=2$ theory can be obtained by decoupling
the adjoint hypermultiplet of the $\mathcal{N}=2^\star$ model. 
More precisely, this decoupling is carried out by
taking the following limit (see for example \cite{Billo:2014bja})
\begin{equation}
\label{qscaling}
m\rightarrow \infty\quad\text{and}\quad q\rightarrow 0\quad\text{such that}\quad
q\, m^{2N}=(-1)^N\Lambda^{2N}\quad\text{is finite,}
\end{equation}
where $\Lambda$ is the strong coupling scale of the pure $\mathcal{N}=2$ theory.
In presence of a surface operator, this limit must be combined with a 
scaling prescription for the continuous variables that characterize the defect. For surface operators of type
$\{p,N-p\}$, which possess only one parameter $x=\rme^{2\pi\ii\,z}$, this scaling is
\begin{equation}
\label{xscaling}
m\rightarrow \infty\quad\text{and}\quad x\rightarrow 0\quad\text{such that}
\quad x\,m^N=(-1)^{p-1}x_0\,\Lambda^{N}\quad\text{is finite.}
\end{equation}
Here $x_0=\rme^{2\pi\ii\,z_0}$ is the parameter that labels the surface operator in the pure theory
\`a la Gukov-Witten \cite{Gukov:2006jk,Gukov:2008sn,Gaiotto:2009fs,Gaiotto:2013sma}.

Performing the limits (\ref{qscaling}) and (\ref{xscaling}) on the localization results described in the previous
sections, we obtain
\begin{equation}
\mathcal{W}^{\,\prime}=\sum_{i=1}^p\mathcal{W}_i^{\,\prime}
\label{superpotpure}
\end{equation}
where
\begin{equation}
\mathcal{W}_i^{\,\prime}
=-a_i-\Lambda^N\Big(x_0+\frac{1}{x_0}\Big) \prod_{j\ne i}^N\frac{1}{a_{ij}}
-\frac{\Lambda^{2N}}{2}\,
\Big(x_0^2+\frac{1}{x_0^2}\Big)\frac{\partial}{\partial a_i}
\Big(\prod_{j\ne i}^N\frac{1}{a_{ij}^2{\phantom{\big|}}}\Big)+ \mathcal{O}\big(\Lambda^{3N}\big)~.
\label{Wprimeipure}
\end{equation}
We have explicitly verified this expression in all cases up to SU(7), and for the low rank groups we have also
computed the higher instanton corrections\,\footnote{For example, for SU(2) and $p=1$
we find
\begin{equation*}
\mathcal{W}_1^{\,\prime}=-a-\frac{\Lambda^2}{2a}\Big(x_0+\frac{1}{x_0}\Big)+
\frac{\Lambda^4}{8a^3}\Big(x_0^2+\frac{1}{x_0^2}\Big)
-\frac{\Lambda^6}{16a^5}\Big(x_0^3+x_0+\frac{1}{x_0}+\frac{1}{x_0^3}\Big)
+\frac{\Lambda^8}{128a^7}\Big(5x_0^4+8x_0^2+\frac{8}{x_0^2}+\frac{5}{x_0^4}\Big)+\mathcal{O}\big(\Lambda^{10}\big)
\end{equation*}
where $a=a_1$.}. With some simple algebra one can check that, up to the order we have worked, 
$\mathcal{W}^{\,\prime}$ is not singular for $a_i\to a_j$ when both $i$ and $j$ are $\le p$ or $>p$. 
Furthermore, one can verify that 
\begin{equation}
\sum_{i=1}^N \mathcal{W}_i^{\,\prime}=0
\label{sumWi}
\end{equation}
as a consequence of the tracelessness condition on the vacuum expectation values.

We now show that this result is completely consistent with the exact twisted chiral ring relation 
obtained in \cite{Gaiotto:2013sma}. For the pure ${\mathcal N}=2$ SU($N$) theory with a 
surface operator parameterized by $x_0$, the twisted chiral ring relation takes the form \cite{Gaiotto:2013sma}
\begin{equation}
\label{twistedchiral}
\mathcal{P}_N(y) - \Lambda^N\,\Big(x_0+ \frac{1}{x_0} \Big)=0
\end{equation}
with
\begin{equation}
\mathcal{P}_N(y)=\prod_{i=1}^N\big(y-e_i\big)
\label{QCGP}
\end{equation}
where $e_i$ are the quantum corrected expectation values of the adjoint scalar. They
reduce to $a_i$ in the classical limit $\Lambda\to 0$ and parameterize the quantum moduli 
space of the theory. The $e_i$, which satisfy the tracelessness condition
\begin{equation}
\sum_{i=1}^N e_i =0~,
\end{equation}
were explicitly computed long ago in the 1-instanton approximation in \cite{D'Hoker:1996nv,Naculich:2002hi} 
by evaluating the period integrals of the Seiberg-Witten differential and read
\begin{equation}
e_i= a_i-\Lambda^{2N}\frac{\partial}{\partial a_i}\Big(\prod_{j\ne i}\frac{1}{a_{ij}^2{\phantom{\big|}}}\Big)
+ \mathcal{O}\big(\Lambda^{4N}\big)~.
\label{eiai}
\end{equation}
The higher instanton corrections can be efficiently computed using localization methods 
\cite{Nekrasov:2012xe, Fucito:2011pn, Fucito:2012xc,Beccaria:2017rfz}, but their expressions
will not be needed in the following. 

Inserting (\ref{eiai}) into (\ref{QCGP}) and systematically working order by order in $\Lambda^N$, 
it is possible to show that the $N$ roots of the chiral ring equation (\ref{twistedchiral}) are
\begin{equation}
y_i=a_i+\Lambda^N\Big(x_0+\frac{1}{x_0}\Big) \prod_{j\ne i}^N\frac{1}{a_{ij}}
+\frac{\Lambda^{2N}}{2}\,
\Big(x_0^2+\frac{1}{x_0^2}\Big)\frac{\partial}{\partial a_i}
\Big(\prod_{j\ne i}^N\frac{1}{a_{ij}^2{\phantom{\big|}}}\Big)+ \mathcal{O}\big(\Lambda^{3N}\big)
\end{equation}
for $i=1,\cdots,N$.
Comparing with (\ref{Wprimeipure}), we see that, up to an overall sign, $y_i$ coincide with
the derivatives of the superpotential $\mathcal{W}_i^{\,\prime}$
we obtained from localization. Therefore, we can rewrite the left hand side of (\ref{twistedchiral}) 
in a factorized form and get
\begin{equation}
\prod_{i=1}^N\big(y+\mathcal{W}_i^{\,\prime}) -\mathcal{P}_N(y) + \Lambda^N\,\Big(x_0+ \frac{1}{x_0} \Big)=0
\label{factorized}
\end{equation}
This shows a perfect match between our localization results and the semi-classical expansion of the chiral ring
relation of \cite{Gaiotto:2013sma}, and provides further non-trivial evidence for the equivalence of 
the two descriptions. Let us elaborate a bit more on this. 
According to \cite{Gaiotto:2013sma}, a surface operator of type $\{p,N-p\}$
has a dual description as a Grassmannian sigma-model coupled to the SU($N$) gauge theory, 
and all information about the twisted chiral ring of the sigma-model 
is contained in two monic polynomials, $Q$ and $\widetilde{Q}$ of degree $p$ and $(N-p)$ respectively, 
given by
\begin{equation}
Q(y) = \sum_{\ell=0}^p y^\ell \,{\mathcal X}_{p-\ell}~,\qquad
\widetilde{Q}(y) = \sum_{k=0}^{N-p} y^k \,{\widetilde{\mathcal X}}_{N-p-k}~.
\label{QtildeQ}
\end{equation}
with ${\mathcal X}_0=\widetilde{{\mathcal X}}_0=1$.
Here,  ${\mathcal X}_\ell$ are the twisted chiral ring elements of the Grassmannian 
sigma-model, and in particular
\begin{equation}
{\mathcal X}_1 = \frac{1}{2\pi\ii} \frac{\p {\mathcal W}}{\p z_0}
\label{Chi1}
\end{equation}
where $\mathcal{W}$ is the superpotential of the surface operator of type $\{p,N-p\}$. 
The polynomial $\widetilde{Q}$ encodes the auxiliary information about 
the ``dual" surface operator obtained by sending $p \rightarrow (N-p)$. The crucial point 
is that, according to the proposal of \cite{Gaiotto:2013sma}, the two 
polynomials $Q$ and $\widetilde{Q}$ satisfy the relation
\begin{equation}
\label{twistedCR}
Q(y)\, \widetilde{Q}(y) - \mathcal{P}_N(y) + \Lambda^N\,\Big(x_0+ \frac{1}{x_0} \Big)=0 ~.
\end{equation}
Comparing with (\ref{factorized}), we are immediately led to the following identifications\,\footnote{We have chosen a specific ordering in which the first $p$ factors correspond to the first $p$ 
vacuum expectation values $a_i$; of course one could as well choose a different ordering by permuting the factors.}
\begin{equation}
\begin{aligned}
Q(y)=\prod_{i=1}^p\big(y+\mathcal{W}_i^{\,\prime}\big)~,\qquad
\widetilde{Q}(y)=\prod_{j=p+1}^{N}\big(y+\mathcal{W}_j^{\,\prime}\big)~.
\end{aligned}
\end{equation}
Thus, using (\ref{Chi1}) and (\ref{superpotpure}), we find
\begin{equation}
\frac{1}{2\pi\ii} \frac{\p {\mathcal W}}{\p z_0}=\sum_{i=1}^p\mathcal{W}_i^{\,\prime}=
\mathcal{W}^{\,\prime}~.
\end{equation}
This equality shows that our localization results for the superpotential of the surface operator 
of type $\{p,N-p\}$ in the pure SU($N$) theory perfectly consistent with the proposal of 
\cite{Gaiotto:2013sma}, thus proving the duality between the two descriptions.
All this is also a remarkable consistency check of the way in which we have extracted the semi-classical 
results for the twisted chiral ring of the Grassmannian sigma-model and of the twisted superpotential 
we have computed. 

\subsection{The ${\mathcal N}=2^\star$ SU($N$) theory}

Inspired by the previous outcome, we now analyze the twisted chiral ring relation for simple 
operators in ${\mathcal N}=2^{\star}$ theories using the Seiberg-Witten curve and compare it with our 
localization results for the undeformed theory. 
To this aim, let us first recall from Section~\ref{subsecn:simpleN} (see in particular (\ref{wellprime}) 
with $\epsilon_1=0$) that for a simple surface
operator corresponding to the following partition of the Coulomb parameters
\begin{equation}
\left\{
\begin{array}{cc}
a_i,&  \undermat{N-1}{ \{a_j~\text{with}~j\ne i\}}
\end{array}
\right\} ~,
\vspace{.5cm}
\end{equation}
the $z$-derivative of the superpotential is
\begin{equation}
\begin{aligned}
\mathcal{W}_i^{\,\prime}&=
-a_i+m^2 \sum_{j\ne i}\frac{h_1^\prime}{a_{ij}}
+\frac{m^3}{2}\sum_{j\ne k \ne i}\frac{h_1^{\prime\prime}}{a_{ij}\,a_{ik}}\\
&\qquad+\frac{m^4}{6}\bigg(\sum_{j\ne i}
\frac{E_2\,h_1^{\prime}-h_1^{\prime\prime\prime}}{a_{ij}^3}+
\!\!\sum_{j\ne k \ne \ell\ne i}\frac{h_1^{\prime\prime\prime}}{a_{ij}\,
a_{ik}\,a_{i\ell}}\bigg)+\mathcal{O}\big(m^5\big)~.
\end{aligned}
\label{Wprimei}
\end{equation}
Let us now see how this information can be retrieved from the Seiberg-Witten curve of the $\mathcal{N}=2^\star$ 
theories. As is well known, in this case there are two possible descriptions 
(see \cite{Ashok:2016ewb} for a review). 
The first one, which we call the Donagi-Witten curve \cite{Donagi:1995cf}, is written naturally in terms 
of the modular covariant coordinates on moduli space, while the second, which we call the d'Hoker-Phong 
curve \cite{DHoker:1997hut}, is written naturally in terms of the quantum corrected coordinates on moduli space. 
As shown in \cite{Ashok:2016ewb}, these two descriptions are linearly related to each other 
with coefficients depending on the second Eisenstein series $E_2$. 

Since our semi-classical results have been resummed into elliptic and quasi-modular forms, we use the 
Donagi-Witten curve, which for the SU$(N)$ gauge theory is an $N$-fold cover of an elliptic curve. 
It is described by the pair of equations:
\begin{equation}
\begin{aligned}
Y^2 = X^3 - \frac{E_4}{48}X + \frac{E_6}{864} ~,\qquad
F_N(y,X,Y) = 0~.
\end{aligned}
\label{DWeq}
\end{equation}
The first equation describes an elliptic curve and thus we can identify $(X, Y)$ with the 
Weierstra\ss~function and its derivative (see (\ref{ellipticeq})). More precisely we have
\begin{equation}
\begin{aligned}
X&=-\widetilde{\wp}=-h_1^\prime+\frac{1}{12}\,E_2~,\\
Y&=\frac{1}{2}\,{\widetilde{\wp}}^{\,\prime}=\frac{1}{2}\,h_1^{\prime\prime}
\end{aligned}
\end{equation}
The second equation in (\ref{DWeq}) contains a polynomial in $y$ of degree $N$ which encodes the 
modular covariant coordinates $A_k$ on the Coulomb moduli space of the gauge theory:
\begin{equation}
F_N(y,X,Y) =\sum_{k=0}^N (-1)^k A_k\,P_{N-k}(y,X,Y)
\end{equation}
where $P_k$ are the modified Donagi-Witten polynomials introduced in \cite{Ashok:2016ewb}. 
The first few of them are\,\footnote{The $E_4$ term in $P_4$ is one of the modifications which 
in \cite{Ashok:2016ewb} were found to be necessary and is crucial also here.}:
\begin{equation}
\begin{aligned}
P_0 &=1~,\qquad \, P_1 = y~,\\
P_2 & = y^2 - m^2\,X~,\qquad \,P_3 = y^3 - 3\,y\, m^2\,X +2 \,m^3\,Y ~,\\
P_4 &=y^4-6\,m^2\,y^2\,X+8\,y\,m^3\,Y-m^4\Big(3\,X^2-\frac{1}{24}\,E_4\Big)~.
\end{aligned}
\label{p123}
\end{equation}
On the other hand, the first few modular covariant coordinates $A_k$ are (see \cite{Ashok:2016ewb}):
\begin{equation}
\begin{aligned}
A_2&=\sum_{i<j}a_i a_j+\frac{m^2}{12}\binom{N}{2}\,E_2+\frac{m^4}{288}\big(E_2^2-E_4\big)
\sum_{i\ne j}\frac{1}{a_{ij}^2 {\phantom{\big|}} }+\mathcal{O}(m^6)~,\\
A_3&=\sum_{i<j<k}a_i a_j a_k-\frac{m^4}{144}\big(E_2^2-E_4\big)\sum_{i}\sum_{j\ne i}
\frac{\!a_i}{a_{ij}^2 {\phantom{\big|}} }+\mathcal{O}(m^6)~,\\
A_4&=\sum_{i<j<k<\ell}a_i a_j a_k a_\ell+\frac{m^2}{12}\binom{N-2}{2}\,E_2\sum_{i<j}a_i a_j
+\frac{m^4}{48}E_2^2\\
&~~\quad+\frac{m^4}{288}\big(E_2^2-E_4\big)\bigg[\sum_{i<j}\sum_{k\ne\ell}
\frac{a_i a_j}{a_{k\ell}^2 {\phantom{\big|}} }+3\sum_{i}\sum_{j\ne i}\frac{\!a_i^2}{a_{ij}^2 {\phantom{\big|}} }
-\binom{N}{2}\bigg]+\mathcal{O}(m^6)~,
\end{aligned}
\end{equation}
and so on.

We now have all the necessary ingredients to proceed. First of all, using the above expressions 
and performing the decoupling limits (\ref{qscaling}) and (\ref{xscaling}), one can check that 
the Donagi-Witten equation $F_N=0$ reduces to the twisted 
chiral ring relation (\ref{twistedchiral}) of the pure theory. Of course this is not a mere coincidence; 
on the contrary it supports the idea that the Donagi-Witten equation
actually encodes also the twisted chiral ring relation of the simple codimension-4 surface 
operators of the $\mathcal{N}=2^\star$ theories. Secondly, working order by order in the hypermultiplet 
mass $m$, one can 
verify that the $N$ roots of the Donagi-Witten equation are given by
\begin{equation}
\begin{aligned}
y_i&=
a_i-m^2 \sum_{j\ne i}\frac{h_1^\prime}{a_{ij}}
-\frac{m^3}{2}\sum_{j\ne k \ne i}\frac{h_1^{\prime\prime}}{a_{ij}\,a_{ik}}\\
&\qquad-\frac{m^4}{6}\bigg(\sum_{j\ne i}
\frac{E_2\,h_1^{\prime}-h_1^{\prime\prime\prime}}{a_{ij}^3}+
\!\!\sum_{j\ne k \ne \ell\ne i}\frac{h_1^{\prime\prime\prime}}{a_{ij}\,
a_{ik}\,a_{i\ell}}\bigg)+\mathcal{O}\big(m^5\big)~.
\end{aligned}
\label{yi}
\end{equation}
Remarkably, this precisely matches, up to an overall sign, the answer (\ref{Wprimei}) 
for the simple codimension-2 surface operator we have obtained using localization. 
Once again, we have exhibited the equivalence of twisted chiral rings calculated for the two 
kinds of surface operators. 
Furthermore, we can rewrite the Donagi-Witten equation in a factorized form as follows
\begin{equation}
\prod_{i=1}^N\big(y+\mathcal{W}^{\,\prime}_i\big)-F_N(y,X,Y)=0
\label{FNfact}
\end{equation}
which is the $\mathcal{N}=2^\star$ equivalent of the pure theory relation (\ref{factorized}). 

At this point 
one is tempted to proceed as in the pure theory and try to deduce also the superpotential for surface 
operators of type $\{p,N-p\}$. However, from our explicit localization results we know that in this case
$\mathcal{W}^{\,\prime}$ is not simply the sum of the superpotentials of type $\{1,N-1\}$, differently from 
what happens in the pure theory (see (\ref{superpotpure})). Thus, a naive 
extension to the $\mathcal{N}=2^\star$ of the proposal of \cite{Gaiotto:2013sma} to describe 
the coupling of a two dimensional Grassmannian sigma-model to the four dimensional gauge theory can not 
work in this case. This problem as well as the coupling of a flag variety to the ${\mathcal N}=2^{\star}$ 
theory, which
is relevant for surface operators of general type, remains an open question which we leave to future investigations.

\subsection{Some remarks on the results}

The result we obtained from the twisted superpotential in the case of simple operators
is totally consistent with the proposal given in the literature for 
simple codimension-4 surface operators labeled by a single continuous parameter 
$z$, whose superpotential has been identified with the line integral 
of the Seiberg-Witten differential of the four-dimensional gauge theory along an open 
path \cite{Alday:2009fs}: 
\begin{equation}
\label{dWislambdaSW}
{\mathcal W}(z)= \int^z_{z*} \lambda_{SW} 
\end{equation}
where $z*$ is an arbitrary reference point. Indeed, in the Donagi-Witten variables, the differential is simply
$\lambda_{SW}(z) = y(z)\, dz$.
Given that the Donagi-Witten curve is an $N$-fold cover of the torus, 
the twisted superpotential with the classical contribution proportional 
to $a_i$ can be obtained by solving for $y(z)$ and writing out the solution 
on the $i$th branch. 

As we have seen in the previous subsection, the general identification in 
(\ref{dWislambdaSW}) works also in the pure 
$\mathcal{N}=2$ theory, once the parameters in the Seiberg-Witten 
differential are rescaled by a factor of $\Lambda^N$ \cite{Gaiotto:2013sma}. 
This rescaling can be interpreted as a renormalization of the continuous parameter that labels 
the surface operator \cite{Gaiotto:2011tf}. 

The agreement we find gives further evidence of the duality between defects realized as 
codimension-2 and codimension-4 operators that we have already discussed in Section~\ref{secn:CFT}, 
where we showed the equality of the twisted effective superpotential computed in the 
two approaches for simple defects in the SU(2) theory. We have extended these checks to defects of
type $\{p,N-p\}$ in pure $\mathcal{N}=2$ theories, and to simple defects in $\mathcal{N}=2^\star$
theories with higher rank gauge groups. All these checks support
the proposal of \cite{Frenkel:2015rda} based on a ``separation of variables'' relation.

\vspace{0.5cm}
\section{Conclusions}
\label{secn:concl}

In this paper we have studied the properties of surface operators on the Coulomb branch 
of the four dimensional ${\mathcal N}=2^{\star}$ theory with gauge group SU$(N)$ focusing on the
superpotential $\mathcal{W}$. This superpotential, describing the effective two-dimensional 
dynamics on the defect world-sheet, receives non-perturbative contributions, which we calculated 
using equivariant localization. 
Furthermore, exploiting the constraints arising from the non-perturbative SL$(2,\mathbb{Z})$ 
symmetry, we showed that in a semi-classical regime in which the mass of the adjoint hypermultiplet 
is much smaller than the classical Coulomb branch parameters, the twisted superpotential satisfies 
a modular anomaly equation that we solved order by order in the mass expansion. 

We would like to remark some interesting properties of our results.
If we focus on the derivatives of the superpotential, 
the coefficients of the various terms in the mass expansion are linear combination of 
elliptic and quasi-modular forms with a given weight. 
The explicit expression for the twisted superpotential can be written 
in a very general and compact form in terms of suitable restricted 
sums over the root lattice of the gauge algebra.

The match of our localization results with the ones obtained
in \cite{Gaiotto:2013sma} by studying the coupling with two-dimensional sigma models 
is a non-trivial check of our methods and provides evidence for the duality between 
the codimension-2 and codimension-4 surface operators proposed in \cite{Frenkel:2015rda}. 
Further evidence is given by the match of the twisted superpotentials in the 
${\mathcal N}=2^{\star}$ theory, which we proved for the simple surface operators
using the Donagi-Witten curve of the model.
A key input for this match is the exact quantum expression of the chiral ring elements calculated 
using localization \cite{Ashok:2016ewb, Beccaria:2017rfz}.  
It would be really important to extend the discussion of
this duality to more general surface operators described by a generic Levi decomposition.

There are several possible extensions of our work. 
A very direct one would be to check that the general expression given for the 
twisted superpotential is actually valid for all simply laced groups, in analogy to what 
happens for the four-dimensional prepotential.
A technically more challenging extension would be to study surface operators for theories with non-simply 
laced gauge groups. The prepotential in these cases has been calculated in 
\cite{Billo':2015jta} using localization methods and expressed in terms of modular forms of suitable
congruence subgroups of SL(2,$\mathbb{Z}$), and it would be very interesting to similarly 
calculate the twisted superpotential in a semi-classical expansion. 

Another interesting direction would be to study surface operators in SQCD theories. 
For SU$(N)$ gauge groups, the prepotential as well as the action of
S-duality on the infrared variables have been calculated in a special locus 
of the Coulomb moduli space that has a $\mathbb{Z}_N$ symmetry 
\cite{Ashok:2015cba, Ashok:2016oyh}. Of special importance was the generalized 
Hecke groups acting on the period integrals and the period matrix of the 
Seiberg-Witten curve. It would be worthwhile to explore if such groups continue 
to play a role in determining the twisted superpotential as well. 

A related development would be to analyze the higher order terms in the $\epsilon_2$ expansion 
of the partition function (see (\ref{FW})) and check whether or not they also obey a modular anomaly
equation like the prepotential and the superpotential do. This would help us in clarifying the properties
of the partition function in the presence of a surface operator in a general $\Omega$ background.

There has been a lot of progress in understanding M2 brane surface operators via 
the $4d$/$2d$ correspondence.
For higher rank theories, explicit results for such surface defects have been obtained 
in various works including \cite{Bonelli:2011fq,Bonelli:2011wx,Gomis:2014eya,Gomis:2016ljm,Pan:2016fbl}.  
In particular in \cite{Gomis:2014eya}, the partition functions of theories with $N_f^2$ free 
hypermultiplets on the deformed 4-sphere in the presence of surface defects have been 
related to specific conformal blocks in Toda conformal field theories. This has been extended in 
\cite{Gomis:2016ljm,Pan:2016fbl} to study gauge theory partition functions in the presence of 
intersecting surface defects. It would be interesting to study such configurations 
directly using localization methods. 

\vskip 1.5cm
\noindent {\large {\bf Acknowledgments}}
\vskip 0.2cm
We would like to thank Dileep Jatkar, Madhusudhan Raman and especially Jan Troost for useful discussions and Matteo Beccaria for comments on the manuscript. 
The work of M.B. and M.F. is partially supported  by the Compagnia di San Paolo 
contract ``MAST: Modern Applications of String Theory'' TO-Call3-2012-0088.
The work of M.B., M.F. and A.L. is partially supported by the MIUR PRIN Contract 2015MP2CX4 ``Non-perturbative Aspects Of Gauge Theories And Strings''.
\vskip 1cm
\appendix

\section{Useful formulas for modular forms and elliptic functions}
\label{useful}
In this appendix we collect some formulas about quasi-modular forms and elliptic functions that are useful
to check the statements of the main text.

\subsection*{Eisenstein series}
We begin with the Eisenstein series $E_{2n}$, which admit a Fourier expansion in 
terms of $q=\rme^{2\pi\ii\tau}$ of the form 
\begin{equation}
E_{2n}=1+\frac{2}{\zeta(1-2n)}\sum_{k=1}^\infty \sigma_{2n-1}(k) q^{k}~,
\label{eis}
\end{equation}
where $\sigma_p(k)$ is the sum of the $p$-th powers of the divisors of $k$. 
More explicitly we have
\begin{equation}
\label{Ekexp}
\begin{aligned}
E_2 & = 1 - 24 \sum_{k=1}^\infty \sigma_1(k) q^{k} = 1 - 24 q - 72 q^2 - 96 q^3 -168q^4+ \cdots~,\\
E_4 & = 1 + 240 \sum_{k=1}^\infty \sigma_3(k) q^{k} = 1 + 240 q + 2160 q^2 + 6720 q^3+17520q^4
 + \cdots~,\\
E_6 & = 1 - 504 \sum_{k=1}^\infty \sigma_5(k) q^{k} =1 - 504 q - 16632 q^2 - 122976 q^3 -532728 q^4+ \cdots~.
\end{aligned}
\end{equation}
Under a modular transformation $\tau\to\frac{a\tau+b}{c\tau+d}$, with $a,b,c,d\in\mathbb{Z}$ 
and $ad-bc=1$, the Eisenstein series transform as
\begin{equation}
E_2\to(c\tau+d)^2\,E_2+\frac{6}{\pi\ii}\,c\,(c\tau+d)~,~~
E_4\to(c\tau+d)^4\,E_4~,~~E_6~\to~(c\tau+d)^6\,E_6~.
\end{equation}
In particular, under S-duality we have
\begin{equation}
\begin{aligned}
&E_2(\tau)\to E_2\Bigl(\!-\frac{1}{\tau}\Bigr)=\tau^2\big(E_2(\tau)+\delta\big)~,\\
&E_4(\tau)\to E_4\Bigl(\!-\frac{1}{\tau}\Bigr)= \tau^4 E_4(\tau)~,\\
&E_6(\tau)\to E_6\Bigl(\!-\frac{1}{\tau}\Bigr)= \tau^6 \,E_6(\tau)
\end{aligned}
\label{SonE}
\end{equation}
where $\delta=\frac{6}{\pi\ii\tau}$.

\subsection*{Elliptic functions}
The elliptic functions that are relevant for this paper can all be obtained from the Jacobi $\theta$-function
\begin{equation}
\theta_1(z|\tau) = \sum_{n=-\infty}^{\infty} q^{\frac{1}{2}(n-\frac{1}{2})^2}\, 
(-x)^{(n-\frac{1}{2})}
\label{theta1}
\end{equation}
where $x = e^{2\pi \ii z}$. {From} $\theta_1$, we first define the function
\begin{equation}
\label{h1defn}
h_1(z|\tau) = \frac{1}{2\pi\ii} \frac{\p}{\p z} \log\theta_1(z|\tau) = x\frac{\p}{\p x} \log\theta_1(z|\tau)~,
\end{equation}
and the Weierstra\ss~$\wp$-function 
\begin{equation}
\label{wpdefn0}
 \wp(z|\tau)=-\frac{\p^2}{\p z^2}\log\theta_1(z|\tau)-\frac{\pi^2}{3}E_2(\tau) ~.
\end{equation}
In most of our formulas the following rescaled $\wp$-function appears:
\begin{equation}
\label{wpdefn}
\widetilde\wp(z|\tau) := \frac{\wp(z,\tau)}{4\pi^2}=
x\frac{\p}{\p x}\Big( x\frac{\p}{\p x}\log\theta_1(z|\tau) \Big)-\frac{1}{12}E_2(\tau)~,
\end{equation}
which we can write also as
\begin{equation}
\widetilde\wp(z|\tau) =h_1^\prime(z|\tau)-\frac{1}{12}E_2(\tau)~.
\label{wph1}
\end{equation}
Another relevant elliptic function is the derivative of the Weierstra\ss~function, namely 
\begin{equation}
\widetilde\wp^{\,\prime}(z|\tau) := 
\frac{1}{2\pi \ii} \frac{\p }{\p z} \widetilde\wp(z|\tau) = x\frac{\p}{\p x} \widetilde\wp(z|\tau) 
=h_1^{\prime\prime}(z|\tau)~.
\label{wprimedefn}
\end{equation}
The Weierstra\ss~function and its derivative satisfy the equation of an elliptic curve, given by
\begin{equation}
\widetilde\wp^{\,\prime}(z|\tau)^2 + 4 \,\widetilde\wp(z|\tau)^3 - \frac{E_4}{12}\,\widetilde
\wp(z|\tau) - \frac{E_6}{216} =0 ~.
\label{ellipticeq}
\end{equation}
By differentiating this equation, we obtain
\begin{equation}
\widetilde\wp^{\,\prime\prime}(z|\tau)=-6 \,\widetilde\wp(z|\tau)^2+ \frac{E_4}{24}
\label{wpdoubleprime}
\end{equation}
which, using (\ref{wph1}) and (\ref{wprimedefn}), we can rewrite as
\begin{equation}
h_1^{\prime\prime\prime}(z|\tau)=-6\,\big(h_1^{\prime}(z|\tau)\big)^2+E_2\,h_1^{\prime}(z|\tau)-\frac{E_2^2-E_4}{24}~.
\end{equation}

The function $h_1$,  $\widetilde\wp$ and $\widetilde\wp^{\,\prime}$ have well-known 
expansions near the point $z=0$. However, a different expansion is needed for our 
purposes, namely the expansion for small $q$ and $x$. To find such an expansion we 
observe that $q$ and $x$ variables must be rescaled differently, 
as is clear from the map (\ref{paramap}) between the gauge theory parameters and 
the microscopic counting parameters. In particular for $M=2$ this map reads (see also (\ref{mapSU2}))
\begin{equation}
q = q_1q_2\quad,\quad x =q_2 ~,
\end{equation}
so that if the microscopic parameters are all scaled equally as $q_i \longrightarrow \lambda q_i$, then
the gauge theory parameters scale as 
\begin{equation}
q \rightarrow \lambda^2 q\qquad x \rightarrow \lambda x ~.
\end{equation}
With this in mind, we now expand the elliptic functions for small $\lambda$ and set $\lambda=1$ in the end, 
since this is the relevant expansion needed to compare with the instanton calculations. Proceeding in this
way, we find\,\footnote{Depending on the context, we denote the arguments of the elliptic functions by either $(z|\tau)$ 
as we did so far, or by their exponentials $(x|q)$ when the expansions are being used.}
\begin{align}
h_1(x|q)&=h_1(\lambda x | \lambda^2 q)\Big|_{\lambda=1} \notag\hspace{10cm}\\
&= \Big[-\frac{1}{2}
+\lambda  \Big(\frac{q}{x}-x\Big)
+\lambda^2 \Big(\frac{q^2}{x^2}-x^2\Big)
+\lambda^3 \Big(\frac{q^3}{x^3}+\frac{q^2}{x}-q x-x^3\Big)\notag\\
&\qquad-\lambda^4\,x^4
+\lambda^5 \left(\frac{q^3}{x}-q^2 x-x^5\right)
-\lambda^6 \left(q^2 x^2+x^6\right) + \cdots\Big]_{\lambda=1}\label{h1exp}\\
&=-\frac{1}{2}-\Big(x+x^2+x^3+x^4+x^5+x^6+\cdots\Big) +\Big(\frac{1}{x}-x\Big)q\notag\\
&\qquad+\Big(\frac{1}{x^2}+\frac{1}{x}-x-x^2\Big)q^2+ \Big(\frac{1}{x^3}+\frac{1}{x}+\cdots\Big)q^3 +\cdots~,
\notag
\end{align}
\begin{align}
\widetilde\wp(x |q)&=\widetilde\wp(\lambda x |\lambda^2q)\Big|_{\lambda=1} \notag\hspace{10.2cm}\\
&= \Big[ -\frac{1}{12}- \lambda\Big
(\frac{q}{x}+x\Big)+\lambda^2 \Big(- \frac{2q^2}{x^2}+2q-2x^2 \Big) \notag \\
&\qquad- \lambda^3 \Big(\frac{3 q^3}{x^3}+\frac{q^2}{x}+q x+3 x^3\Big) +
\lambda^4\big(6q^2-4x^4\big)+ \cdots \Big]_{\lambda=1} \label{wpexp}\\
&=-\frac{1}{12}-\Big(x+2x^2+3x^3+4x^4+\cdots\Big)-\Big(\frac{1}{x}-2+x\Big)q\notag\\
&\qquad-\Big(\frac{2}{x^2}+\frac{1}{x}-6+\cdots\Big)q^2-\frac{3q^3}{x^3}+\cdots~,
\notag
\end{align}
\begin{align}
\widetilde\wp^{\,\prime}(x |q)&=\widetilde\wp^{\,\prime}(\lambda x |\lambda^2q)\Big|_{\lambda=1} 
\notag\\
&= \Big[ \lambda  \Big(\frac{q}{x}-x\Big)+\lambda ^2 \left(\frac{4 q^2}{x^2}-4 x^2\right) \notag \\
&\qquad+\lambda^3 \Big(\frac{9 q^3}{x^3}
+\frac{q^2}{x}-q x-9 x^3\Big)-16 \lambda^4 x^4 + \cdots \Big]_{\lambda=1} \label{wprimeexp}\\
&= -\Big(x+4x^2+9x^3+16x^4+\cdots\!\Big) \!+\!\Big(\frac{1}{x} -x\Big)q 
\!+\!\Big(\frac{4}{x^2}+\frac{1}{x}+\cdots \!
\Big)q^2 \!+\!\frac{9q^3}{x^3}+  \cdots~.\notag
\end{align}
As a consistency check it is possible to verify that, using these expansions and those of the Eisenstein 
series in (\ref{Ekexp}), the elliptic curve equation (\ref{ellipticeq}) is identically satisfied order 
by order in $\lambda$.

As we have seen in Section \ref{SOmicro}, the modular group acts on $(z|\tau )$ as follows:
\begin{equation}
(z|\tau) \rightarrow \Big(\frac{z}{c\tau+d}\,\Big|\, \frac{a\tau+b}{c\tau+d}\Big)
\end{equation}
with $a,b,c,d\in\mathbb{Z}$ and $ad-bc=1$. Under such transformations the 
Weierstra\ss~function and its derivative have, respectively, weight 2 and 3, namely
\begin{equation}
\begin{aligned}
\wp(z | \tau)&\to
\wp\Big(\frac{z}{c\tau+d}\,\Big|\,\frac{a\tau+b}{c\tau+d}\Big) = (c\tau +d)^2\, \wp(z | \tau)~, \\
 \wp^{\,\prime}(z | \tau)&\to\wp^{\,\prime}\Big(\frac{z}{c\tau+d}\,\Big|\,\frac{a\tau+b}{c\tau+d}\Big) = 
(c\tau +d)^3\, \wp^{\,\prime}(z | \tau)~.
\end{aligned}
\end{equation}
Of course, similar relations hold for the rescaled functions $\widetilde \wp$ and $\widetilde\wp^{\,\prime}$.
In particular, under S-duality we have
\begin{equation}
\begin{aligned}
\widetilde\wp(z | \tau)&\to\widetilde\wp\Bigl(-\frac{z}{\tau}\,\Big|-\frac{1}{\tau}\Bigr)=\tau^2\,\widetilde	
\wp(z | \tau)~,\\
\widetilde\wp^{\,\prime}(z | \tau)&
\to\widetilde\wp^{\,\prime}\Bigl(-\frac{z}{\tau}\,\Big|-\frac{1}{\tau}\Bigr)
=-\tau^3\,\widetilde\wp^{\,\prime}(z | \tau)~.
\end{aligned}
\label{SonWP}
\end{equation}

\section{Generalized instanton number in the presence of fluxes}
\label{flux_cartan}
In this Appendix we calculate the second Chern class of the gauge field in the presence of a surface 
operator for a generic Lie algebra $\mathfrak{g}$. 

\subsection*{Surface operator Ansatz}
A surface operator creates a singularity in the gauge field $A$. As discussed in the main text, we 
parametrize the space-time $\mathbb{R}^4\simeq \mathbb{C}^2$ by two complex variables 
$(z_1=\rho\,\rme^{\ii \phi}\,,\,z_2=r\,\rme^{\ii\theta})$,
and consider a two-dimensional defect $D$ located at $z_2=0$ and 
filling the $z_1$-plane. In this set-up, we make the following Ansatz \cite{Alday:2010vg}:
\begin{equation}
\label{sing_ans}
A = \widehat{A} + g(r)\, d\theta~,
\end{equation} 
where $\widehat{A}$ is regular all over $\mathbb{R}^4$ and $g(r)$ is a $\mathfrak{g}$-valued function
regular when $r\to 0$. 
The corresponding field strength is then
\begin{equation}
\label{Fsing}
F := dA-\ii\,A\wedge A\,=\,\widehat{F} + d\bigl(g(r)\, d\theta\bigr)-\ii\, d\theta \wedge \big[g(r),\widehat{A}\,\big]~.
\end{equation}
{From} this expression we obtain
\begin{eqnarray}
\Tr F\wedge F &=& ~
\Tr \widehat{F}\wedge \widehat{F} + 2\, \Tr \left(
d\big(g(r)\, d\theta\big)\wedge\widehat{F}\, \right)
- 2\, \ii \,\Tr\left(d\theta\wedge \big[g(r),\widehat{A}\,\big]\wedge \widehat{F}\,\right)
\phantom{\Big|}\label{trFFsing}\\
&=& ~
\Tr \widehat{F}\wedge \widehat{F}
+ 2\, \Tr 
d\big(g(r)\, d\theta \wedge\widehat{F}\, \big)
+ 2\,\Tr\left(g(r)d\theta\wedge \big( d\widehat{F}- \ii \,\widehat{A}\wedge \widehat{F}
-\ii\, \widehat{F}\wedge\widehat{A}\,\big)\right)~.\nonumber
\end{eqnarray}
The last term vanishes due to the Bianchi identity, and thus we are left with
\begin{equation}
\Tr  F\wedge F=
\Tr \widehat{F}\wedge \widehat{F}
+ 2\, \Tr 
d\big(g(r)\, d\theta \wedge\widehat{F}\, \big)
\label{trFFsinga}
\end{equation}

We now assume that the function $g(r)$ has components only
along the Cartan directions of $\mathfrak{g}$, labeled by an index $i$, such that
\begin{equation}
\label{fi_limits}
\lim_{r\to 0} g_i(r) = -\gamma_i~~~\mbox{and}~~~ \lim_{r\to\infty} g_i(r) = 0~.
\end{equation}
This means that near the defect the gauge connection behaves as
\begin{equation}
\label{Asingappi}
A=A_{\mu}\, dx^{\mu}\,\simeq\,-\,\text{diag}
\left(\gamma_1,\cdots,\gamma_{\mathrm{rank}(\mathfrak{g})} \right)\,d\theta
\end{equation}
for $r\to0$. 
Using this in (\ref{trFFsinga}), we have
\begin{equation}
\label{trFFsing3}
\Tr F\wedge F  = \Tr \widehat{F} \wedge \widehat{F} +  2 \sum_i  d\big(g_i(r)\, d\theta\wedge\widehat{F}_i \big)~.
\vspace{-0.25cm}
\end{equation}
Notice that in the last term we can replace $\widehat{F}_i$ with $F_i$ because the difference lies entirely in
the transverse directions of the surface operator and thus does not contribute 
in the wedge product with $d\theta$.
Since the defect $D$ effectively acts as a boundary in $\mathbb{R}^4$ located at $r=0$, integrating
(\ref{trFFsing3}) over $\mathbb{R}^4$ we have
\begin{equation}
\label{kmod}
\frac{1}{8\pi^2}\int_{\mathbb{R}^4} \Tr F\wedge F = \frac{1}{8\pi^2}\int_{\mathbb{R}^4} 
\Tr  \widehat{F} \wedge \widehat{F}  +\sum_i \frac{\gamma_i}{2\pi}\int_D {F}_i 
= k + \sum_i \gamma_i \,m_i~.
\end{equation}
Here we have denoted by $k$ the instanton number of the smooth connection $\widehat{A}$ and
taken into account a factor of $2\pi$ originating from the integration over $\theta$. Finally, we have defined
\begin{equation}
\label{defmi}
m_i = \frac{1}{2\pi}\int_D {F}_i ~.
\end{equation}
These quantities, which we call fluxes, must satisfy a quantization condition that can be understood 
as follows. All fields of the gauge theory are organized in representations \footnote{Here for simplicity we consider the gauge group $G$ to be 
the universal covering group of $\mathfrak{g}$; in particular for $\mathfrak{g}=A_{N-1}$, we take $G=SU(N)$.} of $\mathfrak{g}$ 
and, in particular, can be chosen to be eigenstates of the Cartan generators $H_i$ 
with eigenvalues $\lambda_i$. 
These eigenvalues define a vector $\vec{\lambda}=\{\lambda_i\}$, which is an element of the weight 
lattice $\Lambda_W$ of $\mathfrak{g}$. Let us now consider a gauge transformation 
in the Cartan subgroup with parameters $\vec\omega=\{\omega_i\}$. 
On a field with weight $\vec\lambda$, this transformation simply acts by a phase factor  $\exp\big(\ii \,\vec\omega\cdot\vec\lambda\big)$. {From} the point of view of the two-dimensional theory on the defect, the Cartan gauge fields ${A}_i$ 
must approach a pure-gauge configuration at infinity so that
\begin{equation}
\label{large_gauge}
{A}_i ~{\sim}~ d\omega_i \quad
\mbox{for}~\rho\to \infty~,
\end{equation}
with $\omega_i$ being a function of $\phi$, the polar angle in the $z_1$-plane. 
In this situation, for the corresponding gauge transformation 
to be single-valued, one finds 
\begin{equation}
\vec\omega(\phi+2\pi)\cdot\vec\lambda = \vec\omega(\phi)\cdot\vec\lambda + 2\pi n
\label{omegalambda}
\end{equation}
with integer $n$. In other words $\vec\omega\cdot\vec\lambda$ must be a map from the circle at infinity 
$S_1^\infty$ into $S_1$ with integer winding number $n$. 
Given this, we have
\begin{equation}
2\pi m_i =\int_D {F}_i = \oint_{S_1^\infty} d\omega_i \,=\,\omega_i(\phi+2\pi)-\omega_i(\phi)~.
\end{equation}
Then, using (\ref{omegalambda}), we immediately deduce that
\begin{equation}
\label{miquant}
\vec m\cdot\vec\lambda\in\mathbb{Z}~.
\end{equation}
For the group SU($N$) this condition amounts to say that $\vec m$ must belong to the dual 
of the weight lattice:
\begin{equation}
\vec m\in(\Lambda_W)^*~.
\label{duallattice}
\end{equation} 

\subsection*{The SU$(N)$ case}
For U$(N)$ the Cartan generators $H_i$ can be taken as the diagonal $(N\times N)$ matrices 
with just a single non-zero entry equal to 1 in the $i$-th place ($i=1,\cdots, N$). 
The restriction to SU$(N)$ can be obtained by choosing a basis of $(N-1)$ traceless generators, 
for instance $(H_i-H_{i+1})/\sqrt{2}$. In terms of the standard 
orthonormal basis $\{\vec{e}_i\}$ of $\mathbb{R}^{N}$, the $(N-1)$ simple roots of SU$(N)$ are then
$\{(\vec e_1-\vec e_2), (\vec e_2 - \vec e_3), \cdots \}$ and the root lattice $\Lambda_R$ is the 
$\mathbb{Z}$-span of these simple roots. 
Note that $\Lambda_R$ lies in a codimension-1 subspace orthogonal to $\sum_i \vec e_i$, and that
the integrality condition for the weights is simply $\vec\alpha\cdot\vec\lambda \in\mathbb{Z}$ for any 
root $\vec{\alpha}$. This shows that the weight lattice is the dual of the root lattice, or 
equivalently that the dual of the weight lattice is the root lattice: $(\Lambda_W)^* = \Lambda_R$.
Therefore, the condition (\ref{duallattice}) implies that the flux vector $\vec{m}$ must be of the form
\begin{equation}
\label{mSUN} 
\vec m = n_1 (\vec e_1 - \vec e_2) + n_2 (\vec e_2 - \vec e_3) + \cdots + n_{N-1} 
(\vec e_{N-1} - \vec e_N)~~~~\text{with\, }
n_i\in \mathbb{Z}~.
\end{equation}  
This simply corresponds to
\begin{equation}
\label{mSUNb}
\vec m = \sum_i m_i \,\vec e_i~~~~\text{with\, } m_i\in\mathbb{Z}~~~\text{and \,} \sum_i m_i = 0~.
\end{equation}
The fact that the fluxes $m_i$ are integers (adding up to zero) has been used in the main text.

\subsection*{Generic surface operator}
The case in which all the $\gamma_i$'s defined in (\ref{fi_limits}) are distinct, corresponds to the 
surface operator of type $[1,1,\ldots, 1]$, also called full surface operator. 
If instead some of the $\gamma_i$'s coincide, the surface operator has a more generic form. 
Let us consider for example the case in which the SU$(N)$ gauge field at the defect takes the form
(see (\ref{Asing})):
\begin{equation}
\label{Asingapp}
A=A_{\mu}\, dx^{\mu}\,\simeq\,-\,\text{diag}
\left(
\begin{array}{cccccccc}
\undermat{n_1}{\gamma_1,\cdots,\gamma_1},\undermat{n_2}{\gamma_2,\cdots,\gamma_2},\cdots, 
\undermat{n_M}{\gamma_M,\cdots,\gamma_M}  
\end{array}
\right)\,d\theta~,
\vspace{.5cm}
\end{equation}
for $r\to0$, which corresponds to splitting the gauge group according to
\begin{equation}
\label{splitapp}
\mathrm{SU}(N) \to \mathrm{S}\big[\mathrm{U}(n_1)\times \mathrm{U}(n_2)\times \cdots \times \mathrm{U}(n_M)\big]~.
\end{equation} 
The calculation of the second Chern class (\ref{kmod}) proceeds as before, but the result can be 
written as follows
\begin{equation}
\label{kmodapp}
\frac{1}{8\pi^2}\int_M \Tr F\wedge F = k + \sum_{I=1}^M \gamma_I \,m_I
\end{equation}
with
\begin{equation}
m_I=\sum_{i=1}^{n_I} m_i = \frac{1}{2\pi}\int_D \sum_{i=1}^{n_I} F_i
= \frac{1}{2\pi}\int_D \Tr F_{\,\mathrm{U}(n_I)}~.
\end{equation}
Here we see that it is the magnetic flux associated with the U$(1)$ factor in each subgroup U$(n_I)$ 
that appears in the expression for the generalized instanton number in the presence of magnetic fluxes.

\section{Ramified instanton moduli and their properties}
\label{secn:moduli}
In this appendix we describe the instanton moduli in the various sectors. Our results are summarized in
Tab.\,\ref{tab:moduli}.

\begin{table}
\begin{center}
{\footnotesize
\begin{tabular}{|c|c|c|c|c|c|}
\hline
$\!\!\!\!\text{Doublet} \!\!\!\!$ & $\!\!(-)^{F_\alpha}\phantom{\Big|}\!\!\!\!\!$ & $\!\!\text{Chan-Paton}\!\!$
 & $\!\!\!\!\mathrm{U}(1)^4\text{charge}\!\!\!$ & $\!\!\!\!Q^2$-eigenvalue $\lambda_\alpha\!\!\!$ & $\!\!\!
\text{Character}\!\!\!$\cr
\hline
$\!\!\phantom{\Big|}\!\!(\bar\chi_I,\bar\eta_I)\!\!$ & $+$ & $(\mathbf{d}_I, \mathbf{\bar d}_I)$ & $\!
\big\{0,0,0,0\big\}\!$ &
$\chi_{I,\sigma} - \chi_{I,\tau}$  & \cr
$\!\!\phantom{\Big|}\!\!(A_I^{z_1},M_I^{z_1})\!\!$ & $+$ & $(\mathbf{d}_I, \mathbf{\bar d}_I)$ & $\!
\big\{\!\!+\!1,0,0,0\big\}\!$ &
$\chi_{I,\sigma} - \chi_{I,\tau} + \epsilon_1$ & $V_I^* V_I T_1$\cr
$\!\!\phantom{\Big|}\!\!(A_I^{z_4},M_I^{z_4})\!\!$ & $+$ & $(\mathbf{d}_I, \mathbf{\bar d}_I)$ & $\!
\big\{0,0,0,+1\big\}\!$ &
$\chi_{I,\sigma} - \chi_{I,\tau} + \epsilon_4$ & $V_I^* V_I T_4 $ \cr
$\!\!\phantom{\Big|}\!\!(\lambda_I,D_I)\!\!$ & $-$ & $(\mathbf{d}_I, \mathbf{\bar d}_I)$ 
& $\!\big\{\!\!+\!\frac 12,\!+\frac 12,\!+\frac 12,\!+\frac 12\big\}\!$  &
$\chi_{I,\sigma} - \chi_{I,\tau}$  & \cr
$\!\!\phantom{\Big|}\!\!(\lambda_I^{z_1},D_I^{z_1})$ & $-$ & $(\mathbf{d}_I, \mathbf{\bar d}_I)$ & 
$\!\big\{\!\!+\!\frac 12,\!-\frac 12,\!-\frac 12,\!+\frac 12\big\}\!$ &
$\chi_{I,\sigma} - \chi_{I,\tau} +\epsilon_1+ \epsilon_4$ 
&  $- V_I^* V_I T_1 T_4 $  \cr
\hline
$\!\!\phantom{\Big|}\!\!(A_I^{z_2},M_I^{z_2})\!\!$ & $+$ & $(\mathbf{d}_I, \mathbf{\bar d}_{I+1})$ & 
$\!\big\{0,+1,0,0\big\}\!$ &
$\chi_{I,\sigma} - \chi_{I+1,\rho} + \hat\epsilon_2$ & $V_{I+1}^* V_I T_2$ \cr
$\!\!\phantom{\Big|}\!\!(\lambda_I^{z_2},D_I^{z_2})\!\!$ & $-$ & $(\mathbf{d}_I, \mathbf{\bar d}_{I+1})$ 
& $\!\big\{\!\!-\!\frac 12,\!+\frac 12,\!-\frac 12,\!+\frac 12\big\}\!$&
$\chi_{I,\sigma} - \chi_{I+1,\rho} +\hat\epsilon_2+\epsilon_4$ & $- V_{I+1}^* V_I T_2 T_4 $\cr
$\!\!\phantom{\Big|}\!\!(\bar{A}_I^{z_3},\bar{M}_I^{z_3})\!\!$ & $+$ & $(\mathbf{d}_{I}, \mathbf{\bar d}_{I+1})$ & 
$\!\big\{0,0,-1,0\big\}\!$ &
$\chi_{I,\sigma} - \chi_{I+1,\rho} -\hat\epsilon_3$ & 
$V_{I+1}^* V_I T_1T_2T_4$ \cr
$\!\!\phantom{\Big|}\!\!(\lambda_I^{z_3},D_I^{z_3})\!\!$ & $-$ & $(\mathbf{d}_I, \mathbf{\bar d}_{I+1})$ 
& $\!\big\{\!\!+\!\frac 12,\!+\frac 12,\!-\frac 12,\!-\frac 12\big\}\!$&
$\chi_{I,\sigma} - \chi_{I+1,\rho} +\epsilon_1+\hat\epsilon_2$ & $- V_{I+1}^* V_I T_1T_2 $\cr
\hline
$\!\!\phantom{\Big|}\!\!(w_I,\mu_I)\!\!$ & $+$ & $(\mathbf{n}_I, \mathbf{\bar{d}}_I)$ & 
$\!\big\{\!\!+\!\frac 12,\!+\frac 12,0,0\big\}\!$ &
$a_{I,s} -\chi_{I,\sigma} + \frac 12 (\epsilon_1+ \hat\epsilon_2)$ & $ V_I^* W_I$\cr
$\!\!\phantom{\Big|}\!\!(\mu_I^\prime,h_I^\prime)\!\!$ & $-$ & $(\mathbf{n}_I, \mathbf{\bar d}_I)$ 
& $\!\big\{0,0,\!-\frac 12,\!+\frac 12\big\}\!$ &
$a_{I,s} - \chi_{I,\sigma} + \frac 12 (\epsilon_1+ \hat\epsilon_2) + \epsilon_4$  & $ - V_I^* W_I T_4$\cr
\hline
$\!\!\phantom{\Big|}\!\!(\hat{w}_I,\hat{\mu}_I)\!\!$ & $+$ & $(\mathbf{d}_I, \mathbf{\bar n}_{I+1})$ 
& $\!\big\{\!\!+\!\frac 12,\!+\frac 12,0,0\big\}\!$  &
$\chi_{I,\sigma} - a_{I+1,t} + \frac 12 (\epsilon_1+ \hat\epsilon_2)$ & $W_{I+1}^* V_I T_1 T_2$\cr
$\!\!\phantom{\Big|}\!\!(\hat{\mu}_I^\prime,\hat{h}_I^\prime)\!\!$ & $-$ & 
$(\mathbf{d}_I, \mathbf{\bar n}_{I+1})$ & $\!\big\{0,0,\!-\frac 12,\!+\frac 12\big\}\!$  &
$\chi_{I,\sigma} - a_{I+1,t} + \frac 12 (\epsilon_1+ \hat\epsilon_2) + \epsilon_4$ 
& $\!-W_{I+1}^* V_I T_1 T_2 T_4\!$ \cr
\hline
\end{tabular}
}
\end{center}
\caption{The spectrum of moduli, organized in doublets of the BRST charge $Q$ (or its conjugate $\bar Q$). 
For each of them, we display their statistics $(-)^{F_\alpha}$, 
the representation of the color and ADHM groups 
in which they transform, their charge vector with respect to the $\mathrm{U}(1)^4$ symmetry, 
the eigenvalue $\lambda_\alpha$ of $Q^2$ and the corresponding contribution to the character. 
The neutral moduli carrying a superscript $z_1$, $z_2$, $z_3$ or $z_4$, and the colored moduli 
in this table are complex. The quantities appearing in the last column, namely
$V_I$, $W_I$, $T_1$,$T_2$ and $T_4$ are defined in (\ref{VIWIdef}) and (\ref{Tidef}).}
\label{tab:moduli}
\end{table}

Let us first consider the neutral states of the strings stretching between two $D$-instantons. 

\paragraph{$\bullet$ $(-1)$/$(-1)$ strings of type $I$-$I$:}  All moduli of this type transform in the 
adjoint representation $(\mathbf{d}_I, \mathbf{\bar d}_I)$ of $\mathrm{U}(d_I)$.
A special role is played by the bosonic states created in the Neveu-Schwarz (NS) sector of such strings by
the complex oscillator $\psi^{v}$ in the last complex space-time direction, which is neutral with respect to 
the orbifold. We denote them by $\chi_I$. They are characterized by a $\mathrm{U}(1)^4$ weight
$\{0,0,0,0\}$ and a charge $(+1)$ with respect to the last U(1).
The complex conjugate moduli ${\bar\chi}_I$, with weight $\{0,0,0,0\}$ and 
charge $(-1)$, are paired in a $Q$-doublet with the fermionic moduli ${\bar\eta}_I$ coming from 
the ground state of the Ramond (R) sector with weight $\big\{\!\!-\ft12,-\ft12,-\ft12,-\ft12\big\}$ 
and charge $(-\frac{1}{2})$.
All other moduli in this sector are arranged in $Q$-doublets. One doublet 
is $(A_I^{z_1},M_I^{z_1})$, where $A_I^{z_1}$ is from the $\psi^{z_1}$ oscillator 
in the NS sector with weight $\{+1,0,0,0\}$ and charge $0$, and $M_I^{z_1}$ is from the 
R ground state $\big\{\!\!+\ft12,-\ft12,-\ft12,-\ft12\big\}$ with charge 
$(+\frac{1}{2})$. Another doublet is $(A_I^{z_4},M_I^{z_4})$, where $A_I^{z_4}$ is from 
the $\psi^{z_4}$ oscillator in the NS sector with weight $\{0,0,0,+1\}$ and charge $0$, and 
$M_I^{z_4}$ is from the R ground state with weight $\big\{\!\!-\ft12,-\ft12,-\ft12,+\ft12\big\}$ and charge 
$(+\frac{1}{2})$. Also the complex conjugate doublets are present. 
Finally, there is a (real) doublet $(\lambda_I,D_I)$ where $\lambda_I$ is from the R ground state 
with weight $\big\{\!\!+\ft12,+\ft12,+\ft12,+\ft12\big\}$ and charge 
$(-\frac{1}{2})$, and $D_I$ is an auxiliary field, and a complex doublet 
$(\lambda_I^{z_1},D_I^{z_1})$ with $\lambda_I^{z_1}$ associated to the R ground state with weight
$\big\{\!\!+\ft12,-\ft12,-\ft12,+\ft12\big\}$ and charge $(-\frac{1}{2})$, and $D_I^{z_1}$ an auxiliary
field.

\paragraph{$\bullet$ $(-1)$/$(-1)$ strings of type $I$-$(I+1)$:}  In this sector the moduli
transform in the bi-fundamental representation $(\mathbf{d}_I, \mathbf{\bar d}_{I+1})$ of 
$\mathrm{U}(\mathbf{d}_I)\times \mathrm{U}(\mathbf{d}_{I+1})$. In order to cancel 
the phase $\omega^{-1}$ due to the different representations on the Chan-Paton indices 
at the two endpoints, the weights under spacetime rotations of the operators creating the states 
in this sector must be such that $l_2 - l_3 = 1$. 
In this way they can survive the $\mathbb{Z}_M$-orbifold projection.
Applying this requirement, we find a doublet $(A_I^{z_2},M_I^{z_2})$, 
$A_I^{z_2}$ is from the $\psi^{z_2}$ oscillator in the NS sector with weight $\{0,+1,0,0\}$ and 
charge $0$, and $M_I^{z_2}$ is from the R ground state $\big\{\!\!-\ft12,+\ft12,-\ft12,-\ft12\big\}$ 
with charge 
$(+\frac{1}{2})$. Another doublet is $(\bar{A}_I^{z_3},\bar{M}_I^{z_3})$ where $\bar{A}_I^{z_3}$ is 
from the $\bar{\psi}^{z_3}$ oscillator in the NS sector with weight $\{0,0,-1,0\}$ and charge $0$, 
and $\bar{M}_I^{z_3}$ is from the R ground state $\big\{\!\!+\ft12,+\ft12,-\ft12,+\ft12\big\}$ with charge 
$(+\frac{1}{2})$\,\footnote{Notice that this last doublet is actually the complex conjugate of a $Q$-doublet of type
$(I+1)$-$I$, which is made of $({A}_I^{z_3},{M}_I^{z_3})$ with ${A}_I^{z_3}$ corresponding to the 
weight $\{0,0,1,0\}$ and ${M}_I^{z_3}$ corresponding to the weight 
$\big\{\!\!-\ft12,-\ft12,+\ft12,-\ft12\big\}$.}. Furthermore, we find two other complex 
$Q$-doublets, $(\lambda_I^{z_2},D_I^{z_2})$ 
and $(\lambda_I^{z_3},D_I^{z_3})$ where $\lambda_I^{z_2}$ and $\lambda_I^{z_3}$ are 
associated to the R ground states with weights $\big\{\!\!-\ft12,+\ft12,-\ft12,+\ft12\big\}$ and
$\big\{\!\!+\ft12,+\ft12,-\ft12,-\ft12\big\}$ and charges $(-\frac{1}{2})$, while $D_I^{z_2}$
and $D_I^{z_3}$ are auxiliary fields. Also the complex conjugate doublets are present in the 
$\mathbb{Z}_M$-invariant spectrum, and arise from strings with the opposite orientation.  

\paragraph{$\bullet$ $3/(-1)$ strings of type $I$-$I$:}
These open strings have mixed Neumann-Dirichlet boundary conditions along the $(z_1,z_2)$-directions 
and thus the corresponding states are characterized by the action of a twist operator 
$\Delta$ \cite{Billo:2002hm}. We assign an orbifold charge $\omega^{-\frac 12}$ to this twist operator, 
so that the states which survive the $\mathbb{Z}_M$-projection are those with weights such 
that $l_2 - l_3 = 1/2$. The moduli in this sector belong to the 
bi-fundamental representation $(\mathbf{n}_I\times \mathbf{\bar{d}}_I)$ of the gauge and ADHM groups, 
and form two complex doublets. One is $(w_I,\mu_I)$ where the NS component
$w_I$ has weight $\big\{\!\!+\ft12,+\ft12,0,0\big\}$ and charge $0$, and the R 
component $\mu_I$ has weight $\big\{0,0,-\ft12,-\ft12\big\}$ and charge $(+\frac{1}{2})$.
The other doublet is $(\mu_I^\prime, h_I^\prime)$ where $\mu_I^\prime$ is 
associated to the R ground state with weight $\big\{0,0,-\ft12,+\ft12\big\}$ and charge $(-\frac{1}{2})$, 
while $h_I^\prime$ is an auxiliary field. Also the complex conjugate doublets, associated to the
$(-1)/3$ strings of type $I$-$I$, are present in the spectrum.

\paragraph{$\bullet$ $(-1)/3$ strings of type $I$-$(I+1)$:}
These open strings have mixed Dirichlet-Neumann boundary conditions along the $(z_1,z_2)$-directions
and transform in the bi-fundamental representation $(\mathbf{d}_I\times \mathbf{\bar{n}}_{I+1})$ of 
the gauge and ADHM groups. 
As compared to the previous case, the states in this sector are characterized by the action of an 
anti-twist operator $\bar{\Delta}$ which carries an orbifold parity $\omega^{+\frac 12}$. Thus 
the $\mathbb{Z}_M$-invariant configurations must have again weights with $l_2-l_3=1/2$ in order to compensate 
for the $\omega^{-1}$ factor carried by the Chan-Paton indices. Taking this into account, we find
two complex doublets: $(\hat{w}_I,\hat{\mu}_I)$ where the NS component
$\hat{w}_I$ has weight $\big\{\!\!+\ft12,+\ft12,0,0\big\}$ and charge $0$, and the R 
component $\hat{\mu}_I$ has weight $\big\{0,0,-\ft12,-\ft12\big\}$ and charge $(+\frac{1}{2})$, and
$(\hat{\mu}_I^\prime, \hat{h}_I^\prime)$ where $\hat{\mu}_I^\prime$ is 
associated to the R ground state with weight $\big\{0,0,-\ft12,+\ft12\big\}$ and charge $(-\frac{1}{2})$, 
while $\hat{h}_I^\prime$ is an auxiliary field. Also the complex conjugate doublets, associated to the
$3/(-1)$ strings of type $(I+1)$-$I$, are present in the spectrum.

Notice that no states from the $3/(-1)$ strings of type $I$-$(I+1)$ or from the $(-1)/3$ strings of 
type $(I+1)$-$I$ survive the orbifold projection. Indeed, in the first case the phases $\omega^{-\frac{1}{2}}$
and $\omega^{-1}$ from the twist operator $\Delta$ and the Chan-Paton factors cannot be compensated
by the NS or R weights; while in the second case the phases $\omega^{+\frac{1}{2}}$ 
and $\omega^{+1}$ from the anti-twist operator $\bar{\Delta}$ and the Chan-Paton factors 
cannot be canceled.

All the above results are summarized in Tab.\,\ref{tab:moduli}, which contains also other relevant information
about the moduli.
As an illustrative example, we now consider in detail the $\mathrm{SU}(2)$ theory.

\subsection{SU(2)}
\label{secn:su2app}
In this case we have $M=2$, and thus necessarily $n_1=n_2=1$. Therefore, in the SU(2) theory
we have only simple surface operators. Furthermore, since the index $s$ takes only one
value, we can simplify the notation and suppress this index in the following. 

Each pair $Y=(Y_1,Y_2)$ of Young tableaux contributes to the instanton partition function with a
weight $q_1^{d_1}\, q_2^{d_2}$ where $d_1$ and $d_2$ are given by (\ref{dIY}), which in this
case take the simple form \cite{Alday:2010vg}
\begin{equation}
d_1 = \sum_j  \left(Y_1^{2j+1} + Y_2^{2j+1}\right)~, \qquad d_2 = \sum_j 
\left(Y_1^{2j+2} + Y_2^{2j+2}\right)~.
\end{equation}
with $Y_I^k$ representing the length of the $k$th column of the tableau $Y_I$. 

Let us begin by considering the case of pairs of Young tableaux with a single box. 
There are two such pairs that can contribute.
One is $Y=\left(\Yfund, \bullet\right)$
corresponding to $d_1=1$ and $d_2=0$. Using these values in (\ref{zexplicit}), we find 
\begin{equation}
z_{\{1,0\}}= \frac{\left(\epsilon_1+\epsilon_4\right)\left(a_{1} - \chi_{1,1} + \frac 12 
(\epsilon_1 + \hat\epsilon_2)+\epsilon_4\right)\left(\chi_{1,1} 
- a_{2} + \frac 12 (\epsilon_1 + \hat\epsilon_2)+\epsilon_4\right)}{\epsilon_1\,\epsilon_4\,
\left(a_{1}-\chi_{1,1} + \frac 12 (\epsilon_1 + \hat\epsilon_2)\right)
\left(\chi_{1,1} - a_{2} + \frac 12 (\epsilon_1 + \hat\epsilon_2)\right)}
\end{equation}
Due to the prescription (\ref{prescription}), only the pole at  
\begin{equation}
\chi_{1,1} = a_{1} + \frac 12 (\epsilon_1 + \hat\epsilon_2)
\label{pole1}
\end{equation}
contributes to the contour integral over $\chi_{1,1}$, yielding
\begin{equation}
Z_{\left(\Yfund, \,\bullet\right)} = \frac{\left(\epsilon_1+\epsilon_4\right)
 \left(a_{12}+\epsilon _1+\hat\epsilon_2+ \epsilon_4\right)}{\epsilon_1\left(a_{12}+\epsilon_1+\hat\epsilon_2\right)}
=\frac{\left(\epsilon_1+\epsilon_4\right) \left(4a+2 \epsilon _1
+\epsilon_2+2 \epsilon_4\right)}{\epsilon_1 \left(4 a+2 \epsilon_1+\epsilon_2\right)}
\end{equation}
where in the last step we used the notation $a_{12} = a_{1} - a_{2}=2a$ and reintroduced 
$\epsilon_2=2\hat{\epsilon}_2$. 
A similar analysis can be done for the second pair of tableaux with one box that contributes, namely 
$Y=\left(\bullet, \Yfund\right)$ corresponding to $d_1=0$ and $d_2=1$. In this case we find
\begin{equation}
Z_{\left(\bullet, \Yfund\right)} = \frac{ \left(\epsilon_1+\epsilon_4\right) 
\left(-4a+2 \epsilon_1+\epsilon_2+2 \epsilon _4\right)}{\epsilon _1\left(-4a+2 \epsilon_1+\epsilon_2\right)}~.
\end{equation}

In the case of two boxes, we have five different pairs of tableaux that can contribute. They are:
$Y=\left(\Yfund,\Yfund\right)$, $Y=\left(\Ysymm,\bullet\right)$, $Y=\left(\bullet,\Ysymm\right)$,
$Y=\left(\Yasymm,\bullet\right)$ and $Y=\left(\bullet,\Yasymm\right)$. The contributions of these
five diagrams are listed below in Tab.\,\ref{ZlistSU2}.

\begin{table}
\begin{center}
\begin{tabular}{|c|c|c|c|}
\hline
$\phantom{\big|}Y$ & $\!\!\text{weight}\!\!$ &  poles & $Z_Y$\cr
\hline
$\!\!\left(\Yfund,\Yfund\right)\!\!$ & $q_1 q_2$ & 
\begin{tabular}{@{}c@{}}$\phantom{\Big|}\chi_{1,1} = a_{1} + \frac 12\left(\epsilon_1 
+ \hat\epsilon_2\right) $ \\ $ \phantom{\Big|}\chi_{2,1} = a_{2} + \frac 12\left(\epsilon_1 + \hat\epsilon_2\right)$
\end{tabular}
& $\frac{\phantom{\big|}(\epsilon_1+ \epsilon_4)^2(4a + \epsilon_2+2\epsilon_4)(-4a 
+ \epsilon_2+2\epsilon_4)}{\phantom{\big|}\epsilon_1^2 (4a +\epsilon_2)(-4a +\epsilon_2)}$ \cr
\hline
$\!\!\left(\Ysymm,\bullet\right)\!\!$ & $q_1q_2$ & 
\begin{tabular}{@{}c@{}}$\phantom{\Big|}\chi_{1,1} = a_{1} + \frac 12\left(\epsilon_1 + \hat\epsilon_2\right)$ \\ 
$\phantom{\Big|}\chi_{2,1} = \chi_{1,1} + \hat\epsilon_2$
\end{tabular}
 &  $ \frac{\phantom{\big|}(\epsilon_1+ \epsilon_4)(\epsilon_2+ \epsilon_4)(4a 
+\epsilon_2-2\epsilon_4)(4a +2\epsilon_1 
 + \epsilon_2+2\epsilon_4)}{\phantom{\big|}\epsilon_1\epsilon_2(4a +\epsilon_2)(4a + 2\epsilon_1 +\epsilon_2)}$ 
\cr
\hline
$\!\!\left(\bullet,\Ysymm\right)\!\!$ & $q_1 q_2$ & 
\begin{tabular}{@{}c@{}}
 $\phantom{\Big|}\chi_{2,1} = a_{2} + \frac 12\left(\epsilon_1 + \hat\epsilon_2\right)$ 
 \\
 $\phantom{\Big|}\chi_{1,1} = \chi_{2,1} + \hat\epsilon_2$ 
 \end{tabular}
 & $ \frac{\phantom{\big|}(\epsilon_1+ \epsilon_4)(\epsilon_2+ \epsilon_4)(-4a 
+\epsilon_2-2\epsilon_4)(-4a +2\epsilon_1 
 + \epsilon_2+2\epsilon_4)}{\phantom{\big|}\epsilon_1\epsilon_2(-4a +\epsilon_2)(-4a + 2\epsilon_1 +\epsilon_2)}$  
\cr
\hline
$\!\!\left(\Yasymm, \bullet\right)\!\!$ & $q_1^2$ & 
\begin{tabular}{@{}c@{}}
$\phantom{\Big|}\chi_{1,1} =a_{1} + \frac 12\left(\epsilon_1 + \hat\epsilon_2\right) $  \\ 
$\phantom{\Big|}\chi_{1,2}=\chi_{1,1}+\epsilon_1$ 
\end{tabular}
& $\frac{\phantom{\big|} \left(\epsilon_1+\epsilon_4\right) 
\left(2 \epsilon_1+\epsilon_4\right) \left(4a+2 \epsilon_1+\epsilon_2+2 \epsilon_4\right) 
\left(4a+4 \epsilon _1+\epsilon_2+2 \epsilon_4\right)}{\phantom{\big|}2 \epsilon_1^2 
\left(4a+2 \epsilon_1+\epsilon_2\right) \left(4a+4 \epsilon_1+\epsilon_2\right)}$ \cr
\hline
$\!\!\left(\bullet, \Yasymm \right)\!\!$ & $q_2^2$ & 
\begin{tabular}{@{}c@{}}
$\phantom{\Big|}\chi_{2,1}=a_{2} + \frac 12\left(\epsilon_1 + \hat\epsilon_2\right) $ \\
$\phantom{\Big|}\chi_{2,2} =\chi_{2,1}+\epsilon_1$ 
\end{tabular}
& $\frac{\phantom{\big|} \left(\epsilon_1+\epsilon_4\right) \left(2 \epsilon_1+\epsilon_4\right) 
\left(-4a+2 \epsilon_1+\epsilon_2+2 \epsilon_4\right) 
\left(-4a+4 \epsilon _1+\epsilon_2+2 \epsilon_4\right)}{\phantom{\big|}2 \epsilon_1^2 
\left(-4a+2 \epsilon_1+\epsilon_2\right) \left(-4a+4 \epsilon_1+\epsilon_2\right)}$  \cr
\hline
\end{tabular}
\end{center}
\caption{We list the tableaux, the weight factors, the pole structure and the contribution 
to the partition function in all five cases with two boxes for the SU(2) theory.} 
\label{ZlistSU2}
\end{table}

Multiplying all contributions with the appropriate weight factor and summing over them, 
we obtain the instanton partition function for the SU(2) gauge theory in the presence of the surface operator:
\begin{align}
\label{Zsu211}
Z_{\text{inst}}[1,1] &= 1+q_1\frac{\left(\epsilon_1+\epsilon_4\right) 
\left(4a+2 \epsilon _1+\epsilon_2+2 \epsilon_4\right)}{\epsilon_1\left(4 a+2 \epsilon_1+\epsilon_2\right)}
+ q_2\frac{ \left(\epsilon_1+\epsilon_4\right) 
\left(-4a+2 \epsilon_1+\epsilon_2+2 \epsilon _4\right)}{\epsilon _1\left(-4a+2 \epsilon_1+\epsilon_2\right)}\cr
&~~ +q_1^2\,\frac{\left(\epsilon_1+\epsilon_4\right) \left(2 \epsilon_1+\epsilon_4\right) 
\left(4a+2 \epsilon_1+\epsilon_2+2 \epsilon_4\right) 
\left(4a+4 \epsilon _1+\epsilon_2+2 \epsilon_4\right)}{2 \epsilon_1^2 
\left(4a+2 \epsilon_1+\epsilon_2\right) \left(4a+4 \epsilon_1+\epsilon_2\right)}\cr
&~~+ q_2^2\,\frac{\left(\epsilon_1+\epsilon_4\right) \left(2 \epsilon_1+\epsilon_4\right) 
\left(-4a+2 \epsilon_1+\epsilon_2+2 \epsilon_4\right) 
\left(-4a+4 \epsilon _1+\epsilon_2+2 \epsilon_4\right)}{2 \epsilon_1^2 
\left(-4a+2 \epsilon_1+\epsilon_2\right) \left(-4a+4 \epsilon_1+\epsilon_2\right)}\cr
&~~+q_1q_2\left( \frac{(\epsilon_1+ \epsilon_4)(\epsilon_2+ \epsilon_4)(4a +\epsilon_2-2\epsilon_4)(4a +2\epsilon_1 
 + \epsilon_2+2\epsilon_4)}{\epsilon_1\epsilon_2(4a +\epsilon_2)(4a + 2\epsilon_1 +\epsilon_2)}\right.\cr
&\hspace{2cm} 
+\frac{(\epsilon_1+ \epsilon_4)(\epsilon_2+ \epsilon_4)(-4a +\epsilon_2-2\epsilon_4)(-4a +2\epsilon_1 
 + \epsilon_2+2\epsilon_4)}{\epsilon_1\epsilon_2(-4a +\epsilon_2)(-4a + 2\epsilon_1 +\epsilon_2)} \cr
&\hspace{3cm} +\left. \frac{(\epsilon_1+ \epsilon_4)^2(4a + 
\epsilon_2+2\epsilon_4)(-4a + \epsilon_2+2\epsilon_4)}{\epsilon_1^2 (4a +\epsilon_2)(-4a +\epsilon_2)} \right) 
+\cdots
\end{align}
where the ellipses stand for the contributions originating from tableaux with higher number of boxes,
which can be easily generated with a computer program. We have explicitly computed these terms up 
six boxes, but we do not write them here since the raw expressions are very long and not particularly
illuminating. To the extent it is possible to make comparisons, we observe that the above result agrees
with the instanton partition function reported in eq. (B.6) of \cite{Alday:2010vg} under the following 
change of notation
\begin{equation}
q_1\to y~,\quad q_2\to x~,\quad \epsilon_4\to -m~,\quad 2a\to 2a+\ft{\epsilon_2}{2}~. 
\end{equation}
Note then that the mass $m$ appearing in \cite{Alday:2010vg} is the equivariant mass of the hypermultiplet
\cite{Okuda:2010ke}, which differs by $\epsilon$-corrections from the mass we have used in this paper (see
(\ref{e4})).

\section{Prepotential coefficients for the SU($N$) gauge theory}
\label{Freview}
The prepotential $\mathcal{F}$ of the ${\mathcal N}=2^{\star}$ SU$(N)$ 
gauge theory has been determined in terms of quasi-modular forms 
in \cite{Billo:2014bja,Billo':2015ria}. Expanding $\mathcal{F}$ as in (\ref{FandW}), the first 
few non-zero coefficients $f_\ell$ in the NS limit turn out to be
\begin{align}
f_2 &= \frac{1}{4}\Big(m^2-\frac{\epsilon_1^2}{4}\Big)\sum_{u\neq v}
\log\frac{(a_u-a_v)^2}{\Lambda^2}+N\Big(m^2-\frac{\epsilon_1^2}{4}\Big)\log\widehat\eta~,
\hspace{3.5cm}\label{f2app}
\end{align}
\begin{align}
f_4 &=-\frac{1}{24}\Big(m^2-\frac{\epsilon_1^2}{4}\Big)^2\,E_2\, C_2~,
\hspace{8.4cm}\label{f4app}
\end{align}
\begin{align}
f_6 &= -\frac{1}{288}\Big(m^2-\frac{\epsilon_1^2}{4}\Big)^2\Bigg\{\bigg[\frac{2}{5}
\Big(m^2-\frac{\epsilon_1^2}{4}\Big)
\big(5E_2^2+E_4\big)-6\,\epsilon_1^2\,E_4\bigg]C_4\notag\hspace{3.5cm}\\
&\qquad+\frac{1}{2}\Big(m^2-\frac{\epsilon_1^2}{4}\Big)\big(E_2^2-E_4\big)\,C_{2;1,1}\Bigg\}~,
\label{f6app}
\end{align}
\begin{align}
f_8 &=-\frac{1}{1728} \Big(m^2-\frac{\epsilon_1^2}{4}\Big)^2\Bigg\{\bigg[
\frac{2}{105}\Big(m^2-\frac{\epsilon_1^2}{4}\Big)^2\big(175 E_2^3+84E_2E_4+11E_6\big)
\notag\\
&\qquad-\frac{24\,\epsilon^2}{35} \Big(m^2-\frac{\epsilon_1^2}{4}\Big) \big(7E_2E_4+3E_6\big)
+\frac{24\,\epsilon^4}{7} \,E_6 \bigg]C_6 \notag\\
&\qquad-\frac{1}{5}\Big(m^2-\frac{\epsilon_1^2}{4}\Big)\bigg[\Big(m^2-\frac{\epsilon_1^2}{4}\Big)
\big(5 E_2^3-3E_2E_4-2E_6\big)-6 \,\epsilon^2\big(E_2E_4-E_6\big)\bigg]C_{4;2}  \notag \\
&\qquad
-\frac{1}{5}\Big(m^2-\frac{\epsilon_1^2}{4}\Big)\bigg[\frac{1}{12}\Big(m^2-\frac{\epsilon_1^2}{4}\Big)
\big(5 E_2^3-3E_2E_4-2E_6\big)-3\,\epsilon^2\big(E_2E_4-E_6\big) \bigg]C_{3;3} \nonumber \\
&\qquad
+\frac{1}{24}\Big(m^2-\frac{\epsilon_1^2}{4}\Big)^2 \big(E_2^3-3E_2E_4+2E_6\big)
\, C_{2;1,1,1,1}\Bigg\}~.\label{f8app}
\end{align}
Here $E_2$, $E_4$ and $E_6$ are the Eisenstein series and
\begin{equation}
 \log\widehat\eta= -\sum_{k=1}^\infty \frac{\sigma_1(k)}{k}\,q^k=-\frac{1}{24}\log q+\log \eta
\label{etahat}
\end{equation}
with $\eta$ being the Dedekind $\eta$-function. Finally, the root lattice sums are defined by
\begin{equation}
C_{n;m_1,m_2,\cdots, m_k}  = \sum_{\vec{\alpha}\in\Phi}\,\sum_{\vec{\beta}_1\not=\vec{\beta}_2
\not=\cdots \not=\vec{\beta}_k\in\Phi(\vec{\alpha})} 
\frac{1}{(\vec{\alpha}\cdot \vec{a})^n(\vec{\beta}_1\cdot \vec{a})^{m_1}(\vec{\beta}_2
\cdot \vec{a})^{m_1}\cdots (\vec{\beta}_k\cdot \vec{a})^{m_k}\ }
\end{equation}
where $\Phi$ is the root system of SU($N$) and
\begin{equation}
\Phi(\vec{\alpha}) =\{\vec{\beta}\in\Phi~\big|~\vec{\alpha}\cdot\vec{\beta}=1\} ~.
\label{Psialphaapp}
\end{equation}
We refer to \cite{Billo':2015ria} for the details and the derivation of these results. Notice, however, that we
have slightly changed our notation, since $f_{2\ell}^{\text{here}}=f_{\ell}^{\text{there}}$.
By expanding the modular functions in powers of $q$ and selecting SU(2) as gauge group, it is easy to show that
the above formulas reproduce both the perturbative part and the instanton contributions, reported
respectively in (\ref{Fpertexp}) and (\ref{fellinst}) of the main text.

\providecommand{\href}[2]{#2}\begingroup\raggedright
\endgroup


\begin{thebibliography}{10}
\bibitem{Gukov:2014gja}
S.~Gukov, \emph{{Surface Operators}},
\href{http://arxiv.org/abs/1412.7127}{{\tt arXiv:1412.7127 [hep-th]}}.

\bibitem{Gukov:2006jk}
S.~Gukov and E.~Witten, \emph{{Gauge Theory, Ramification, And The Geometric
  Langlands Program}},
\href{http://arxiv.org/abs/hep-th/0612073}{{\tt arXiv:hep-th/0612073
  [hep-th]}}.

\bibitem{Gukov:2008sn}
S.~Gukov and E.~Witten, \emph{{Rigid Surface Operators}},
  \href{http://dx.doi.org/10.4310/ATMP.2010.v14.n1.a3}{Adv. Theor. Math. Phys.
  {\bf 14} (2010) 1, 87--178},
\href{http://arxiv.org/abs/0804.1561}{{\tt arXiv:0804.1561 [hep-th]}}.

\bibitem{Gaiotto:2009fs}
D.~Gaiotto, \emph{{Surface Operators in N = 2 4d Gauge Theories}},
  \href{http://dx.doi.org/10.1007/JHEP11(2012)090}{JHEP {\bf 11} (2012)  090},
\href{http://arxiv.org/abs/0911.1316}{{\tt arXiv:0911.1316 [hep-th]}}.

\bibitem{Gaiotto:2013sma}
D.~Gaiotto, S.~Gukov, and N.~Seiberg, \emph{{Surface Defects and Resolvents}},
  \href{http://dx.doi.org/10.1007/JHEP09(2013)070}{JHEP {\bf 09} (2013)  070},
\href{http://arxiv.org/abs/1307.2578}{{\tt arXiv:1307.2578 [hep-th]}}.

\bibitem{Witten:1997sc}
E.~Witten, \emph{{Solutions of four-dimensional field theories via M theory}},
  \href{http://dx.doi.org/10.1016/S0550-3213(97)00416-1}{Nucl.Phys. {\bf B500}
  (1997)  3--42},
\href{http://arxiv.org/abs/hep-th/9703166}{{\tt arXiv:hep-th/9703166
  [hep-th]}}.

\bibitem{Gaiotto:2009we}
D.~Gaiotto, \emph{{N=2 dualities}},
  \href{http://dx.doi.org/10.1007/JHEP08(2012)034}{JHEP {\bf 1208} (2012)
  034},
\href{http://arxiv.org/abs/0904.2715}{{\tt arXiv:0904.2715 [hep-th]}}.

\bibitem{Alday:2010vg}
L.~F. Alday and Y.~Tachikawa, \emph{{Affine SL(2) conformal blocks from 4d
  gauge theories}}, \href{http://dx.doi.org/10.1007/s11005-010-0422-4}{Lett.
  Math. Phys. {\bf 94} (2010)  87--114},
\href{http://arxiv.org/abs/1005.4469}{{\tt arXiv:1005.4469 [hep-th]}}.

\bibitem{Alday:2009aq}
L.~F. Alday, D.~Gaiotto, and Y.~Tachikawa, \emph{{Liouville correlation
  functions from four-dimensional gauge theories}},
  \href{http://dx.doi.org/10.1007/s11005-010-0369-5}{Lett. Math. Phys. {\bf 91}
  (2010)  167--197},
\href{http://arxiv.org/abs/0906.3219}{{\tt arXiv:0906.3219 [hep-th]}}.

\bibitem{Kozcaz:2010yp}
C.~Kozcaz, S.~Pasquetti, F.~Passerini, and N.~Wyllard, \emph{{Affine sl(N)
  conformal blocks from N=2 SU(N) gauge theories}},
  \href{http://dx.doi.org/10.1007/JHEP01(2011)045}{JHEP {\bf 01} (2011)  045},
\href{http://arxiv.org/abs/1008.1412}{{\tt arXiv:1008.1412 [hep-th]}}.

\bibitem{Alday:2009fs}
L.~F. Alday, D.~Gaiotto, S.~Gukov, Y.~Tachikawa, and H.~Verlinde, \emph{{Loop
  and surface operators in N=2 gauge theory and Liouville modular geometry}},
  \href{http://dx.doi.org/10.1007/JHEP01(2010)113}{JHEP {\bf 1001} (2010)
  113},
\href{http://arxiv.org/abs/0909.0945}{{\tt arXiv:0909.0945 [hep-th]}}.

\bibitem{Taki:2009zd}
M.~Taki, \emph{{On AGT Conjecture for Pure Super Yang-Mills and W-algebra}},
  \href{http://dx.doi.org/10.1007/JHEP05(2011)038}{JHEP {\bf 05} (2011)  038},
\href{http://arxiv.org/abs/0912.4789}{{\tt arXiv:0912.4789 [hep-th]}}.

\bibitem{Marshakov:2010fx}
A.~Marshakov, A.~Mironov, and A.~Morozov, \emph{{On AGT Relations with Surface
  Operator Insertion and Stationary Limit of Beta-Ensembles}},
  \href{http://dx.doi.org/10.1016/j.geomphys.2011.01.012}{J.Geom.Phys. {\bf 61}
  (2011)  1203--1222},
\href{http://arxiv.org/abs/1011.4491}{{\tt arXiv:1011.4491 [hep-th]}}.

\bibitem{Kozcaz:2010af}
C.~Kozcaz, S.~Pasquetti, and N.~Wyllard, \emph{{A $\&$ B model approaches to
  surface operators and Toda theories}},
  \href{http://dx.doi.org/10.1007/JHEP08(2010)042}{JHEP {\bf 08} (2010)  042},
\href{http://arxiv.org/abs/1004.2025}{{\tt arXiv:1004.2025 [hep-th]}}.

\bibitem{Dimofte:2010tz}
T.~Dimofte, S.~Gukov, and L.~Hollands, \emph{{Vortex Counting and Lagrangian
  3-manifolds}}, \href{http://dx.doi.org/10.1007/s11005-011-0531-8}{Lett. Math.
  Phys. {\bf 98} (2011)  225--287},
\href{http://arxiv.org/abs/1006.0977}{{\tt arXiv:1006.0977 [hep-th]}}.

\bibitem{Maruyoshi:2010iu}
K.~Maruyoshi and M.~Taki, \emph{{Deformed Prepotential, Quantum Integrable
  System and Liouville Field Theory}},
  \href{http://dx.doi.org/10.1016/j.nuclphysb.2010.08.008}{Nucl. Phys. {\bf
  B841} (2010)  388--425},
\href{http://arxiv.org/abs/1006.4505}{{\tt arXiv:1006.4505 [hep-th]}}.

\bibitem{Taki:2010bj}
M.~Taki, \emph{{Surface Operator, Bubbling Calabi-Yau and AGT Relation}},
  \href{http://dx.doi.org/10.1007/JHEP07(2011)047}{JHEP {\bf 07} (2011)  047},
\href{http://arxiv.org/abs/1007.2524}{{\tt arXiv:1007.2524 [hep-th]}}.

\bibitem{Bullimore:2014awa}
M.~Bullimore, H.-C.~Kim, and P.~Koroteev,\emph{{Defects and Quantum Seiberg-Witten Geometry}},
  \href{http://dx.doi.org/10.1007/JHEP05(2015)095}{JHEP {\bf 05} (2015)  095},
\href{http://arxiv.org/abs/1412.6081}{{\tt arXiv:1412.6081 [hep-th]}}.

\bibitem{Assel:2016wcr}
B.~Assel and S.~Sch\"afer-Nameki, \emph{{Six-dimensional Origin of N=4 SYM with Duality Defects}},
\href{http://dx.doi.org/10.1007/JHEP12(2016)058}{JHEP {\bf 12} (2016)  058},
\href{http://arxiv.org/abs/1610.03663}{{\tt arXiv:1610.03663 [hep-th]}}.
  
\bibitem{Nekrasov:2002qd}
N.~Nekrasov, \emph{{Seiberg-Witten prepotential from instanton counting}}, Adv.
  Theor. Math. Phys. {\bf 7} (2004)  831--864,
\href{http://arxiv.org/abs/hep-th/0206161}{{\tt arXiv:hep-th/0206161}}.

\bibitem{Billo:2013fi}
M.~Billo, M.~Frau, L.~Gallot, A.~Lerda, and I.~Pesando, \emph{{Deformed N=2
  theories, generalized recursion relations and S-duality}},
  \href{http://dx.doi.org/10.1007/JHEP04(2013)039}{JHEP {\bf 1304} (2013)
  039},
\href{http://arxiv.org/abs/1302.0686}{{\tt arXiv:1302.0686 [hep-th]}}.

\bibitem{Billo:2013jba}
M.~Billo, M.~Frau, L.~Gallot, A.~Lerda, and I.~Pesando, \emph{{Modular anomaly
  equation, heat kernel and S-duality in $N=2$ theories}},
  \href{http://dx.doi.org/10.1007/JHEP11(2013)123}{JHEP {\bf 1311} (2013)
  123},
\href{http://arxiv.org/abs/1307.6648}{{\tt arXiv:1307.6648 [hep-th]}}.

\bibitem{Minahan:1997if}
J.~Minahan, D.~Nemeschansky, and N.~Warner, \emph{{Instanton expansions for
  mass deformed N=4 superYang-Mills theories}},
  \href{http://dx.doi.org/10.1016/S0550-3213(98)00314-9}{Nucl.Phys. {\bf B528}
  (1998)  109--132}, \href{http://arxiv.org/abs/hep-th/9710146}{{\tt
  arXiv:hep-th/9710146 [hep-th]}}.

\bibitem{Bershadsky:1993ta}
M.~Bershadsky, S.~Cecotti, H.~Ooguri, and C.~Vafa, \emph{{Holomorphic anomalies
  in topological field theories}},
  \href{http://dx.doi.org/10.1016/0550-3213(93)90548-4}{Nucl.Phys. {\bf B405}
  (1993)  279--304},
\href{http://arxiv.org/abs/hep-th/9302103}{{\tt arXiv:hep-th/9302103
  [hep-th]}}.

\bibitem{Witten:1993ed}
E.~Witten, \emph{{Quantum background independence in string theory}},
\href{http://arxiv.org/abs/hep-th/9306122}{{\tt arXiv:hep-th/9306122
  [hep-th]}}.

\bibitem{Aganagic:2006wq}
M.~Aganagic, V.~Bouchard, and A.~Klemm, \emph{{Topological Strings and (Almost)
  Modular Forms}},
  \href{http://dx.doi.org/10.1007/s00220-007-0383-3}{Commun.Math.Phys. {\bf
  277} (2008)  771--819},
\href{http://arxiv.org/abs/hep-th/0607100}{{\tt arXiv:hep-th/0607100
  [hep-th]}}.

\bibitem{Gunaydin:2006bz}
M.~Gunaydin, A.~Neitzke, and B.~Pioline, \emph{{Topological wave functions and
  heat equations}}, \href{http://dx.doi.org/10.1088/1126-6708/2006/12/070}{JHEP
  {\bf 0612} (2006)  070},
\href{http://arxiv.org/abs/hep-th/0607200}{{\tt arXiv:hep-th/0607200
  [hep-th]}}.

\bibitem{Huang:2006si}
M.-x. Huang and A.~Klemm, \emph{{Holomorphic Anomaly in Gauge Theories and
  Matrix Models}}, \href{http://dx.doi.org/10.1088/1126-6708/2007/09/054}{JHEP
  {\bf 0709} (2007)  054},
\href{http://arxiv.org/abs/hep-th/0605195}{{\tt arXiv:hep-th/0605195
  [hep-th]}}.

\bibitem{Grimm:2007tm}
T.~W. Grimm, A.~Klemm, M.~Marino, and M.~Weiss, \emph{{Direct Integration of
  the Topological String}},
  \href{http://dx.doi.org/10.1088/1126-6708/2007/08/058}{JHEP {\bf 0708} (2007)
   058},
\href{http://arxiv.org/abs/hep-th/0702187}{{\tt arXiv:hep-th/0702187
  [HEP-TH]}}.

\bibitem{Huang:2009md}
M.-x. Huang and A.~Klemm, \emph{{Holomorphicity and Modularity in
  Seiberg-Witten Theories with Matter}},
  \href{http://dx.doi.org/10.1007/JHEP07(2010)083}{JHEP {\bf 1007} (2010)
  083},
\href{http://arxiv.org/abs/0902.1325}{{\tt arXiv:0902.1325 [hep-th]}}.

\bibitem{Huang:2010kf}
M.-x. Huang and A.~Klemm, \emph{{Direct integration for general $\Omega$
  backgrounds}},
\href{http://arxiv.org/abs/1009.1126}{{\tt arXiv:1009.1126 [hep-th]}}.

\bibitem{Galakhov:2012gw}
D.~Galakhov, A.~Mironov, and A.~Morozov, \emph{{S-duality as a beta-deformed
  Fourier transform}}, \href{http://dx.doi.org/10.1007/JHEP08(2012)067}{JHEP
  {\bf 1208} (2012)  067},
\href{http://arxiv.org/abs/1205.4998}{{\tt arXiv:1205.4998 [hep-th]}}.

\bibitem{Nemkov:2013qma}
N.~Nemkov, \emph{{S-duality as Fourier transform for arbitrary
  $\epsilon_1,\epsilon_2$}},
\href{http://arxiv.org/abs/1307.0773}{{\tt arXiv:1307.0773 [hep-th]}}.

\bibitem{Billo:2014bja}
M.~Billo, M.~Frau, F.~Fucito, A.~Lerda, J.~Morales, F.~Fucito, R.~Poghossian and D.~Ricci Pacifici,
  \emph{{Modular anomaly equations in $ \mathcal{N} =2^*$ theories and their
  large-$N$ limit}}, \href{http://dx.doi.org/10.1007/JHEP10(2014)131}{JHEP {\bf
  1410} (2014)  131},
\href{http://arxiv.org/abs/1406.7255}{{\tt arXiv:1406.7255 [hep-th]}}.

\bibitem{KashaniPoor:2012wb}
A.-K. Kashani-Poor and J.~Troost, \emph{{The toroidal block and the genus
  expansion}}, \href{http://dx.doi.org/10.1007/JHEP03(2013)133}{JHEP {\bf 1303}
  (2013)  133},
\href{http://arxiv.org/abs/1212.0722}{{\tt arXiv:1212.0722 [hep-th]}}.

\bibitem{Kashani-Poor:2013oza}
A.-K. Kashani-Poor and J.~Troost, \emph{{Transformations of Spherical Blocks}},
\href{http://arxiv.org/abs/1305.7408}{{\tt arXiv:1305.7408 [hep-th]}}.

\bibitem{Kashani-Poor:2014mua}
A.-K. Kashani-Poor and J.~Troost, \emph{{Quantum geometry from the toroidal
  block}}, \href{http://dx.doi.org/10.1007/JHEP08(2014)117}{JHEP {\bf 1408}
  (2014)  117},
\href{http://arxiv.org/abs/1404.7378}{{\tt arXiv:1404.7378 [hep-th]}}.

\bibitem{Ashok:2015cba}
S.~K. Ashok, M.~Billo, E.~Dell'Aquila, M.~Frau, A.~Lerda, and M.~Raman,
  \emph{{Modular anomaly equations and S-duality in $ \mathcal{N}=2 $ conformal
  SQCD}}, \href{http://dx.doi.org/10.1007/JHEP10(2015)091}{JHEP {\bf 10} (2015)
   091},
\href{http://arxiv.org/abs/1507.07476}{{\tt arXiv:1507.07476 [hep-th]}}.

\bibitem{Ashok:2016oyh}
S.~K. Ashok, E.~Dell'Aquila, A.~Lerda, and M.~Raman, \emph{{S-duality, triangle
  groups and modular anomalies in $ \mathcal{N}=2 $ SQCD}},
  \href{http://dx.doi.org/10.1007/JHEP04(2016)118}{JHEP {\bf 04} (2016)  118},
\href{http://arxiv.org/abs/1601.01827}{{\tt arXiv:1601.01827 [hep-th]}}.

\bibitem{Beccaria:2016vxq}
M.~Beccaria, A.~Fachechi, G.~Macorini, and L.~Martina, \emph{{Exact partition
  functions for deformed $\mathcal{N}=2$ theories with $N_{f}=4$ flavours}},
  \href{http://dx.doi.org/10.1007/JHEP12(2016)029}{JHEP {\bf 12} (2016)  029},
\href{http://arxiv.org/abs/1609.01189}{{\tt arXiv:1609.01189 [hep-th]}}.

\bibitem{Billo':2015ria}
M.~Billo, M.~Frau, F.~Fucito, A.~Lerda, and J.~F. Morales, \emph{{S-duality and
  the prepotential in $ \mathcal{N}={2}^{\star } $ theories (I): the ADE
  algebras}}, \href{http://dx.doi.org/10.1007/JHEP11(2015)024}{JHEP {\bf 11}
  (2015)  024},
\href{http://arxiv.org/abs/1507.07709}{{\tt arXiv:1507.07709 [hep-th]}}.

\bibitem{Billo':2015jta}
M.~Billo, M.~Frau, F.~Fucito, A.~Lerda, and J.~F. Morales, \emph{{S-duality and
  the prepotential of $ \mathcal{N}={2}^{\star } $ theories (II): the
  non-simply laced algebras}},
  \href{http://dx.doi.org/10.1007/JHEP11(2015)026}{JHEP {\bf 11} (2015)  026},
\href{http://arxiv.org/abs/1507.08027}{{\tt arXiv:1507.08027 [hep-th]}}.

\bibitem{Beccaria:2016nnb}
M.~Beccaria and G.~Macorini, \emph{{Exact partition functions for the
  Ω-deformed $ \mathcal{N}={2}^{\ast } $ SU(2) gauge theory}},
  \href{http://dx.doi.org/10.1007/JHEP07(2016)066}{JHEP {\bf 07} (2016)  066},
\href{http://arxiv.org/abs/1606.00179}{{\tt arXiv:1606.00179 [hep-th]}}.

\bibitem{Ashok:2016ewb}
S.~K. Ashok, M.~Billo, E.~Dell'Aquila, M.~Frau, A.~Lerda, M.~Moskovic, and
  M.~Raman, \emph{{Chiral observables and S-duality in N = 2* U(N) gauge
  theories}}, \href{http://dx.doi.org/10.1007/JHEP11(2016)020}{JHEP {\bf 11}
  (2016)  020},
\href{http://arxiv.org/abs/1607.08327}{{\tt arXiv:1607.08327 [hep-th]}}.

\bibitem{Kanno:2011fw}
H.~Kanno and Y.~Tachikawa, \emph{{Instanton counting with a surface operator
  and the chain-saw quiver}},
  \href{http://dx.doi.org/10.1007/JHEP06(2011)119}{JHEP {\bf 06} (2011)  119},
\href{http://arxiv.org/abs/1105.0357}{{\tt arXiv:1105.0357 [hep-th]}}.

\bibitem{mehta1980}
V.~Mehta and C.~Seshadri, \emph{Moduli of vector bundles on curves with
  parabolic structures}, Mathematische Annalen {\bf 248}  205.

\bibitem{biswas1997}
I.~Biswas,
  \href{http://dx.doi.org/10.1215/S0012-7094-97-08812-8}{\emph{{Parabolic
  bundles as orbifold bundles}},Duke Math. J. 06 305--325}.

\bibitem{Feigin}
B.~Feigin, M.~Finkelberg, A.~Negut, and L.~Rybnikov, \emph{Yangians and cohomology rings of
Laumon spaces},
\href{http://dx.doi.org/10.1007/s00029-011-0059-x}{Selecta Mathematica {\bf 17}
 (2008) 1}, \href{http://arxiv.org/abs/0812.4656}{{\tt arXiv:0812.4656 [math.AG]}}.

\bibitem{Douglas:1995bn}
M.~R. Douglas, \emph{{Branes within branes}},
\href{http://arxiv.org/abs/hep-th/9512077}{{\tt arXiv:hep-th/9512077}}.

\bibitem{Billo:2002hm}
M.~Billo, M.~Frau, I.~Pesando, F.~Fucito, A.~Lerda, and A.~Liccardo,
  \emph{{Classical gauge instantons from open strings}}, JHEP {\bf 02} (2003)
  045,
\href{http://arxiv.org/abs/hep-th/0211250}{{\tt arXiv:hep-th/0211250}}.

\bibitem{Nekrasov:2009rc}
N.~Nekrasov and S.~Shatashvili, \emph{{Quantization of Integrable Systems and
  Four Dimensional Gauge Theories}},
\href{http://arxiv.org/abs/0908.4052}{{\tt arXiv:0908.4052 [hep-th]}}.

\bibitem{Frenkel:2015rda}
E.~Frenkel, S.~Gukov, and J.~Teschner, \emph{{Surface Operators and Separation
  of Variables}}, \href{http://dx.doi.org/10.1007/JHEP01(2016)179}{JHEP {\bf
  01} (2016)  179},
\href{http://arxiv.org/abs/1506.07508}{{\tt arXiv:1506.07508 [hep-th]}}.

\bibitem{Moore:1998et}
G.~W. Moore, N.~Nekrasov, and S.~Shatashvili, \emph{{D-particle bound states
  and generalized instantons}},
  \href{http://dx.doi.org/10.1007/s002200050016}{Commun. Math. Phys. {\bf 209}
  (2000)  77--95},
\href{http://arxiv.org/abs/hep-th/9803265}{{\tt arXiv:hep-th/9803265}}.

\bibitem{Bonelli:2013mma}
G.~Bonelli, A.~Sciarappa, A.~Tanzini, and P.~Vasko, \emph{{Vortex partition
  functions, wall crossing and equivariant Gromov-Witten invariants}},
  \href{http://dx.doi.org/10.1007/s00220-014-2193-8}{Commun. Math. Phys. {\bf
  333} (2015) no.~2, 717--760},
\href{http://arxiv.org/abs/1307.5997}{{\tt arXiv:1307.5997 [hep-th]}}.

\bibitem{Nawata:2014nca}
S.~Nawata, \emph{{Givental J-functions, Quantum integrable systems, AGT
  relation with surface operator}},
  \href{http://dx.doi.org/10.4310/ATMP.2015.v19.n6.a4}{Adv. Theor. Math. Phys.
  {\bf 19} (2015)  1277--1338},
\href{http://arxiv.org/abs/1408.4132}{{\tt arXiv:1408.4132 [hep-th]}}.

\bibitem{Doroud:2012xw}
N.~Doroud, J.~Gomis, B.~Le~Floch, and S.~Lee, \emph{{Exact Results in D=2
  Supersymmetric Gauge Theories}},
  \href{http://dx.doi.org/10.1007/JHEP05(2013)093}{JHEP {\bf 05} (2013)  093},
\href{http://arxiv.org/abs/1206.2606}{{\tt arXiv:1206.2606 [hep-th]}}.

\bibitem{Benini:2012ui}
F.~Benini and S.~Cremonesi, \emph{{Partition Functions of ${\mathcal{N}=(2,2)}$
  Gauge Theories on S$^{2}$ and Vortices}},
  \href{http://dx.doi.org/10.1007/s00220-014-2112-z}{Commun. Math. Phys. {\bf
  334} (2015) no.~3, 1483--1527},
\href{http://arxiv.org/abs/1206.2356}{{\tt arXiv:1206.2356 [hep-th]}}.

\bibitem{Billo:2006jm}
M.~Billo, M.~Frau, F.~Fucito, and A.~Lerda, \emph{{Instanton calculus in R-R
  background and the topological string}}, JHEP {\bf 11} (2006)  012,
\href{http://arxiv.org/abs/hep-th/0606013}{{\tt arXiv:hep-th/0606013}}.

\bibitem{Wyllard:2010vi}
N.~Wyllard, \emph{{Instanton partition functions in N=2 SU(N) gauge theories
  with a general surface operator, and their W-algebra duals}},
  \href{http://dx.doi.org/10.1007/JHEP02(2011)114}{JHEP {\bf 02} (2011)  114},
\href{http://arxiv.org/abs/1012.1355}{{\tt arXiv:1012.1355 [hep-th]}}.

\bibitem{Awata:2010bz}
H.~Awata, H.~Fuji, H.~Kanno, M.~Manabe, and Y.~Yamada, \emph{{Localization with
  a Surface Operator, Irregular Conformal Blocks and Open Topological String}},
  \href{http://dx.doi.org/10.4310/ATMP.2012.v16.n3.a1}{Adv. Theor. Math. Phys.
  {\bf 16} (2012) no.~3, 725--804},
\href{http://arxiv.org/abs/1008.0574}{{\tt arXiv:1008.0574 [hep-th]}}.

\bibitem{Wyllard:2010rp}
N.~Wyllard, \emph{{W-algebras and surface operators in N=2 gauge theories}},
  \href{http://dx.doi.org/10.1088/1751-8113/44/15/155401}{J. Phys. {\bf A44}
  (2011)  155401},
\href{http://arxiv.org/abs/1011.0289}{{\tt arXiv:1011.0289 [hep-th]}}.

\bibitem{Witten:1993yc}
E.~Witten, \emph{{Phases of N=2 theories in two-dimensions}},
  \href{http://dx.doi.org/10.1016/0550-3213(93)90033-L}{Nucl. Phys. {\bf B403}
  (1993)  159--222},
\href{http://arxiv.org/abs/hep-th/9301042}{{\tt arXiv:hep-th/9301042
  [hep-th]}}.

\bibitem{Hanany:1997vm}
A.~Hanany and K.~Hori, \emph{{Branes and N=2 theories in two-dimensions}},
  \href{http://dx.doi.org/10.1016/S0550-3213(97)00754-2}{Nucl. Phys. {\bf B513}
  (1998)  119--174},
\href{http://arxiv.org/abs/hep-th/9707192}{{\tt arXiv:hep-th/9707192
  [hep-th]}}.

\bibitem{D'Hoker:1996nv}
E.~D'Hoker, I.~M. Krichever, and D.~H. Phong, \emph{{The effective prepotential
  of N = 2 supersymmetric SU(N(c)) gauge theories}},
  \href{http://dx.doi.org/10.1016/S0550-3213(97)00035-7}{Nucl. Phys. {\bf B489}
  (1997)  179--210},
\href{http://arxiv.org/abs/hep-th/9609041}{{\tt arXiv:hep-th/9609041}}.

\bibitem{Naculich:2002hi}
S.~G. Naculich, H.~J. Schnitzer, and N.~Wyllard, \emph{{The N = 2 U(N) gauge
  theory prepotential and periods from a perturbative matrix model
  calculation}}, \href{http://dx.doi.org/10.1016/S0550-3213(02)01120-3}{Nucl.
  Phys. {\bf B651} (2003)  106--124},
\href{http://arxiv.org/abs/hep-th/0211123}{{\tt arXiv:hep-th/0211123
  [hep-th]}}.

\bibitem{Nekrasov:2012xe}
N.~Nekrasov and V.~Pestun, \emph{{Seiberg-Witten geometry of four dimensional
  N=2 quiver gauge theories}},
\href{http://arxiv.org/abs/1211.2240}{{\tt arXiv:1211.2240 [hep-th]}}.

\bibitem{Fucito:2011pn}
F.~Fucito, J.~Morales, D.~R. Pacifici, and R.~Poghossian, \emph{{Gauge theories
  on $\Omega$-backgrounds from non commutative Seiberg-Witten curves}},
  \href{http://dx.doi.org/10.1007/JHEP05(2011)098}{JHEP {\bf 1105} (2011)
  098},
\href{http://arxiv.org/abs/1103.4495}{{\tt arXiv:1103.4495 [hep-th]}}.

\bibitem{Fucito:2012xc}
F.~Fucito, J.~F. Morales, and D.~R. Pacifici, \emph{{Deformed Seiberg-Witten
  Curves for ADE Quivers}},
  \href{http://dx.doi.org/10.1007/JHEP01(2013)091}{JHEP {\bf 1301} (2013)
  091},
\href{http://arxiv.org/abs/1210.3580}{{\tt arXiv:1210.3580 [hep-th]}}.

\bibitem{Beccaria:2017rfz}
M.~Beccaria, A.~Fachechi, and G.~Macorini, \emph{{Chiral trace relations in
  $\Omega$-deformed $\mathcal N=2$ theories}},
\href{http://arxiv.org/abs/1702.01254}{{\tt arXiv:1702.01254 [hep-th]}}.

\bibitem{Donagi:1995cf}
R.~Donagi and E.~Witten, \emph{{Supersymmetric Yang-Mills theory and integrable
  systems}}, \href{http://dx.doi.org/10.1016/0550-3213(95)00609-5}{Nucl. Phys.
  {\bf B460} (1996)  299--334},
\href{http://arxiv.org/abs/hep-th/9510101}{{\tt arXiv:hep-th/9510101
  [hep-th]}}.

\bibitem{DHoker:1997hut}
E.~D'Hoker and D.~H. Phong, \emph{{Calogero-Moser systems in SU(N)
  Seiberg-Witten theory}},
  \href{http://dx.doi.org/10.1016/S0550-3213(97)00763-3}{Nucl. Phys. {\bf B513}
  (1998)  405--444},
\href{http://arxiv.org/abs/hep-th/9709053}{{\tt arXiv:hep-th/9709053
  [hep-th]}}.

\bibitem{Gaiotto:2011tf}
D.~Gaiotto, G.~W. Moore, and A.~Neitzke, \emph{{Wall-Crossing in Coupled 2d-4d
  Systems}}, \href{http://dx.doi.org/10.1007/JHEP12(2012)082}{JHEP {\bf 12}
  (2012)  082},
\href{http://arxiv.org/abs/1103.2598}{{\tt arXiv:1103.2598 [hep-th]}}.

\bibitem{Bonelli:2011fq}
G.~Bonelli, A.~Tanzini, and J.~Zhao, \emph{{Vertices, Vortices and Interacting
  Surface Operators}}, \href{http://dx.doi.org/10.1007/JHEP06(2012)178}{JHEP
  {\bf 06} (2012)  178},
\href{http://arxiv.org/abs/1102.0184}{{\tt arXiv:1102.0184 [hep-th]}}.

\bibitem{Bonelli:2011wx}
G.~Bonelli, A.~Tanzini, and J.~Zhao, \emph{{The Liouville side of the Vortex}},
  \href{http://dx.doi.org/10.1007/JHEP09(2011)096}{JHEP {\bf 09} (2011)  096},
\href{http://arxiv.org/abs/1107.2787}{{\tt arXiv:1107.2787 [hep-th]}}.

\bibitem{Gomis:2014eya}
J.~Gomis and B.~Le~Floch, \emph{{M2-brane surface operators and gauge theory
  dualities in Toda}}, \href{http://dx.doi.org/10.1007/JHEP04(2016)183}{JHEP
  {\bf 04} (2016)  183},
\href{http://arxiv.org/abs/1407.1852}{{\tt arXiv:1407.1852 [hep-th]}}.

\bibitem{Gomis:2016ljm}
J.~Gomis, B.~Le~Floch, Y.~Pan, and W.~Peelaers, \emph{{Intersecting Surface
  Defects and Two-Dimensional CFT}},
\href{http://arxiv.org/abs/1610.03501}{{\tt arXiv:1610.03501 [hep-th]}}.

\bibitem{Pan:2016fbl}
Y.~Pan, and W.~Peelaers, \emph{{Intersecting Surface Defects and Instanton 
Partition Functions}},
\href{http://arxiv.org/abs/1612.04839}{{\tt arXiv:1612.04839 [hep-th]}}.

\bibitem{Okuda:2010ke}
T.~Okuda and V.~Pestun, \emph{{On the instantons and the hypermultiplet mass of
  N=2* super Yang-Mills on $S^{4}$}},
  \href{http://dx.doi.org/10.1007/JHEP03(2012)017}{JHEP {\bf 1203} (2012)
  017},
\href{http://arxiv.org/abs/1004.1222}{{\tt arXiv:1004.1222 [hep-th]}}.
\end{thebibliography}
\end{document}